\documentclass[twocolumn]{aastex62}

\usepackage[varg]{txfonts}
\usepackage{natbib}
\usepackage{morefloats} 
\usepackage[utf8]{inputenc}
\usepackage{ textcomp }
\usepackage{ gensymb }  
\usepackage{hyperref}   
\usepackage{float}
\usepackage{multirow}
\usepackage{blindtext}
\usepackage{enumitem}
\usepackage{scrextend}

\begin{document}

	\title{On the occurrence of thermal non-equilibrium in coronal loops}

	\author[0000-0001-5315-2890]{C.~Froment}    
	\affiliation{Rosseland Centre for Solar Physics, University of Oslo, P.O. Box 1029 Blindern, NO-0315 Oslo, Norway} 	
	\affiliation{Institute of Theoretical Astrophysics, University of Oslo, P.O Box 1029, Blindern, NO-0315, Oslo, Norway}
    \newcommand{\iasaffil}{Institut d'Astrophysique Spatiale, CNRS, Univ. Paris-Sud, Universit\'e Paris-Saclay, B\^at. 121, 91405 Orsay, France}
	\affiliation{\iasaffil}

	\author[0000-0003-0972-7022]{F.~Auch\`ere}
	\affiliation{\iasaffil}

	\author[0000-0002-3164-930X]{Z.~Miki\'c}
	\affiliation{Predictive Science, Inc., San Diego, CA 92121, USA}
	
	\author[0000-0001-5810-1566]{G.~Aulanier}
	\affiliation{LESIA, Observatoire de Paris, PSL Research University, CNRS, Sorbonne Universit\'es, UPMC Univ. Paris 06, Univ. Paris Diderot, Sorbonne Paris Cit\'e, 5 place Jules Janssen, F-92195 Meudon, France}
	
	\author[0000-0001-9426-8558]{K.~Bocchialini}
	\affiliation{\iasaffil}
	
	\author[0000-0003-4290-1897]{E.~Buchlin}
	\affiliation{\iasaffil}
	
	\author{J.~Solomon}
	\affiliation{\iasaffil}
	
	\author[0000-0001-9295-1863 ]{E. Soubri\'e}
	\affiliation{\iasaffil}
	\affiliation{Institute of Applied Computing \& Community Code, Universitat de les Illes Balears, 07122 Palma de Mallorca, Spain}
	
	\email{clara.froment@astro.uio.no} 
	
	\received{January 11, 2018}
	\revised{February 09, 2018}
	\accepted{February 11, 2018}
	\submitjournal{The Astrophysical Journal}
	
\begin{abstract}

Long-period EUV pulsations, recently discovered to be common in active regions, are understood to be the coronal manifestation of thermal non-equilibrium (TNE). The active regions previously studied with EIT/SOHO and AIA/SDO indicated that long-period intensity pulsations are localized in only one or two loop bundles. The basic idea of this study is to understand why. For this purpose, we tested the response of different loop systems, using different magnetic configurations, to different stratifications and strengths of the heating. We present an extensive parameter-space study using 1D hydrodynamic simulations (1,020 in total) and conclude that the occurrence of TNE requires specific combinations of parameters.
Our study shows that the TNE cycles are confined to specific ranges in parameter space. This naturally explains why only some loops undergo constant periodic pulsations over several days: since the loop geometry and the heating properties generally vary from one loop to another in an active region, only the ones in which these parameters are compatible exhibits TNE cycles. Furthermore, these parameters (heating and geometry) are likely to vary significantly over the duration of a cycle, which potentially limits the possibilities of periodic behavior. This study also confirms that long-period intensity pulsations and coronal rain are two aspects of the same phenomenon: both phenomena can occur for similar heating conditions and can appear simultaneously in the simulations.

\end{abstract}
	
	
\section{Introduction}   

Solving the coronal heating problem remains one of the biggest challenge in astrophysics. How can the tenuous plasma that constitutes the highest layer of the solar atmosphere be maintained at temperatures two orders of magnitude higher than that of the solar surface? One of the fundamental facets of this problem is to determine the spatial and temporal distribution of the heating.

Thermal non-equilibrium (TNE) is a phenomenon that can occur in the solar atmosphere when the heating is highly-stratified \citep[e.g.,][]{mendoza-briceno2005, karpen2008,mok2008, susino2010, antolin2010, lionello2013, mok2016}. 
This particular localization of the heating produces chromospheric evaporative upflows that supply the coronal structure with dense and hot material. A thermal runaway is eventually triggered when the radiative losses overcome the limited heating at coronal heights. Condensations are formed locally in the corona and fall down to the loop footpoints along the magnetic field lines. 
Furthermore, if the heating is quasi-steady, i.e. with a high heating frequency compared to the typical cooling time, this phenomenon can be cyclic. Such a system has no existing thermal equilibrium and will undergo evaporation and condensation cycles with periods of a few hours. This highly nonlinear behavior is what we call TNE. The limit cycle solutions in coronal loops were first explored by \citet{kuin1982}. 

TNE has received an increasing interest in the recent years. The thermal runaway triggered by a local excess of density and leading to cool condensations is one of the standard explanations for the existence of cool materials in the corona. Such catastrophic cooling events can, for example, end up in the formation of prominences \citep{antiochos1991,antiochos1999,antiochos2000, karpen2006, xia_formation_2011,xia_simulating_2014} or coronal rain \citep{schrijver2001, degroof2004, degroof2005, muller2003,muller2004,muller2005,antolin2010, antolin2012, vashalomidze2015}. 
Coronal rain is widely observed in off-limb active regions \citep[e.g.,][]{antolin_observing_2012,antolin2015}, even if a proper quantification of the proportion of loops experiencing episodes of coronal rain is still lacking.

Nevertheless, the widespread existence of TNE in loops and, consequently, the widespread contribution of quasi-steady footpoint heating, has been questioned \citep{klimchuk2010}.
However, recent modeling studies have shown that the role of TNE in the dynamics of loops may need to be revisited \citep{lionello2013, mikic2013, winebarger2014, lionello2016, mok2016}. 
These modeling studies support the idea that TNE is probably common in coronal loops, with two main types of condensations involved. 
\citet{mikic2013} in particular suggest that different regimes of TNE cycles could exist in loops. They differentiate cycles with complete condensations (CCs) where the temperature, locally in the corona, drops to chromospheric temperatures to form dense (up to $\sim10^{17}~\mathrm{m}^{-3}$) and cool blobs, related to the observed coronal rain, and cycles with incomplete condensations (ICs). For this other regime of TNE, the temperature stays at coronal temperatures, and the density remains relatively low ($\sim 5 \times10^{15}~\mathrm{m}^{-3}$). These two different regimes of evaporation and condensation cycles are obtained with different combinations of parameters of the loop geometry and of the heating strength and spatial distribution.

The early statistical study of long-period intensity pulsations by \citet{auchere2014} brings new impetus to this debate.
Using 13 years of data in the 195~\AA~channel of the Extreme-ultraviolet Imaging Telescope \citep[EIT;][]{eit1995} on board the \textit{Solar and Heliospheric Observatory} \citep[SOHO;][]{soho1995}, the authors found that at least half of the active regions likely undergo these intensity pulsations with periods ranging from 2 to 16~hours. 
In particular, these pulsations are very common in coronal loops. They have also been observed with the coronal channels of the Atmospheric Imaging Assembly \citep[AIA;][]{boerner2012sdo,lemen2012sdo} on board the \textit{Solar Dynamics Observatory} \citep[SDO;][] {pesnell2012sdo}, during the six first years of the AIA archive \citep{froment_pulsations_2016}. 

\citet{froment2015} studied three examples of such events in detail, with respectively periods of 3.8, 5.6 and 9.0 hours. The authors concluded that these events are related to TNE cycles. This study was focused on the thermal structure evolution of these loops, using simultaneously the six coronal passbands of AIA. The authors used both Differential Emission Measure (DEM), with diagnostics developed by \citet{guennou2012_1, guennou2012_2, guennou2013}, and time lag analysis, using the same method as presented in \citet{viall&klimchuk2012}. \citet{auchere2016a} have recently critically re-examined and confirmed the statistical significance of the detections used in \citet{froment2015}.
Furthermore, the pulse-train nature of the observed signals highlighted by \citet{auchere2016b} reinforced the conclusion that TNE is the cause of the long-period intensity pulsations observed.
Finally, we recently conducted a modeling study in order to compare directly the simulations with these observations. In this 1D hydrodynamic simulation study \citep[hereafter Paper~I]{froment_long-period_2017}, we showed that with a highly-stratified and steady heating we can reproduce the main characteristics (average behavior integrated along the line-of-sight and long-term temporal variations) of the thermal behavior of the loop bundle of event 1 studied in \citet{froment2015}.

As already mentioned in \citet{froment2015}, in the active regions where long-period intensity pulsations are detected, only one loop bundle (or in some rare  cases two) shows this kind of behavior. The automatic detection algorithm used may have missed events due to the rather strict detection thresholds that we used. It is also likely that some events are missed by the Fourier detection if they are not strictly periodic. Furthermore the background and foreground emission could mask the time-varying signal in certain cases. However, some properties, such as the geometry of the magnetic field lines and the heating characteristics, could favor the TNE cycles for only some loop bundles.
In the present paper we explore the sensitivity of TNE occurrence as a function of the heating strength and stratification, for several loop geometries.

The heating parameters used in the simulation presented in Paper~I have been chosen among hundreds of heating combinations tested for a single loop geometry, which corresponds to a pulsating loop bundle observed with AIA. We present here an extensive parameter-space study which justifies this particular choice. To extend our analysis, we also pick another loop geometry, corresponding to a non-pulsating loop from a neighboring region, and use a semi-circular one as a test sample to do the same analysis and thus test the influence of the loop geometry.

This parameter-space exploration is presented in Section~\ref{scan}. In Section~\ref{discussion} we discuss our results regarding the occurrence of TNE cycles within the parameter space explored. Then, in Section~\ref{sec:comparaison_obs} we examine the properties of the different types of loop systems produced in our simulations (with TNE or not) in regard of the observed loops properties. Finally, we summarize our results in Section~\ref{summary}.

\section{Parameter-space scan}\label{scan}	

For this parameter-space study, we use the same 1D hydrodynamic code as in \citet{mikic2013} and in Paper~I.
The 1D description is particularly suited for this kind of study as multiple configurations of loops can be easily tested. The loop geometries used in these simulations, except for one loop (loop~A, see Section~\ref{sec:loop_geometry}), are from the Linear Force-Free (LFFF) extrapolations presented in Section~2.1 of Paper~I. These extrapolations are made using magnetograms from the Helioseismic and Magnetic Imager \citep[HMI;][]{scherrer2012} corresponding to the active regions of event~1 of \citet{froment2015}, i.e. NOAA~AR~11499, and NOAA~AR~11501, a smaller adjacent active region. Some selected extrapolated field lines are presented in Figure~\ref{fig:carte_extrapol}. We detected intensity pulsations (with a period of 9.0~hr) in a large loop bundle of NOAA~AR~11499, with a very clear signal; the probability that this detection was caused by noise is $1.7 \times 10^{-8}$ \citep{auchere2016a}. This area is delineated by the orange contour in Figure~\ref{fig:carte_extrapol}.

In the field-of-view presented in Figure~\ref{fig:carte_extrapol}, outside of the orange-contoured bundle, no other loop bundle shows a long-period pulsating behavior. The evolution of the pulsating loops has been followed from June 03, 2012 18:00~UT to June 10, 2012 04:29~UT using AIA data.  

	\subsection{Method and parameters explored}   
	
We choose to focus on three different loop geometries and to scan various heating configurations for these loops.	
In addition of the loop geometry that matches the pulsating loop bundle already used in Paper~I (noted here as loop~B), we use a semi-circular geometry as a control sample (loop~A) and we picked another loop geometry from the LFFF extrapolation that matches with a non-pulsating loop bundle as observed with AIA (loop~C). The field lines corresponding to loop~B and loop~C, and matching visible loops bundles in the 171~\AA~AIA image, are indicated in Figure~\ref{fig:carte_extrapol}. 
Loop~A is an ad-hoc loop and is therefore not present in the field-of-view.

The 1D hydrodynamic simulations are made using the same initial conditions and assumptions as in Paper~I (see description in Section~2.2). The only difference here is that instead of using the Spitzer thermal conductivity as in Paper~I, we use an option present in the code \citep{mikic2013} to artificially broaden the transition region at low temperatures. This modification of $\kappa_{\parallel}$ allows to reduce the steep gradients below a cut-off temperature (chosen as $T_c = 250,000$~K here) with a minimal effect on the coronal solutions, as described in \citet{lionello2009} and \citet{mikic2013}. In that way we can afford to use bigger mesh cells and thus fewer mesh points, i.e. 10,000 points, than in Paper I. This technique is particularity suited for this study, since we wish to scan a large area of the space of parameters. Some of the runs presented in this paper were repeated with the classic, unmodified, Spitzer conductivity using more mesh points (typically 100,000 points). 
It is the case for the particular simulations that we choose to present in detail in Section~\ref{sec:comparaison_obs} (see Figure~\ref{fig:detailed_simu}) and some examples presented in \citet{froment_pulsations_2016}.	
The overall pattern is not affected by this technique but some differences related to the precise nature of condensations can appear between the simulations using the Spitzer thermal conductivity and the ones using modified conductivity. 

\medskip

We choose a simple heating function that can be tuned with three free parameters. This heating function is the same as the one used in Paper~I (see Equation~2):
\begin{equation}
H(s) = H_0 +  H_1(e^{-g(s)/\lambda_{1}}+e^{-g(L-s)/\lambda_{2}})   
\label{eq:heating_model}
\end{equation}

where $g(s)=\mathrm{max}(s-\Delta,0)$  and $\Delta=5$~Mm is the thickness of the chromosphere, where the heating is constant.

$H(s)$ is the volumetric heating rate, expressed in $\mathrm{W~m}^{-3}$. 
$H_0$ is the value to which $H(s)$ tends at the apex and $(H_0+H_1)$ is the value of the heating in the chromosphere.
$\lambda_1$ and $\lambda_2$ are the scale lengths for the energy deposition at the eastern and western leg of the loop, respectively.

For each loop geometry we test several values of $H_1$, $\lambda_1$ and $\lambda_2$. We choose to fix the $H_0$ value for each loop geometry, to a value that allows us to obtain a static loop (we set $H_1=0$ to use a uniform heating) with an apex temperature around 1~MK. 

For each loop geometry, we explored values of $H_1$ in increments of a factor of 2. In that way we can easily compare the simulations. Note that the value of the factor itself is arbitrary. For each value of $H_1$ explored, we test a large set of combinations of $\lambda_1$ and $\lambda_2$, specified in percentage of the total length $L$ of each loop. The scan cube is then $(H_1, \lambda_1, \lambda_2)$.

For each simulation, we can define the heat flux, i.e. the total heat that the loop receives over its length, normalized to the first loop footpoint cross-section area :

\begin{equation}
\label{eq:int_heating_power}
 Q_0= \frac{1}{A_0} \int_{0}^{L} H(s) \times A(s)\, ds  \hspace{1cm}[\mathrm{W} \, \mathrm{m}^{-2}]
\end{equation}

with $A(s)$ the cross-sectional area of the loop which is~$\sim 1/B(s)$.
As we will see in Section~\ref{sec:loop_geometry} (Figures~\ref{fig:geometries_a} and~\ref{fig:geometries_b_c}), the magnetic field strength is similar at both end of each loop geometry chosen. The heat flux normalised at the second footpoint, $Q_1$, is thus always similar to $Q_0$. We thus only consider this latest value.
The $Q_0$ heat flux value does not give information about the asymmetry of the heating. However, looking at Equations~\ref{eq:heating_model} and \ref{eq:int_heating_power}, we can see that the heating in a leg will be dominant when the scale height of energy deposition is larger, and the more the loop area expansion is important in this leg. 

It is worth noting that the way we choose to explore the parameter space of the heating strength and stratification implies that $Q_0$ changes between two slices through the scan cube, i.e. between simulations with the same stratification ($\lambda_1$, $\lambda_2$) but with different values of the heating at the footpoints ($H_1$). Indeed, $Q_0$ changes from one simulation to another. We could have chosen a different parametrization, fixing $Q_0$ instead of $H_1$ for each exploration of $\lambda_1$ and $\lambda_2$. However, we confirmed that this different parametrization did not change our conclusions. In addition, whatever the parametrization, since  $Q_0$ and $H_1$ are linked, the results can be explored at $Q_0$ constant. 
We tested as well an alternate way of creating asymmetry in the heating along the loop by applying a different amount of heating at each footpoint. We verified that this did not change the conclusions of this paper either.

	\subsection{Loop geometries} \label{sec:loop_geometry}
		   
	
	\begin{figure}
		\centering
                 \includegraphics[width=\linewidth]{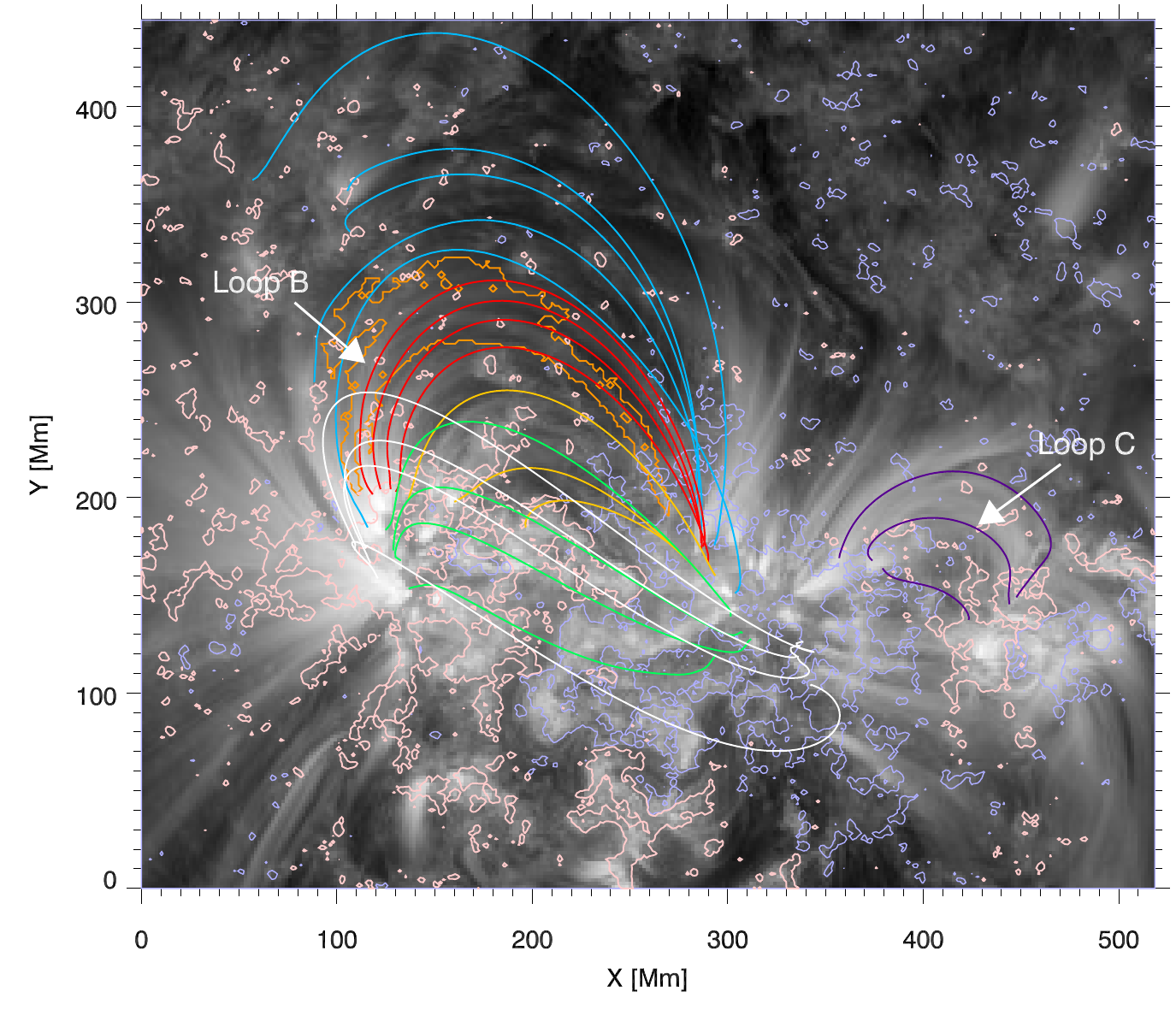} 
                 \caption{Some field lines extrapolated for NOAA AR 11499 and 11501 with a LFFF model. Contours of magnetic field ($B_z$ at $z = 0$) from the HMI magnetogram are in light red for positive values and in light blue for negative ones for $\pm$30 G. The AIA 171~\AA~image and the magnetogram are both taken on June 06, 2012 at 23:12 UT. The orange contour delimits the area of the pulsations (9.0~hr of period) detected in the 335~\AA~passband of AIA \citep[see Figure~4 in][]{froment2015}, for a sequence of data between June 05, 2012 11:14~UT and June 08, 2012 11:16~UT. The red field lines match this contour. The white arrows indicate  the loops~B and C chosen for the parameter-space study. See Section~2.1 of Paper I for further details regarding these extrapolations.}
                 \label{fig:carte_extrapol}
	\end{figure}
	


	\begin{figure}
		\centering

                 \includegraphics[width=\linewidth, trim= 0.8cm 0cm 0cm 0cm]{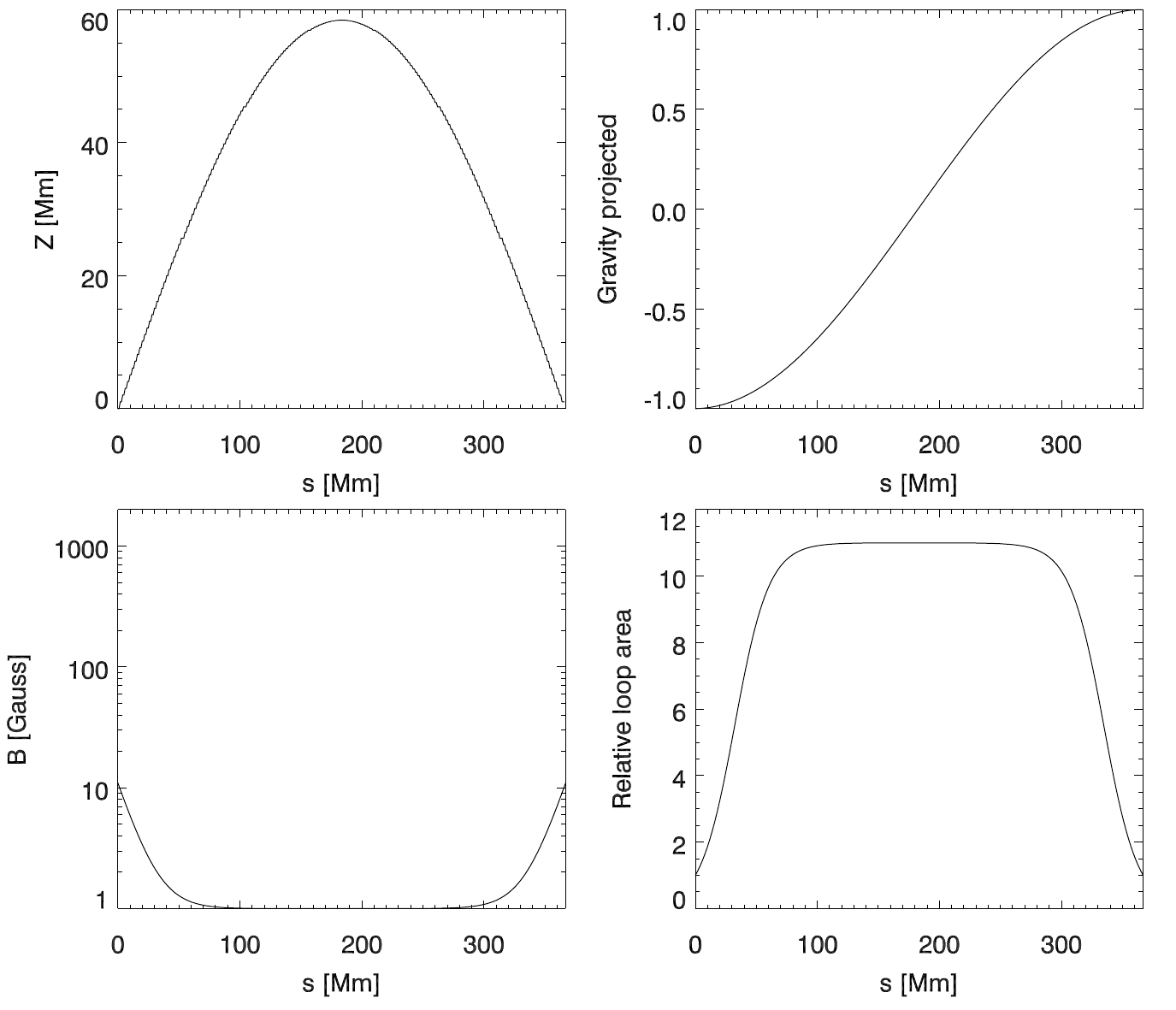}
	        
                 \caption{Geometry profiles for loop~A. This loop geometry is not from the LFFF extrapolations. Top left: loop profile, altitude in Mm of each point along the loop. Top right: the normalized acceleration of the gravity projected along the loop (see Equation~1 of Paper~I). Bottom left: the strength of the magnetic field $B(s)$ in Gauss. Bottom right: the loop expansion given by the evolution of the cross-sectional area A(s), normalized to the first footpoint.}
                 \label{fig:geometries_a}
	\end{figure}    


	\begin{figure}
		\centering
		
		\gridline{\fig{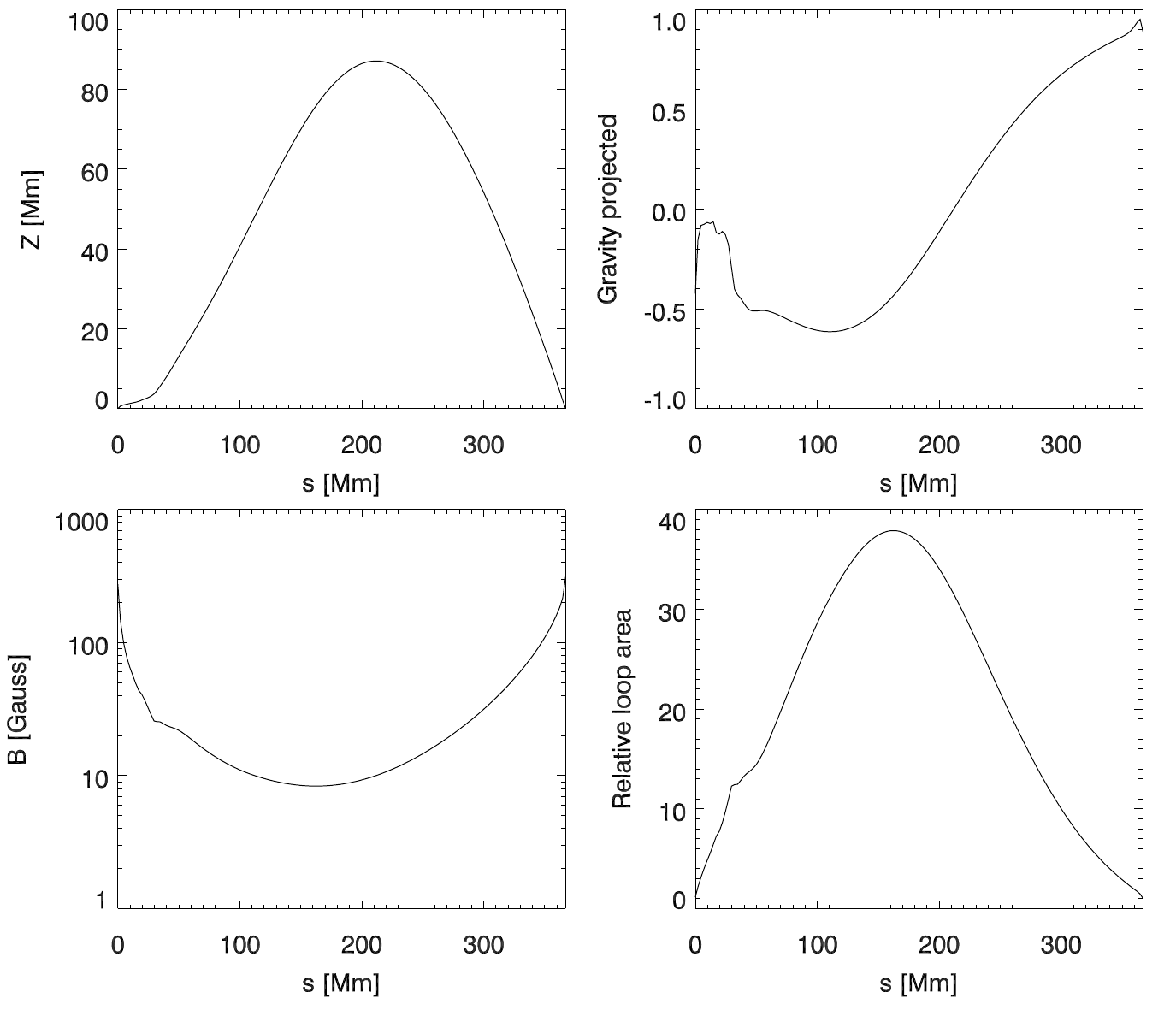}{0.5\textwidth}{Loop~B}
			    }
         	\gridline{\fig{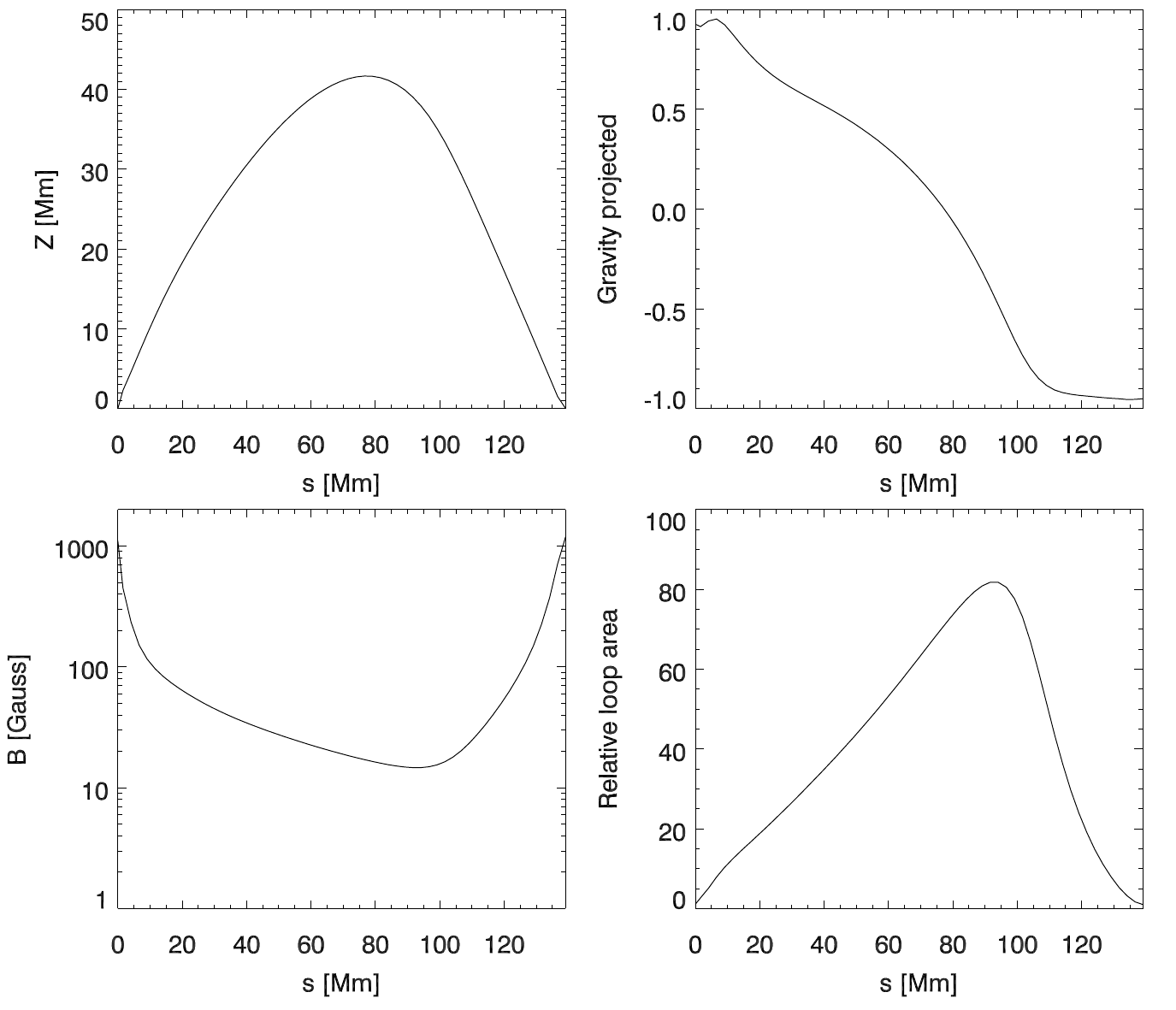}{0.5\textwidth}{Loop~C}
			      }
	
                 \caption{Geometry profiles for loops~B and~C. These loop geometries are from the LFFF extrapolations of NOAA AR 11499 and 11501, introduced in Paper~I and presented in Figure~\ref{fig:carte_extrapol}. These two loops are indicated by white arrows in Figure~\ref{fig:carte_extrapol}. For each panel, same as Figure~\ref{fig:geometries_a}.}
                 \label{fig:geometries_b_c}
	\end{figure}    

In Figure~\ref{fig:geometries_a}, we present the loop geometry of loop~A, the test case semi-circular geometry we use for the first set of simulations.
In Figure~\ref{fig:geometries_b_c}, we present the loop geometry of loops~B and~C which are from two field lines extracted from the LFFF extrapolations of the active regions presented in Figure~\ref{fig:carte_extrapol}.
These loop geometries, even if from single field-lines, are selected to model the behavior of the corresponding loop bundles. When we compare the simulations with observations, a simulated loop represents the average behavior of a loop bundle, whose individual threads all have similar properties.

In these figures, we show the quantities that are input to the simulations, i.e., the loop profile (the altitude of each point of the loop), the gravity projected along the loop (see Equation~1 of Paper~I), the magnetic field strength along the loop and the loop area $A(s)$, normalized to the first footpoint. 
For both loops, $s=0$ corresponds to the eastern footpoint.

These three loop  geometries have the following characteristics:
\paragraph{Loop~A} 
This loop is a semi-circular loop  with the same length as loop~B, i.e $L=367 \, \mathrm{Mm}$. The magnetic field along the loop is given by:

\begin{equation}
	\label{eq:mag_field_model}
	B(s) = B_0 +  B_1(e^{-s/l}+e^{-(L-s)/l}) 
\end{equation}

with $B_0 = 1 \, \mathrm{G}$, $B_1 =10 \, \mathrm{G}$ et $l=14 \, \mathrm{Mm}$ \citep[see Equation~4 of][]{mikic2013}. The loop expansion factor reaches a value of 11 at the loop apex (i.e. at 58~Mm of height). 

In input of the simulations we select the mesh spacing to be $\Delta s=19$~km in the chromosphere and transition region, increasing to $\Delta s=190$~km in the corona.

\paragraph{Loop~B} 
This is the same loop that we studied in Paper~I. It corresponds to the pulsating loop bundle detected in AIA data. It is quite a large and asymmetric loop with $L=367 \, \mathrm{Mm}$. As in Paper~I, $s=0$ corresponds to the eastern footpoint, while $s=L$ corresponds to the western footpoint. The apex is at an altitude of 87~Mm at $s=212$~Mm (i.e. at $0.58 \, L$), i.e. the loop is skewed towards one footpoint. The magnetic field strength is about the same at $s=0$ ($\sim315$~Gauss), where $B_{z0}>0$, as at $s=L$  ($\sim275$~Gauss), where $B_{z0}<0$.  The cross-sectional area at each footpoint is thus similar. The loop expansion factor reaches a value of 38 at $s=162$~Mm, i.e. to the east of the loop apex. 
As seen in Paper~I, the large low-lying portion at the eastern footpoint is due to the magnetic topology in this area, i.e. a low-lying null-point and many bald patches. As for loop~A, the mesh spacing is $\Delta s=19$~km in the chromosphere and transition region, and $\Delta s=190$~km in the corona.

\paragraph{Loop~C} This field-line corresponds to a non-pulsating loop bundle in the AIA data. It is located in the small active region at the West of NOAA AR 11499. We did not find any intensity pulsations in any of the loops of this region. The loop chosen is shorter than the previous ones, with $L=139 \, \mathrm{Mm}$. As for loop~B, $s=0$ corresponds to the eastern footpoint. The apex of this loop has an altitude of 42~Mm at $s=77 \, \mathrm{Mm}$, i.e. $s=0.55 \, \mathrm{L}$. The loop expansion is quite large, with a maximum value of 82 at $s=92 \, \mathrm{Mm}$, i.e. to the west of the loop apex. The magnetic field strength is about the same at each footpoint with respectively 1100~G and 1200~G. Using the same number of mesh points as for the two other loops, the mesh spacing is smaller for this loop: $\Delta s=7$~km in the chromosphere and transition region, increasing to $\Delta s=70$~km in the corona.

	\subsection{Exploration of heating parameter space}\label{sec:main_simu}

For each loop geometry, we present hundreds of simulations, which is still only a fraction of the parameter space we explored. We choose here to focus only on the area of the parameter space surrounding the simulations showing TNE cycles. 

The results of the exploration for each loop geometry are displayed using grid plots, each individual plot showing the temperature evolution along the loop for three days of simulation time. Each figure represents a scan of the $\lambda_1$ and $\lambda_2$ values for a given value of $H_1$, i.e. a cut through the scan cube. By this means we can analyze the global behavior of the simulations within the parameter space.   

We choose arbitrarily to show only the temperature, although the density or the velocity would be also suitable to present a different view of the loop behaviors. 
Nevertheless, the density and the velocity profiles are shown for a few examples analyzed in detail (see Figure~\ref{fig:detailed_simu}). Temperature, density, and velocity averaged around the apex are also shown in Section~\ref{sec:main_simu}. Throughout the present paper the apex area is defined, as in Paper~I, as the part of the loop above 90\% of the apex height. Temperature, density, and velocity are averaged in this area, to avoid quoting values at a single location.

	\subsubsection{Physical limitations on the domains explored}\label{sec:discussion_simu}
	
As mentioned earlier, we explored a very large range of heating for each loop geometry, in particular in terms of scale heights: $\lambda_{1}$ and $\lambda_{2}$. We did not limit the study to specific ranges that would be appropriate to the magnetic field configuration of each loop system. As a consequence, it is likely that the domain of the parameter space presented in this paper is larger than a \textquotedblleft realistic\textquotedblright~one that would be  constrained by the magnetic field strength, for example. This was done deliberately, to separate the heating mechanism from the magnetic field, so as not to make any strong assumptions about the heating. Thus, we are maybe scanning scale heights that are not possible in reality for a given loop geometry.

The choice of the scanned $H_1$ values is justified by the temperature and density of the simulated loops, i.e. we aimed to produce loops with coronal temperature (close to 1~MK and at most 4~MK for the temperature and $10^{14}~\mathrm{m}^{-3}$ to $10^{15}~\mathrm{m}^{-3}$ for the density).
We explored as well $H_1$ producing cooler loops but for these cuts through the scan cube, not enough density was injected, so no TNE was produced. We tested also higher $H_1$ but then the loops are extremely hot which does not correspond to our observations.
Note that given the large loop expansion of loops~B and~C (respectively 38 and 82), some of the values of $H_1$ give really high $Q_0$ (up to $20 \times 10^{4}~\mathrm{W} \, \mathrm{m}^{-2}$ for loop C, see Table~\ref{table:scan_loop}). However, it can be delicate to interpret the absolute values of $Q_0$. In our simulations the chromosphere and low-transition region are not well modeled. The part of the heating that is radiated away in these layers may be unrealistic and further studies would be needed to quantify it. 
Therefore, we are not interpreting the absolute values of the strength of the heating but rather the evolution of the size of the TNE domain when $H_1$ is doubled (relative heating strength).	

	\subsubsection{Criteria to distinguish between the different behaviors}\label{sec:criteria_TNE}
	
Within the parameter-space, we detect the TNE cases and determine the nature of the condensations, using only the temperature profiles.

\paragraph{TNE events} They are detected within the parameter-space using Fourier analysis. We look at periods between 2 and 16 hours as we did for the AIA observations. For each simulation, we look at the evolution of the temperature averaged around the loop apex. We do not consider the beginning of these temperatures curves, i.e. the first 10~hours of the simulations, to minimize effects related to the initialization of the simulation, and concentrate on the asymptotic behavior only. Simulations are labeled as TNE events when the Fourier power, for at least one frequency bin, is $20\sigma$ above an estimate of the average local power. Moreover we discard the simulations with amplitudes\footnote{Temperature difference between the maximum and the minimum.} smaller than $0.2$~MK in the second half of the simulations to avoid having to much loops with damped cycles in the TNE domain. 

\paragraph{Distinction between ICs and CCs} 
We look at the nature of the condensations not only around the apex but all along the coronal part of the loops. In fact if some CCs occur low enough in one of the loop legs, the evolution of temperature around the loop apex is not very different from an IC case (see e.g. the first, noted as IC and the third, noted as CC~2, simulations presented in Figure~\ref{fig:detailed_simu}). Testing if the condensations are complete or incomplete only around the apex is then insufficient. The coronal part of the loop is defined as the parts of the loops above 10~Mm of altitude.
The CC cases are then detected if the temperature drops locally under 0.5~MK. The other TNE cases are labeled as IC.

	\subsubsection{Loop A}


\begin{figure*}
	\centering
	\gridline{\fig{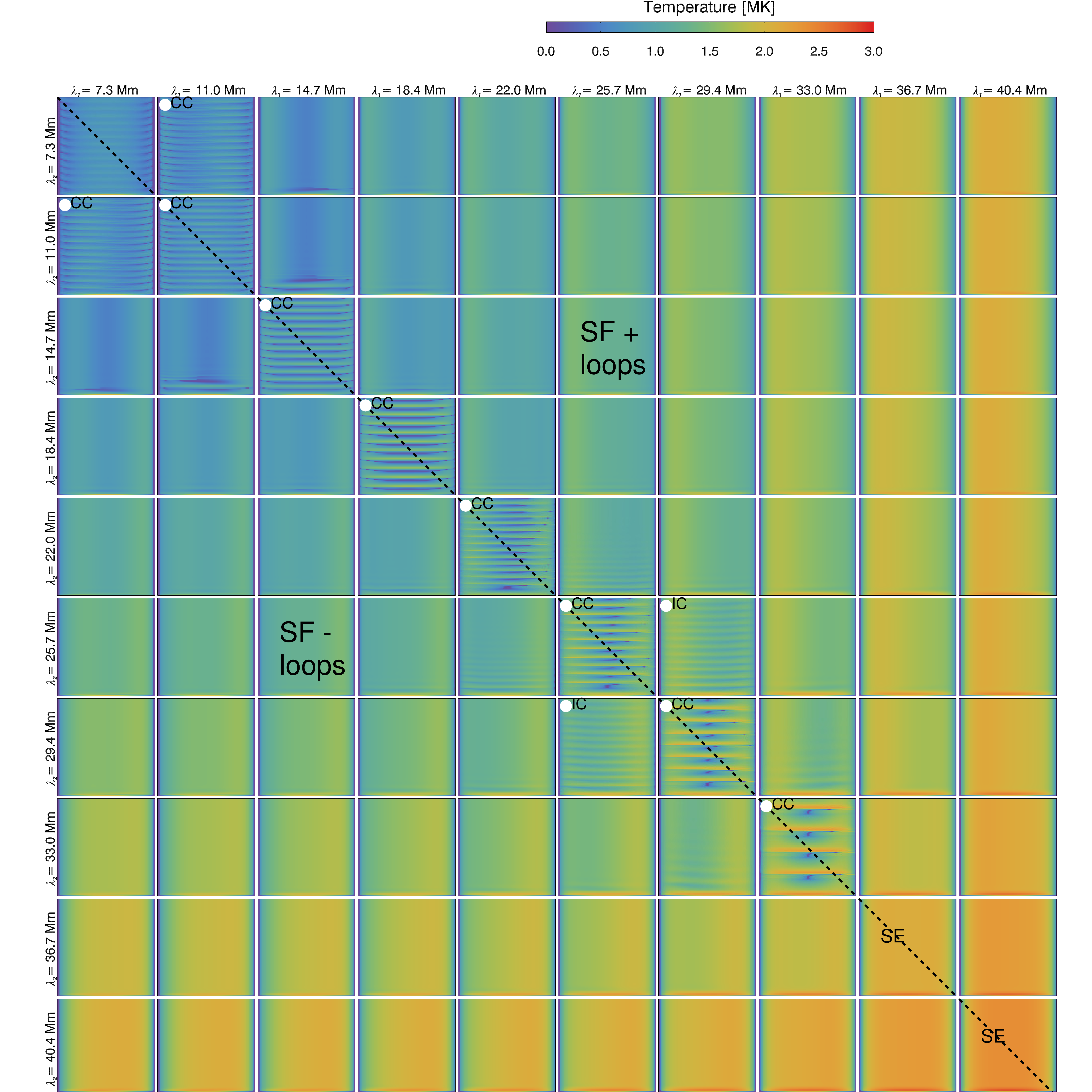}{0.5\textwidth}{$H_1=6.4 \times 10^{-5} \, \mathrm{W} \, \mathrm{m}^{-3}$}
		\fig{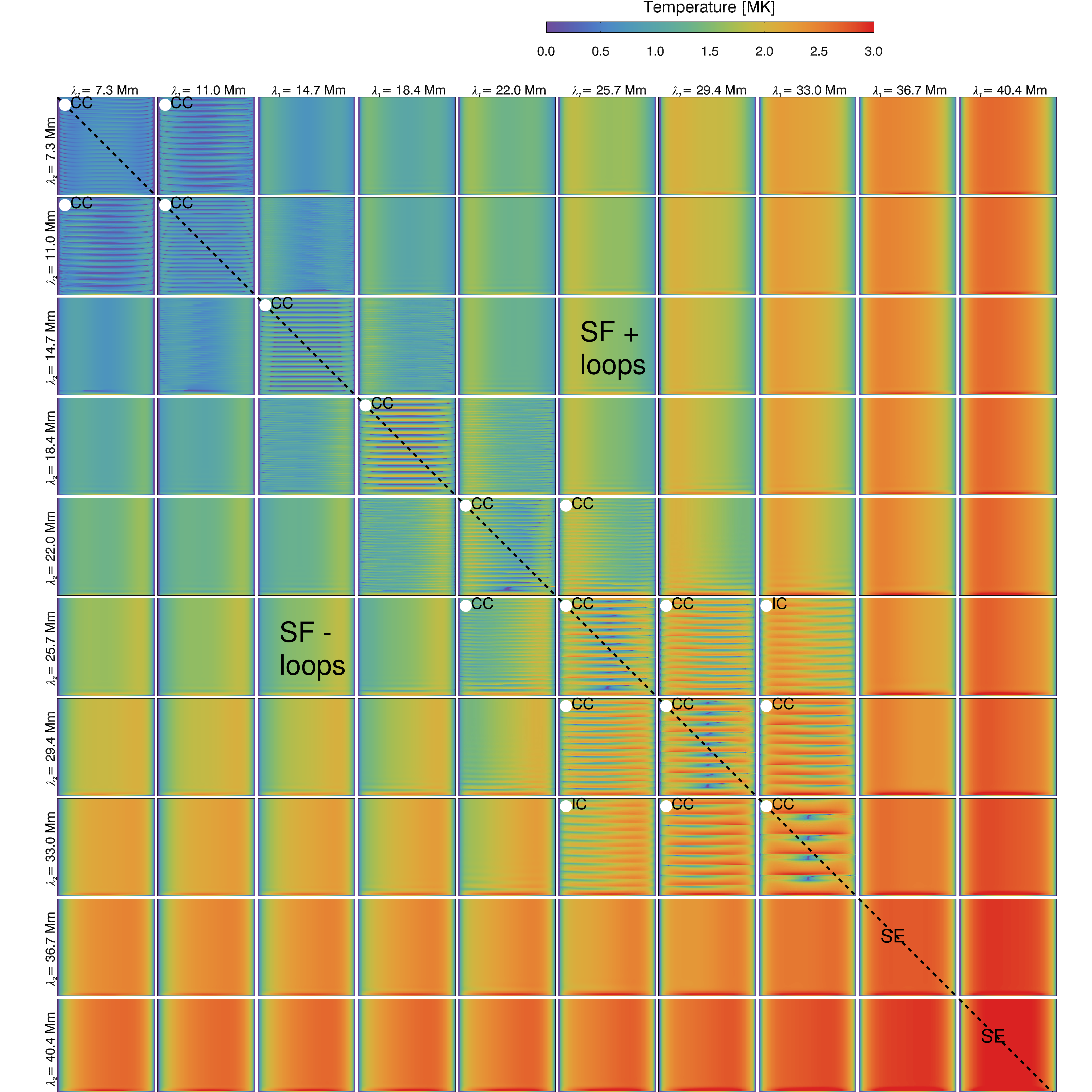}{0.5\textwidth}{$H_1=12.8 \times 10^{-5} \, \mathrm{W} \, \mathrm{m}^{-3}$}
	}
	\gridline{\fig{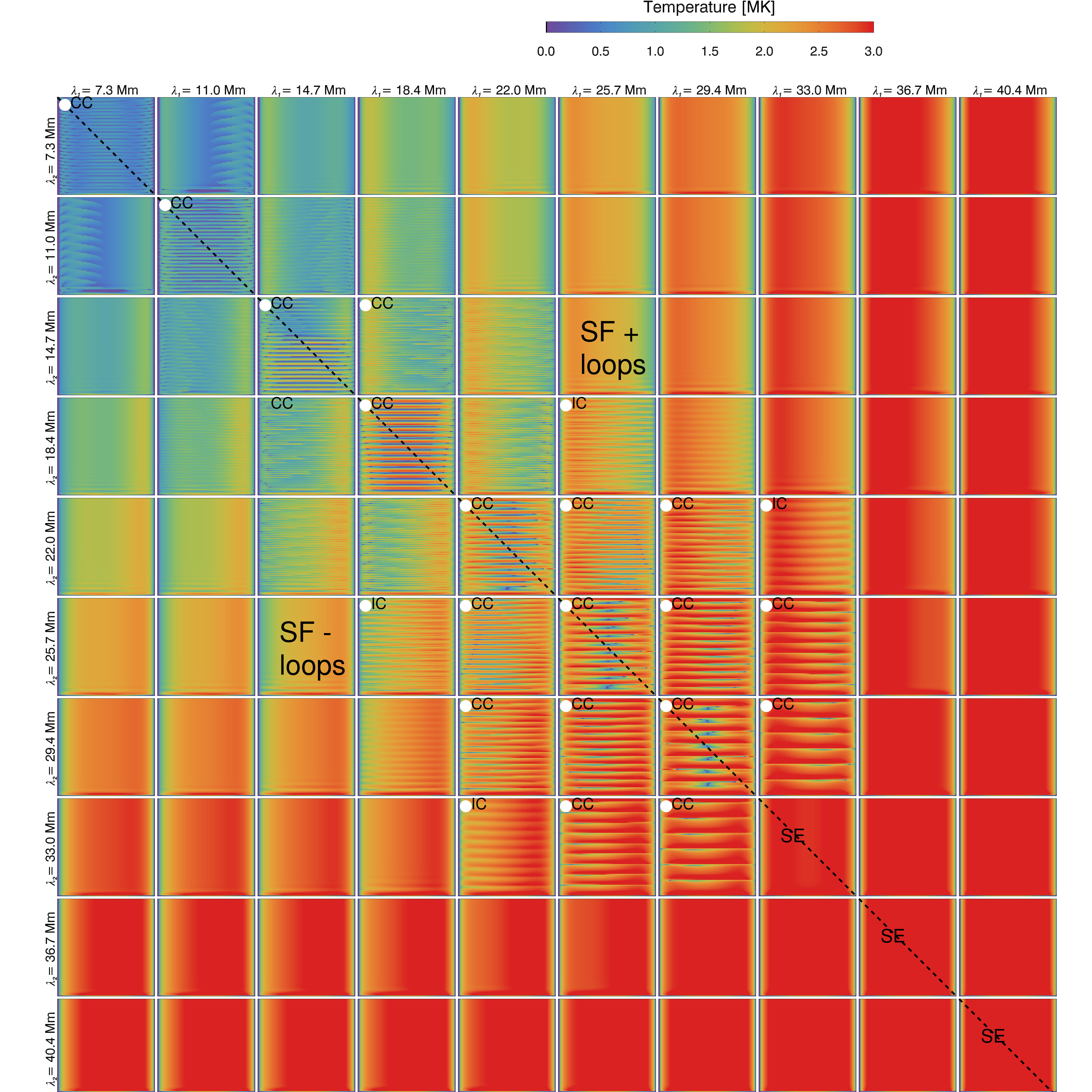}{0.5\textwidth}{$H_1=25.6 \times 10^{-5} \, \mathrm{W} \, \mathrm{m}^{-3}$}}
	\caption{Temperature evolution for 300 loops from 1D hydrodynamic simulations using the loop~A geometry (semi-circular, see Figure~\ref{fig:geometries_a}). Each grid plot shows a cut through the heating parameter scan cube, i.e. each grid plot corresponds to a different value of $H_1$ (the heating imposed at the footpoints). $\lambda_1$ and $\lambda_2$ are scanned between $2\%$ and $11\%$ of $L$, i.e. between 7.3~Mm and 44.0~Mm. The black dashed lines indicate the symmetric heating simulations (i.e. the diagonal of the squared grid plots for this loop geometry). Each small 2D plot shows the evolution of the temperature for one single simulation, along the loop (horizontal direction) and during the 72 hr of the simulation (vertical direction). The white dots indicate cases of TNE (see \ref{sec:criteria_TNE}). For each TNE case we distinguish ICs (incomplete condensations) from CCs (complete condensations). The approximated areas where the simulations are dominated by continuous siphon flows (surrounding the TNE area) are indicated by \textquotedblleft~SF~+~loops\textquotedblright~or \textquotedblleft~SF~-~loops\textquotedblright, respectively for left to right siphon flows and right to left ones. \textquotedblleft~SE~\textquotedblright designate the loops in static equilibrium. The color scale is saturated at 3~MK for every panel.}
	\label{fig:scan_loop_a}

\end{figure*}


\begin{figure*}
	\centering
	\gridline{\fig{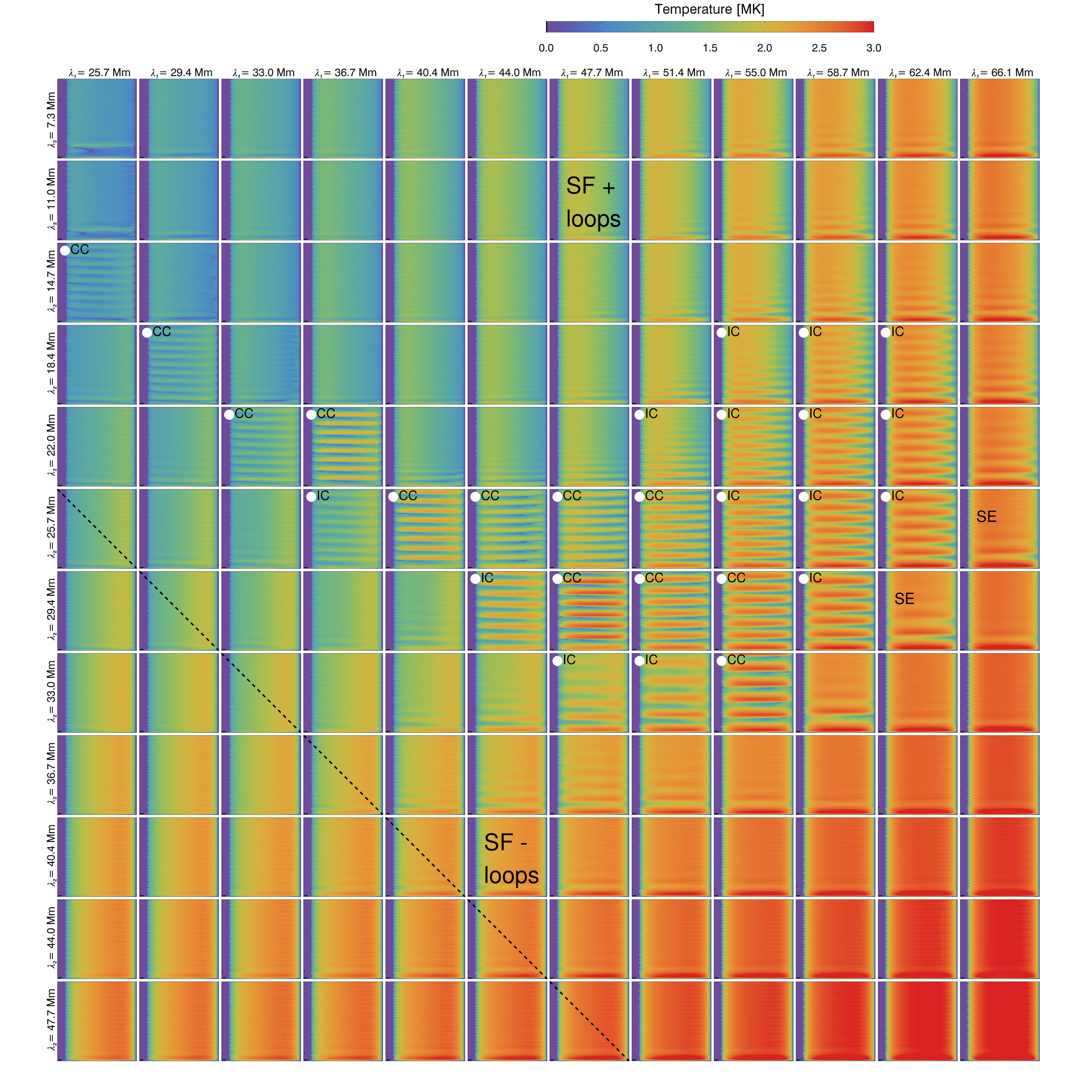}{\textwidth}{$H_1=6.4 \times 10^{-5} \, \mathrm{W} \, \mathrm{m}^{-3}$}}
	\caption{Same as Figure~\ref{fig:scan_loop_a} for the temperature evolution for 144 loops, using the loop~B geometry (see Figure~\ref{fig:geometries_b_c}). Parameter scan with $H_1=6.4 \times 10^{-5} \, \mathrm{W} \, \mathrm{m}^{-3}$. $\lambda_1$ is scanned between $7\%$ and $18\%$ of $L$, i.e. 25.7~Mm and 66.1~Mm and $\lambda_2$ is scanned between $2\%$ and $13\%$ of $L$, i.e. between 7.3~Mm and 47.7~Mm. The black dashed line indicates the symmetric heating simulations, which does not corresponds to the diagonal of the grid plots for this loop geometry because the loop shape is not symmetric. The color scale is saturated at 3~MK.} 
	\label{fig:scan_loop_b_0}
\end{figure*}    


\begin{figure*}
	\centering
	\gridline{\fig{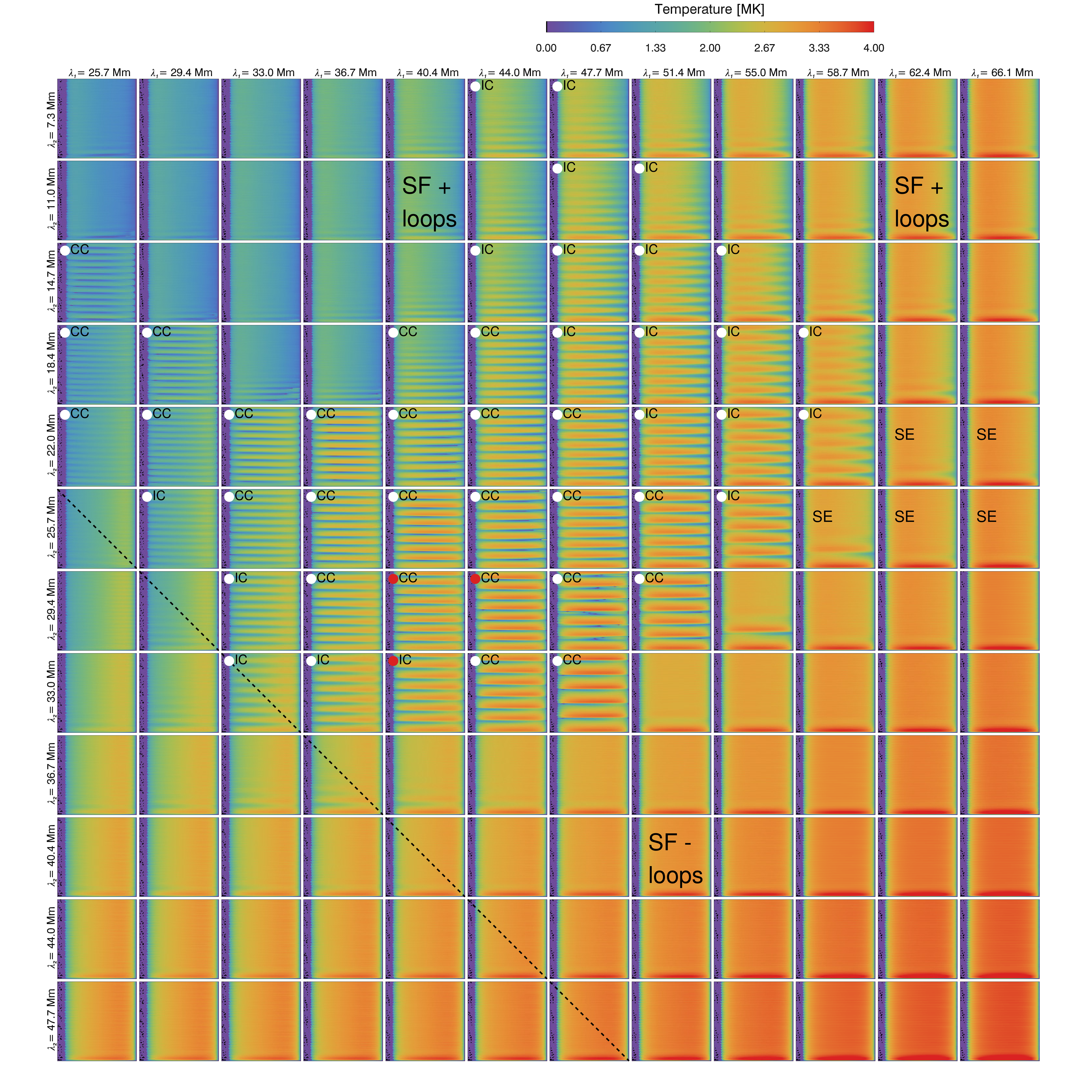}{\textwidth}{$H_1=12.8 \times 10^{-5} \, \mathrm{W} \, \mathrm{m}^{-3}$}}
	\caption{Same as Figure~\ref{fig:scan_loop_b_0} with $H_1=12.8 \times 10^{-5} \, \mathrm{W} \, \mathrm{m}^{-3}$. The color scale is saturated at 4~MK. The red dots indicate TNE cases studied in detail in Section~\ref{sec:condensations_details} and shown at higher resolution in Figure~\ref{fig:detailed_simu}.} 
	\label{fig:scan_loop_b_1}
\end{figure*}    


\begin{figure*}
	\centering
	\gridline{\fig{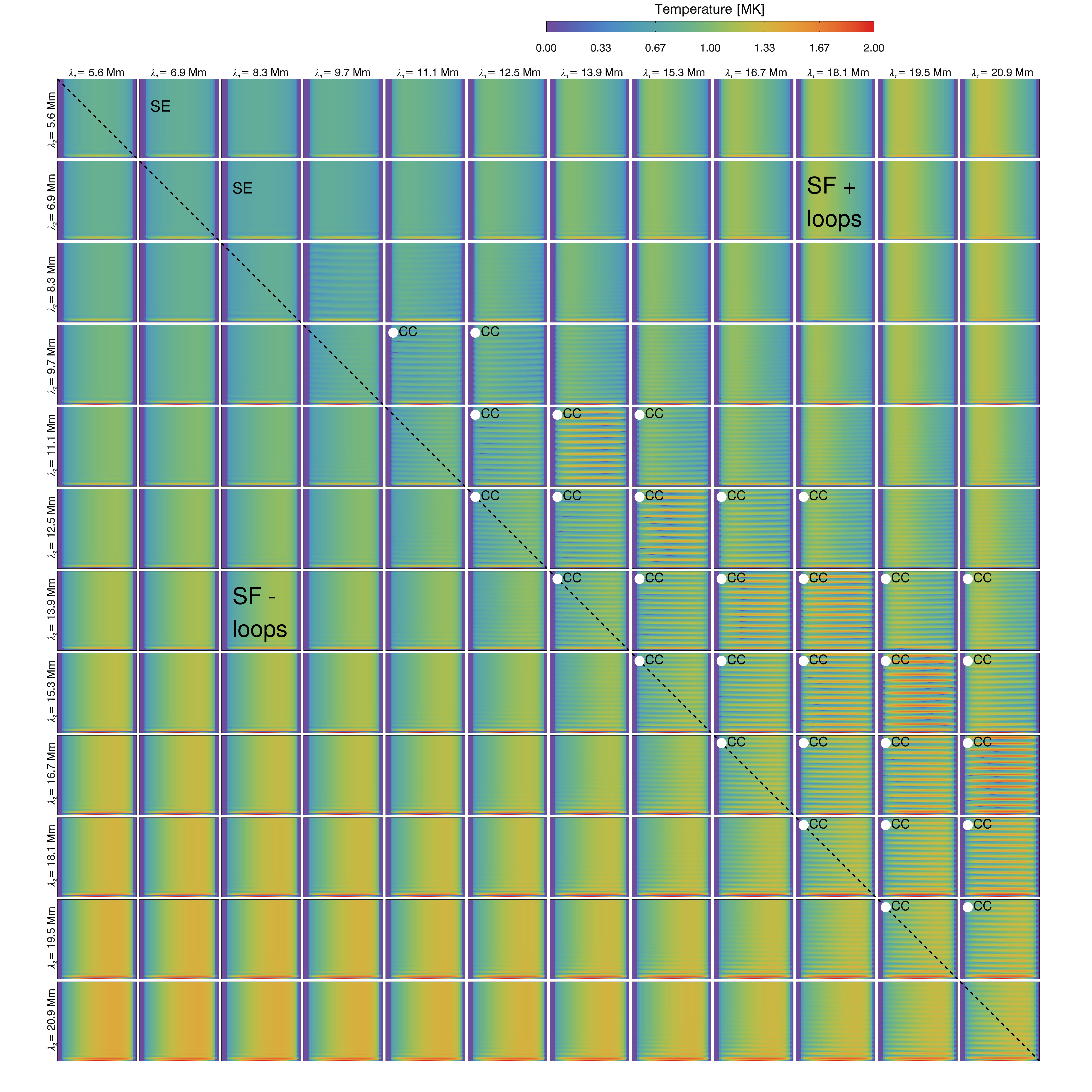}{\textwidth}{$H_1=4.0 \times 10^{-5} \, \mathrm{W} \, \mathrm{m}^{-3}$}}
	\caption{Same as Figure~\ref{fig:scan_loop_a} for the temperature evolution for 144 loops, using the loop~C geometry (see Figure~\ref{fig:geometries_b_c}). Parameter scan with $H_1=4.0 \times 10^{-5} \, \mathrm{W} \, \mathrm{m}^{-3}$. $\lambda_1$ and $\lambda_2$ are scanned between $4\%$ and $15\%$ of $L$, i.e. between 5.6~Mm and 20.9~Mm. The black dashed line indicate the symmetric heating simulations ($\lambda_1=\lambda_2$), which corresponds to the diagonal of the grid plots for this loop geometry. The color scale is saturated at 2~MK.} 
	\label{fig:scan_loop_c_0}
\end{figure*}    


\begin{figure*}
	\centering
	\gridline{\fig{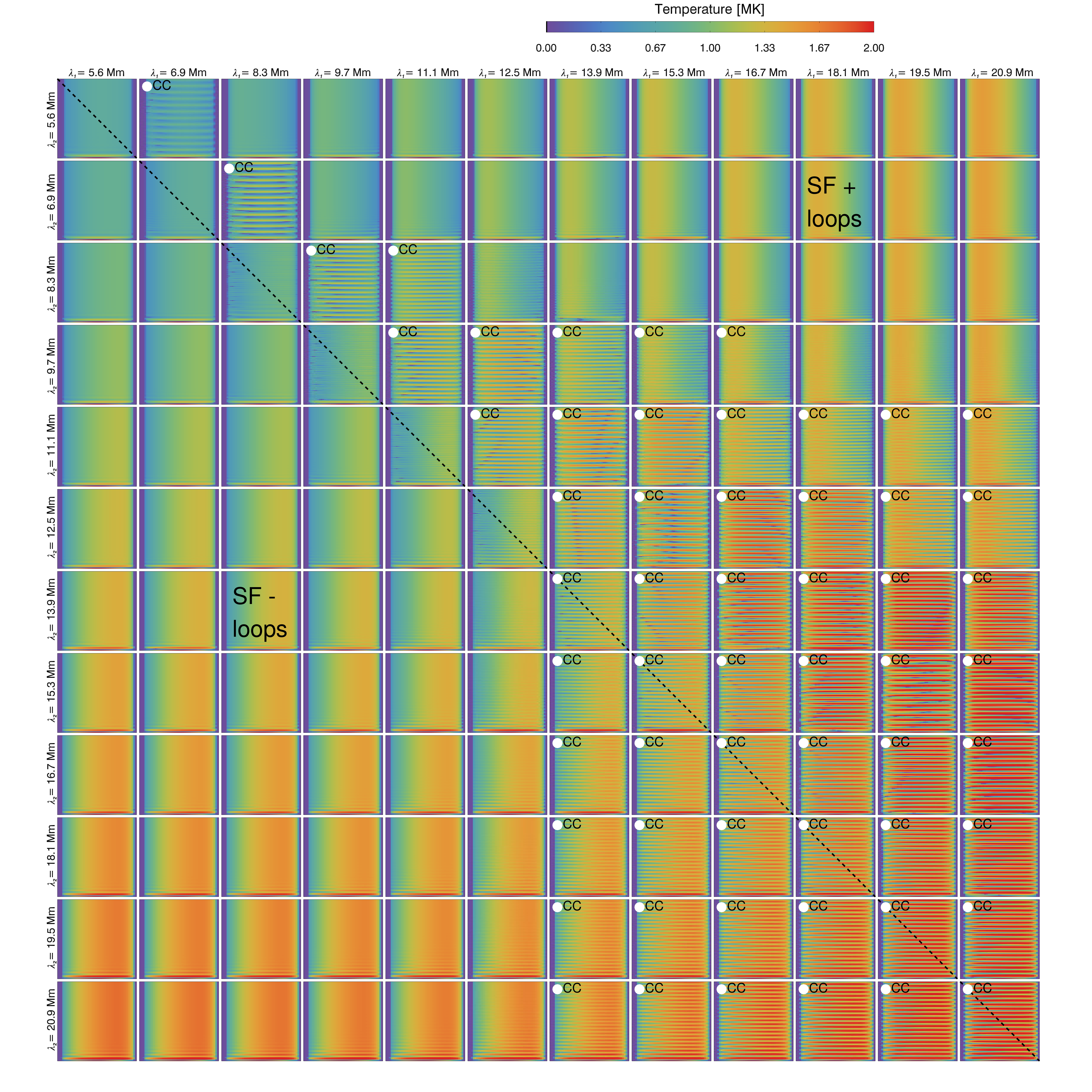}{\textwidth}{$H_1=8.0 \times 10^{-5} \, \mathrm{W} \, \mathrm{m}^{-3}$}}
	\caption{Same as Figure~\ref{fig:scan_loop_c_0} with $H_1=8.0 \times 10^{-5} \, \mathrm{W} \, \mathrm{m}^{-3}$. The color scale is saturated at 2~MK.} 
	\label{fig:scan_loop_c_1}
\end{figure*}    


\begin{figure*}
	\centering
	\gridline{\fig{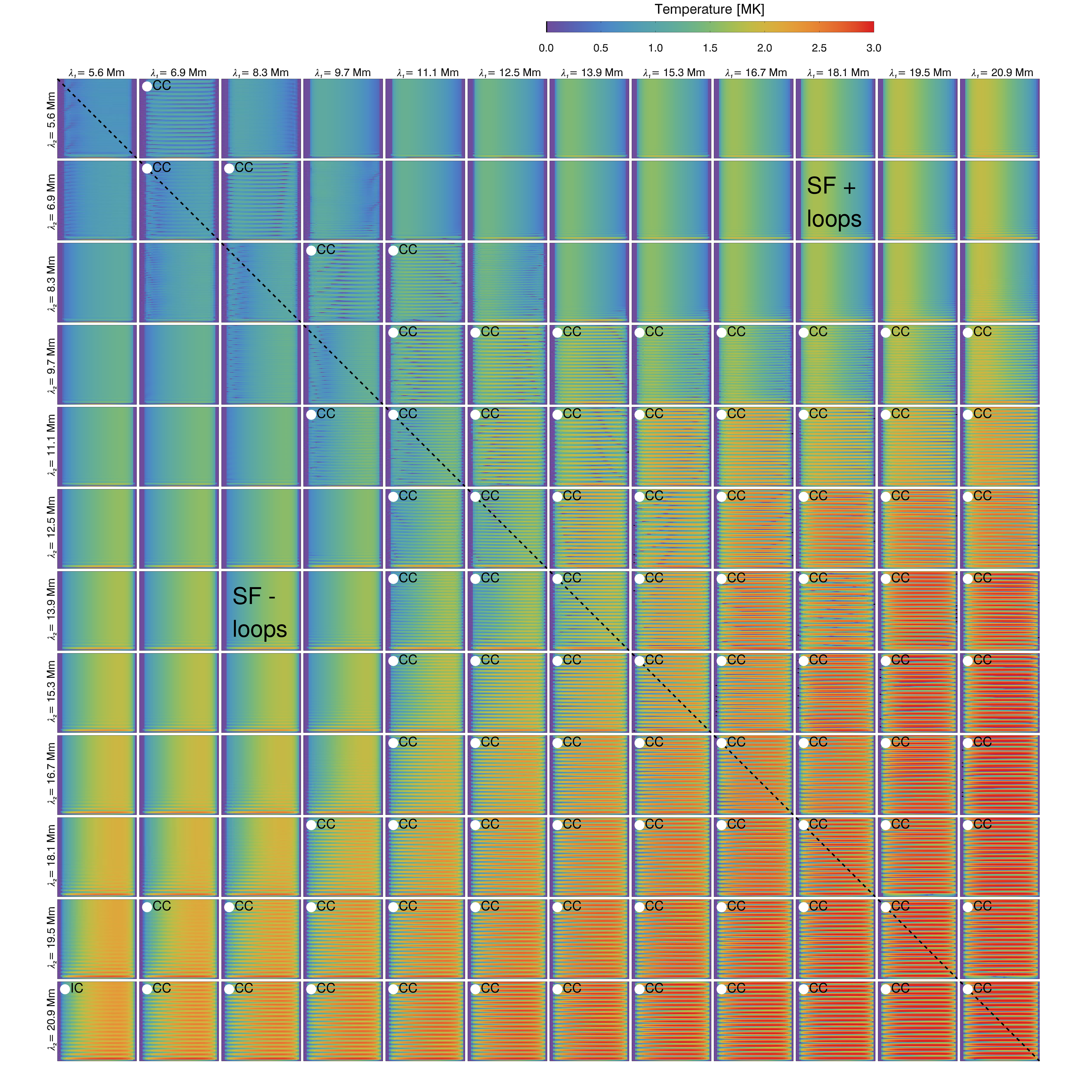}{\textwidth}{$H_1=16.0 \times 10^{-5} \, \mathrm{W} \, \mathrm{m}^{-3}$}}
	\caption{Same as Figure~\ref{fig:scan_loop_c_0} with $H_1=16.0 \times 10^{-5} \, \mathrm{W} \, \mathrm{m}^{-3}$. The color scale is saturated at 3~MK.} 
	\label{fig:scan_loop_c_2}
\end{figure*}    

For this group of simulations $H_0=1 \times 10^{-7} \, \mathrm{W} \, \mathrm{m}^{-3}$. We scan three values of $H_1$: $640 H_0$, $1280 H_0$, and $2560 H_0$.
For each value of $H_1$, $\lambda_1$ and $\lambda_2$ are scanned between $2\%$ and $11\%$ of $L$, i.e. we test 10 values between 7.3~Mm and 44.0~Mm. Every combination is tested, so we have eventually $3 \times 10 \times 10$ different heating configurations, i.e. 300 simulations.
All these simulations are presented in Figure~\ref{fig:scan_loop_a}.

Some of the simulations show cyclic cases of evaporation and condensation.
Around a restricted domain of TNE cycles, the simulations produce either continuous siphon flows or loops reaching a static equilibrium. For each simulation, we indicate:
\begin{itemize}
	\item if it is a TNE case with a white dot, and either CC (for Complete Condensation) or IC (for Incomplete Condensation). CC cases can be visually identified by the dark blue and purple drops in the temperature evolution (temperature $\leq 0.5$~MK);
	\item SF is stated for continuous Siphon Flows; 
	\item and SE for Static Equilibrium. 
\end{itemize} 

In order to determine whether a simulation exhibits TNE cycles and what the nature of the condensations are, we use the criteria presented in Section~\ref{sec:criteria_TNE}.
We see that TNE cycles are encountered only around the diagonal of each of these square grid plots, i.e. for simulations for which a symmetric heating function is applied. 
The upper limit for theses cycles is $\lambda_1 = \lambda_2 = 33.0$~Mm (i.e. $9\%$ of $L$), i.e., the solutions with  $\lambda_1$ or $\lambda_2$ larger than this value are stable.
We also notice that the more heating is applied ($H_1$ high, and consequently $Q_0$ high), the more the TNE domain extends away from the diagonal. 

\begin{figure}
	\centering

	\gridline{\fig{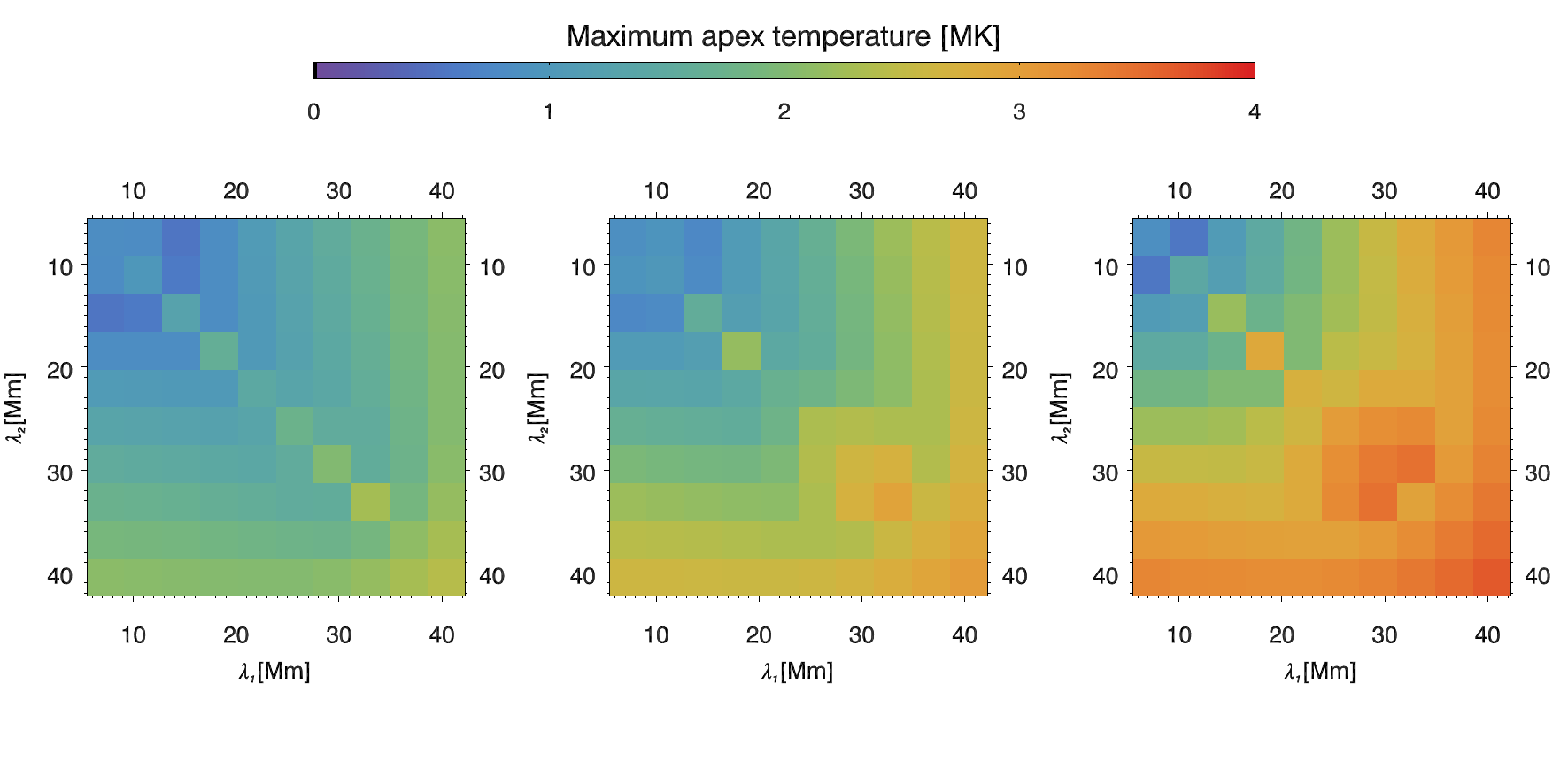}{0.5\textwidth}{$H_1=6.4 \times 10^{-5} \, \mathrm{W} \, \mathrm{m}^{-3}$; $H_1=12.8 \times 10^{-5} \, \mathrm{W} \, \mathrm{m}^{-3}$; $H_1=25.6 \times 10^{-5} \, \mathrm{W} \, \mathrm{m}^{-3}$}
	}
	\gridline{\fig{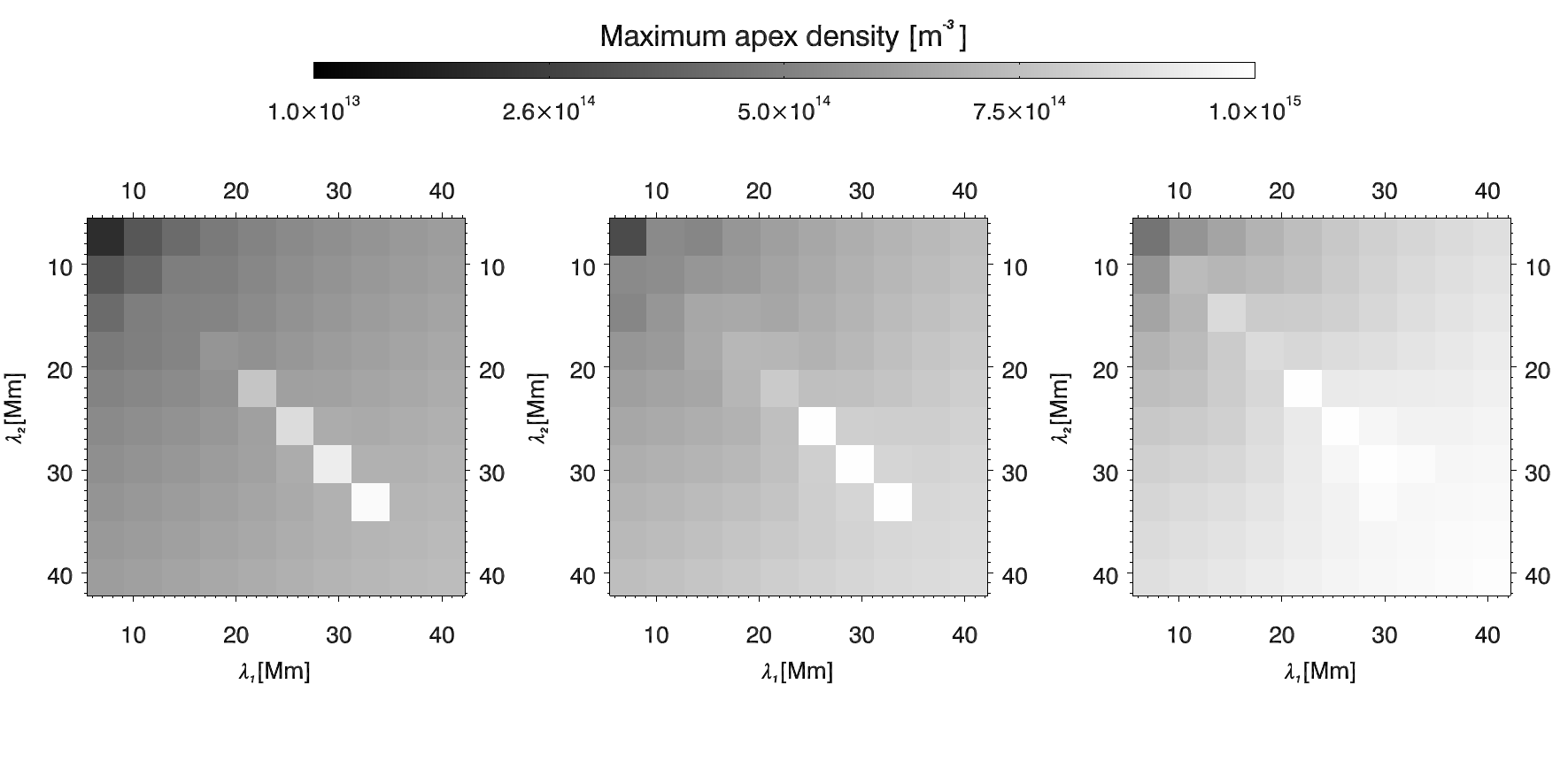}{0.5\textwidth}{$H_1=6.4 \times 10^{-5} \, \mathrm{W} \, \mathrm{m}^{-3}$; $H_1=12.8 \times 10^{-5} \, \mathrm{W} \, \mathrm{m}^{-3}$; $H_1=25.6 \times 10^{-5} \, \mathrm{W} \, \mathrm{m}^{-3}$}
	}
	\gridline{\fig{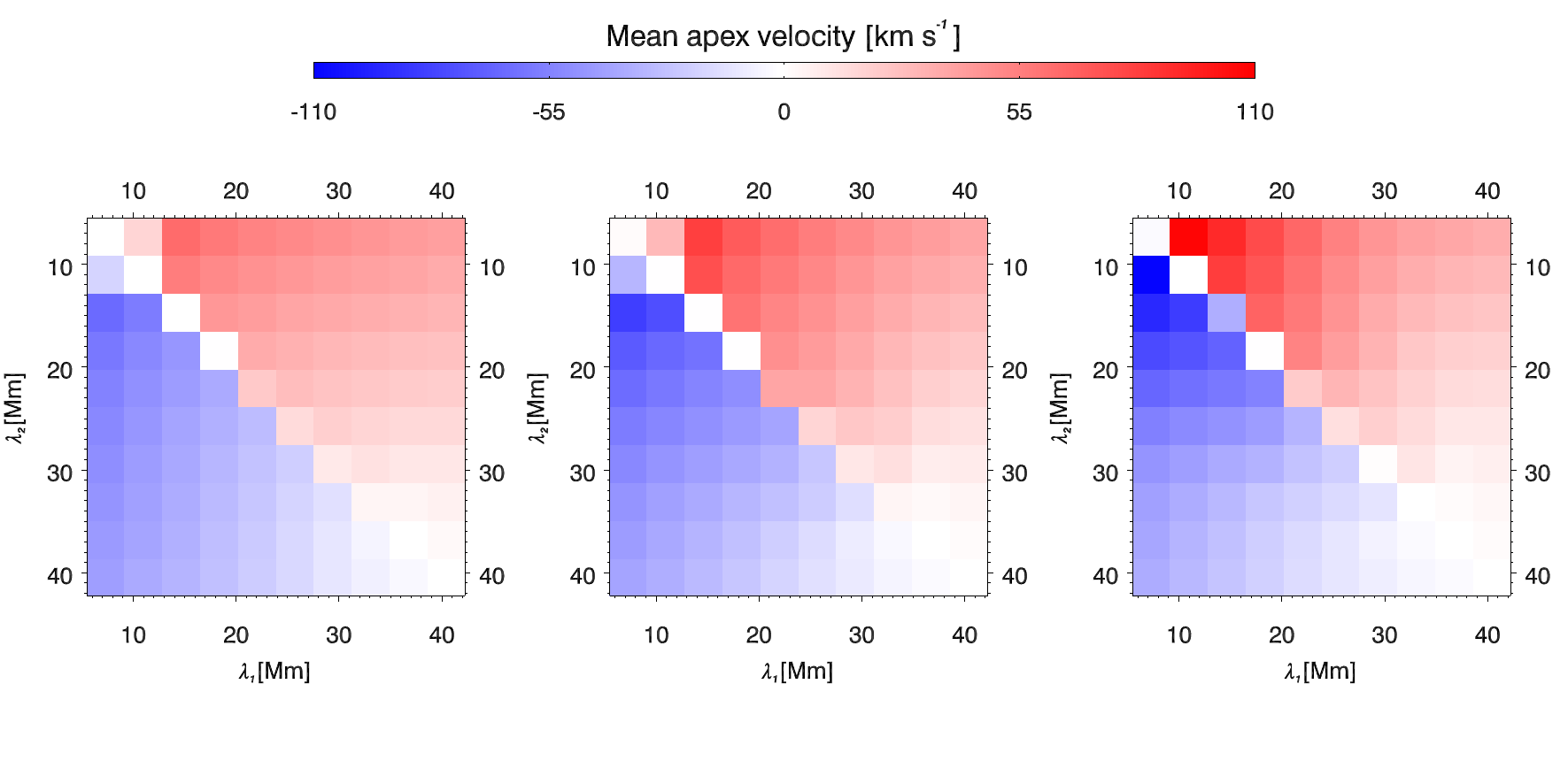}{0.5\textwidth}{$H_1=6.4 \times 10^{-5} \, \mathrm{W} \, \mathrm{m}^{-3}$; $H_1=12.8 \times 10^{-5} \, \mathrm{W} \, \mathrm{m}^{-3}$; $H_1=25.6 \times 10^{-5} \, \mathrm{W} \, \mathrm{m}^{-3}$}
	}
	
	\caption{Evolution of the maximum temperature and density, and mean velocity around the loop apex within the heating parameter-space ($\lambda_{1}$, $\lambda_{2}$, and $H_1$) for the simulations with the loop~A geometry. Temperature and density are averaged around the apex before determining the maximum. The velocity is displayed between $\pm110 \, \mathrm{km} \, \mathrm{s}^{-1}$.} 
	\label{fig:te_ne_simu_loop_a}
\end{figure}

In Figure~\ref{fig:te_ne_simu_loop_a}, we show the maximum temperature and density averaged near the loop apex and the averaged velocity over the loop apex. This temperature is most of the time coronal (with very few simulations below $\sim 0.6$~MK). It increases to 4~MK for large values of $H_1$, $\lambda_1$ and $\lambda_2$. We see a clear signature of heating for symmetric heating cases (with $\lambda_1 \sim \lambda_2$). Indeed, comparing with Figure~\ref{fig:scan_loop_a}, we can identify that within the TNE domain (indicated by the field of white dots in Figure~\ref{fig:scan_loop_a}), high temperatures are reached more easily. The maximum density plot shows also clearly a condensation pattern for the TNE simulations. We notice that this maximum density is up to $\sim 10^{15} \, \mathrm{m}^{-3}$ which is a reasonable value for a large coronal loop \citep{reale_coronal_2014}. However these values are quite low for CCs cases but we have to bear in mind that we average these quantities around the apex before determining the maximum, and as we will see for other loop geometries, the density peak is not necessarily reached around the apex.

We notice that the velocity around the loop apex is quite high ($\sim 100 \, \mathrm{km} \, \mathrm{s}^{-1}$)  when $\lambda_1 + \lambda_2 < 40 \, \mathrm{Mm}$, when the heating is very stratified. The velocities are lower close to TNE cases.

We use the velocity maps\footnote{We do not show the velocity maps in this paper for conciseness. However, Figure~\ref{fig:te_ne_simu_loop_a} allows us to identify the SE cases, without looking at the velocity maps.} to analyze the loops without cycles. For each $H_1$, in the region where $\lambda_1 > \lambda_2$, we encounter loops whose evolution is dominated by siphon flows to the right footpoint (i.e. the less heated footpoint, see SF~+ in Figure~\ref{fig:scan_loop_a}). We witness the reversed siphon flows when $\lambda_1 <\lambda_2$ (see SF~- in Figure~\ref{fig:scan_loop_a}). We have thus continuous siphon flows to the less heated footpoint when the heating is strongly asymmetric. 
The last main behavior encountered is static equilibrium (velocity close to zero along the loop, see SE in Figure~\ref{fig:scan_loop_a}), for the loops along the diagonal and with $\lambda_1, \lambda_2 > 33.0$~Mm.
Note that the TNE cases show periodic siphon flows (see Figure~\ref{fig:detailed_simu}) but that the cases pointed out as SFs here are simulations dominated by continuous siphon flows for several days. For each values of $H_1$, the TNE, SF~+, SF~- and SE domains do not overlap.

To conclude for this semi-circular loop geometry, we found that the majority of TNE cases produced CC cycles: between $82 \%$ and $89 \%$ of the TNE cases are CCs, depending on the $H_1$ used. This is consistent with the results of  \citet[see e.g. Case 7]{mikic2013}. A few ICs are encountered at the boundaries of the TNE domain (i.e. when the heating is asymmetric) when the total heating is increasing (i.e. for higher $H_1$). 
The domain within the parameter space in which loops are undergoing TNE cycles is rather restricted to symmetric (or close to) heating cases. We notice that there is more dispersion around the diagonal when the total heating is increased.

	\subsubsection{Loop B}


\begin{figure}
	\centering
	
	\gridline{\fig{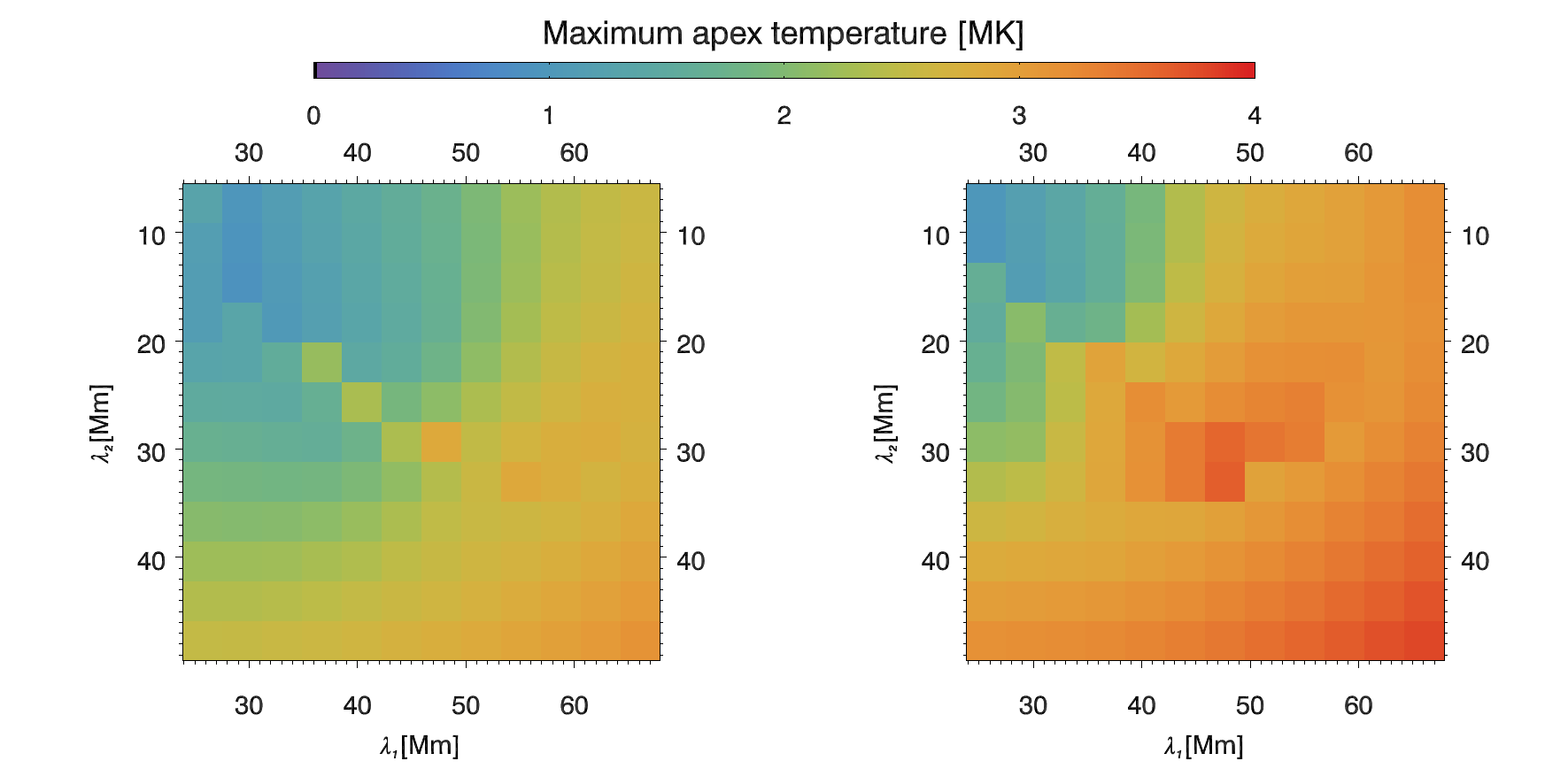}{0.5\textwidth}{$H_1=6.4 \times 10^{-5} \, \mathrm{W} \, \mathrm{m}^{-3}$; $H_1=12.8 \times 10^{-5} \, \mathrm{W} \, \mathrm{m}^{-3}$}
	}
	\gridline{\fig{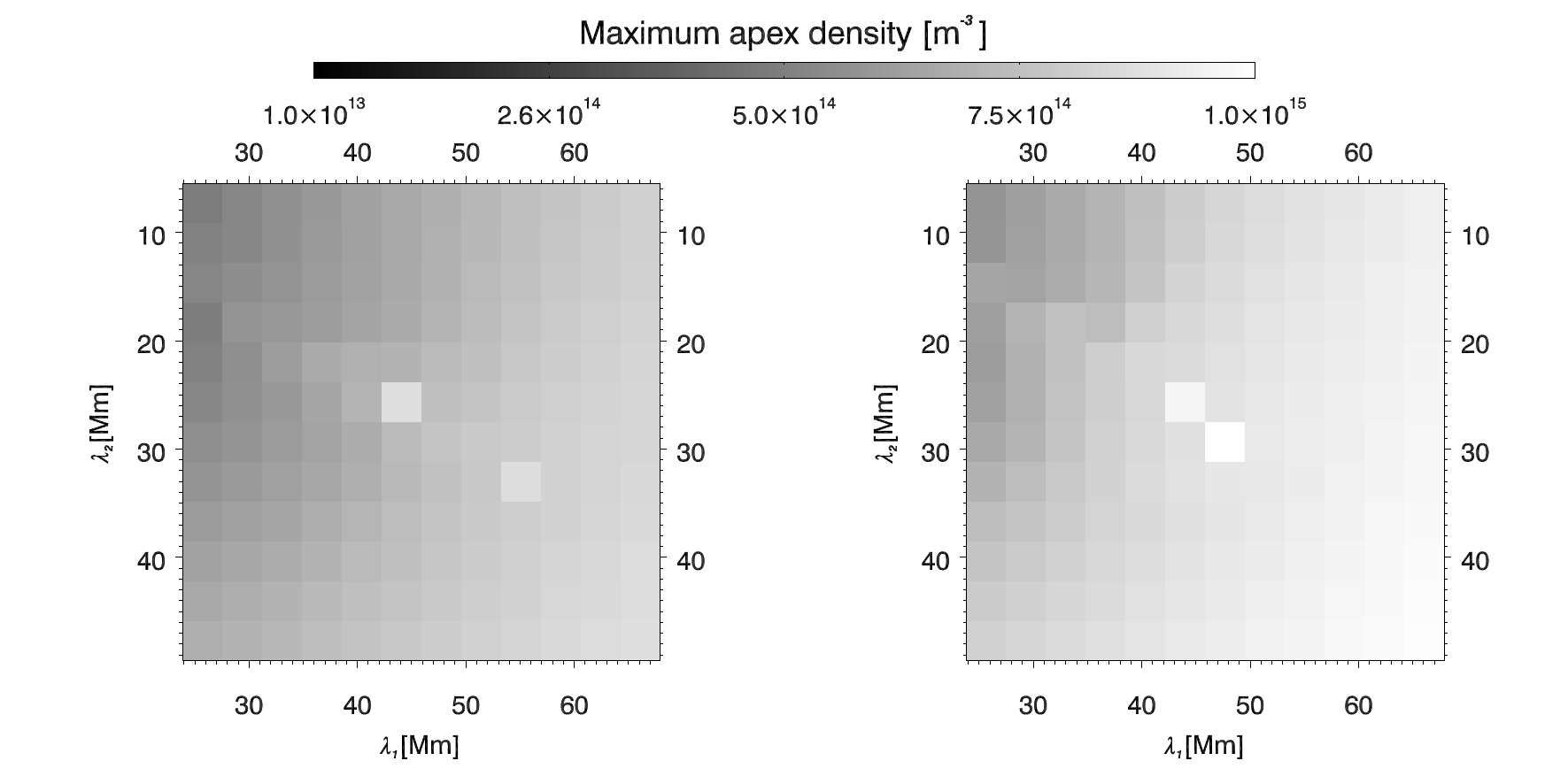}{0.5\textwidth}{$H_1=6.4 \times 10^{-5} \, \mathrm{W} \, \mathrm{m}^{-3}$; $H_1=12.8 \times 10^{-5} \, \mathrm{W} \, \mathrm{m}^{-3}$}
	}
	\gridline{\fig{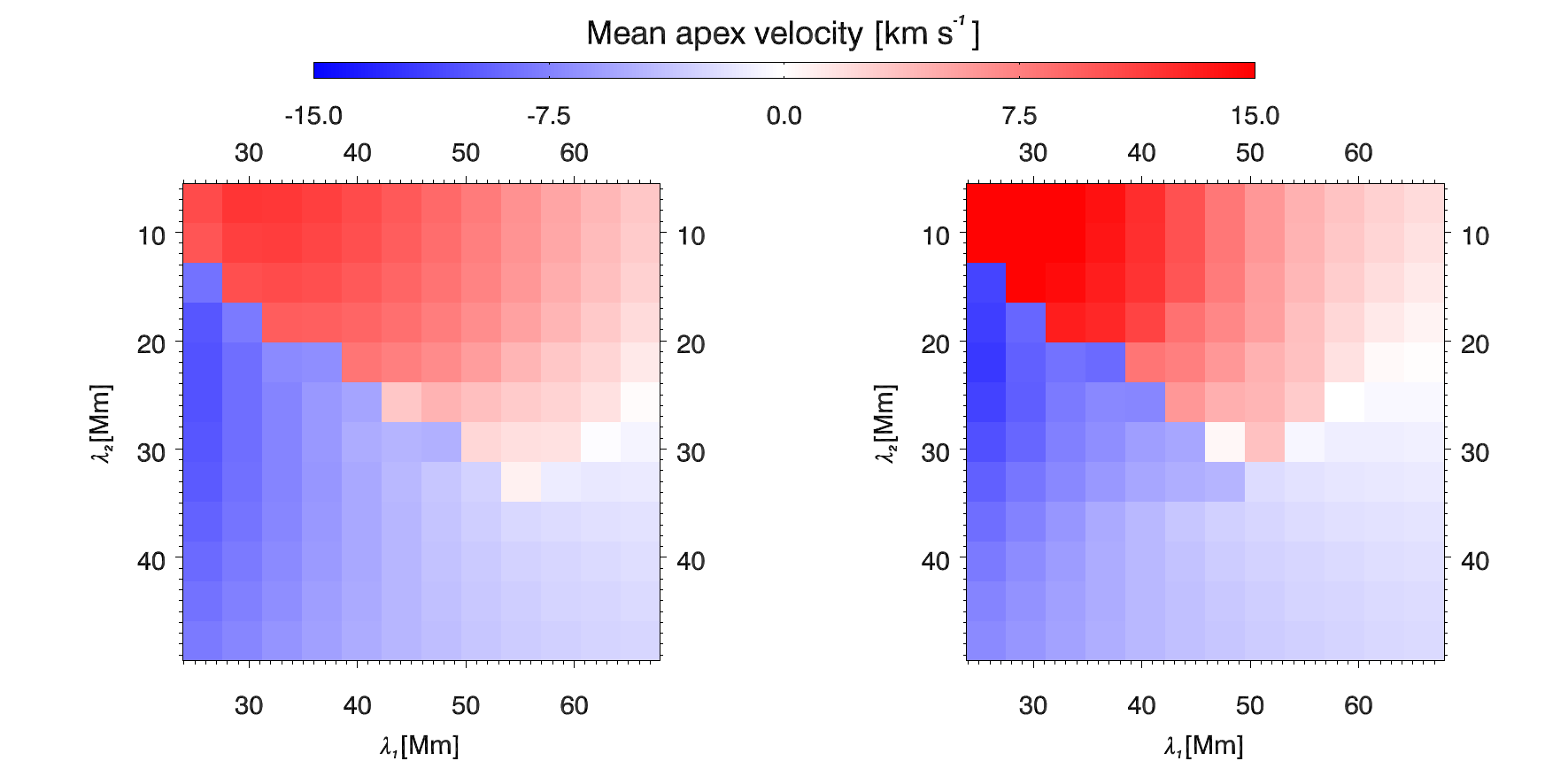}{0.5\textwidth}{$H_1=6.4 \times 10^{-5} \, \mathrm{W} \, \mathrm{m}^{-3}$; $H_1=12.8 \times 10^{-5} \, \mathrm{W} \, \mathrm{m}^{-3}$}
	}	
	\caption{Same as Figure~\ref{fig:te_ne_simu_loop_a} for the simulations using the loop~B geometry. The velocity is displayed between $\pm15 \, \mathrm{km} \, \mathrm{s}^{-1}$.} 
	\label{fig:te_ne_simu_loop_b}
\end{figure}    


\begin{figure}
	\centering
	
	\gridline{\fig{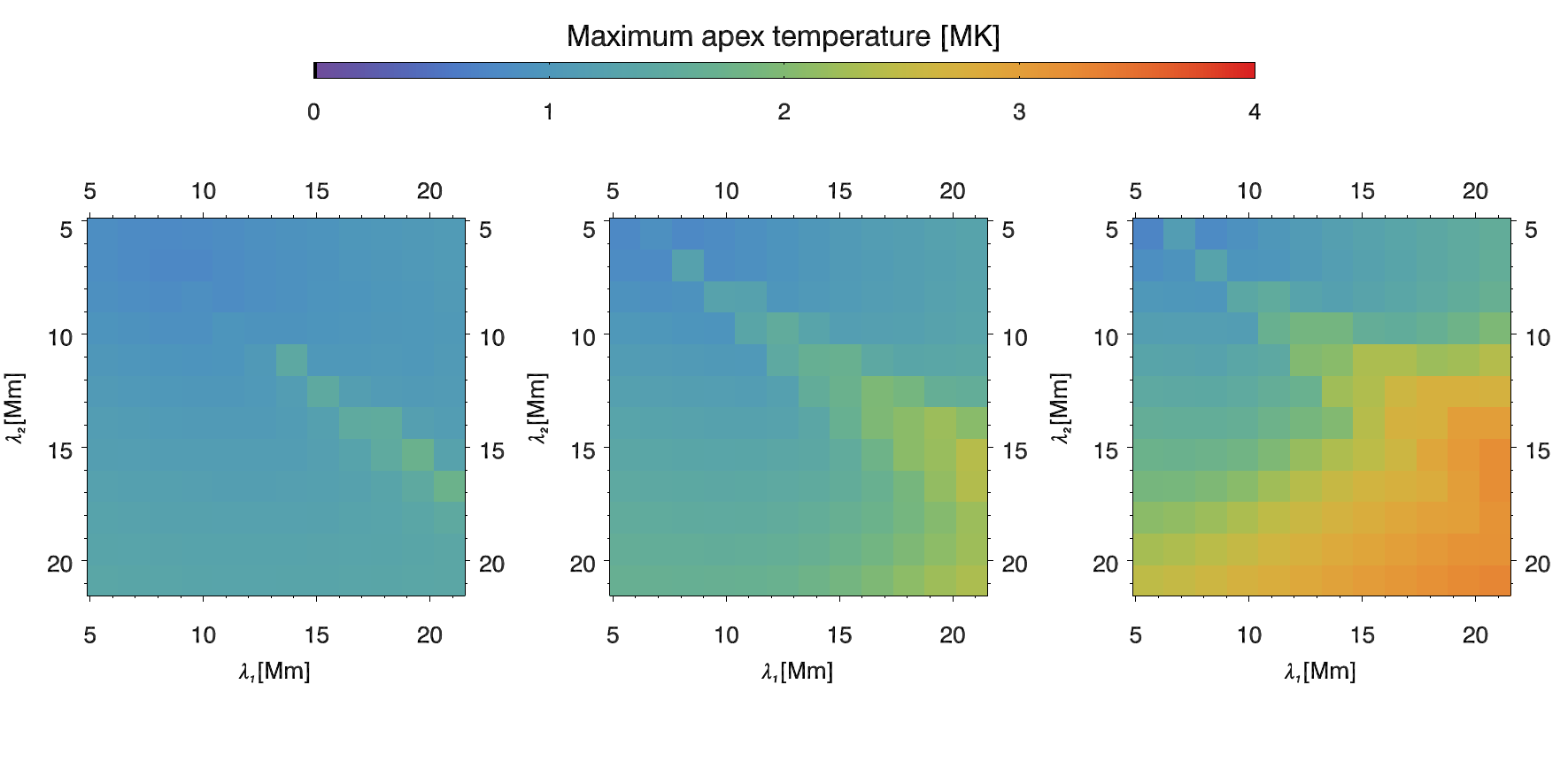}{0.5\textwidth}{$H_1=4. \times 10^{-5} \, \mathrm{W} \, \mathrm{m}^{-3}$; $H_1=8. \times 10^{-5} \, \mathrm{W} \, \mathrm{m}^{-3}$; $H_1=16. \times 10^{-5} \, \mathrm{W} \, \mathrm{m}^{-3}$}
	}
	\gridline{\fig{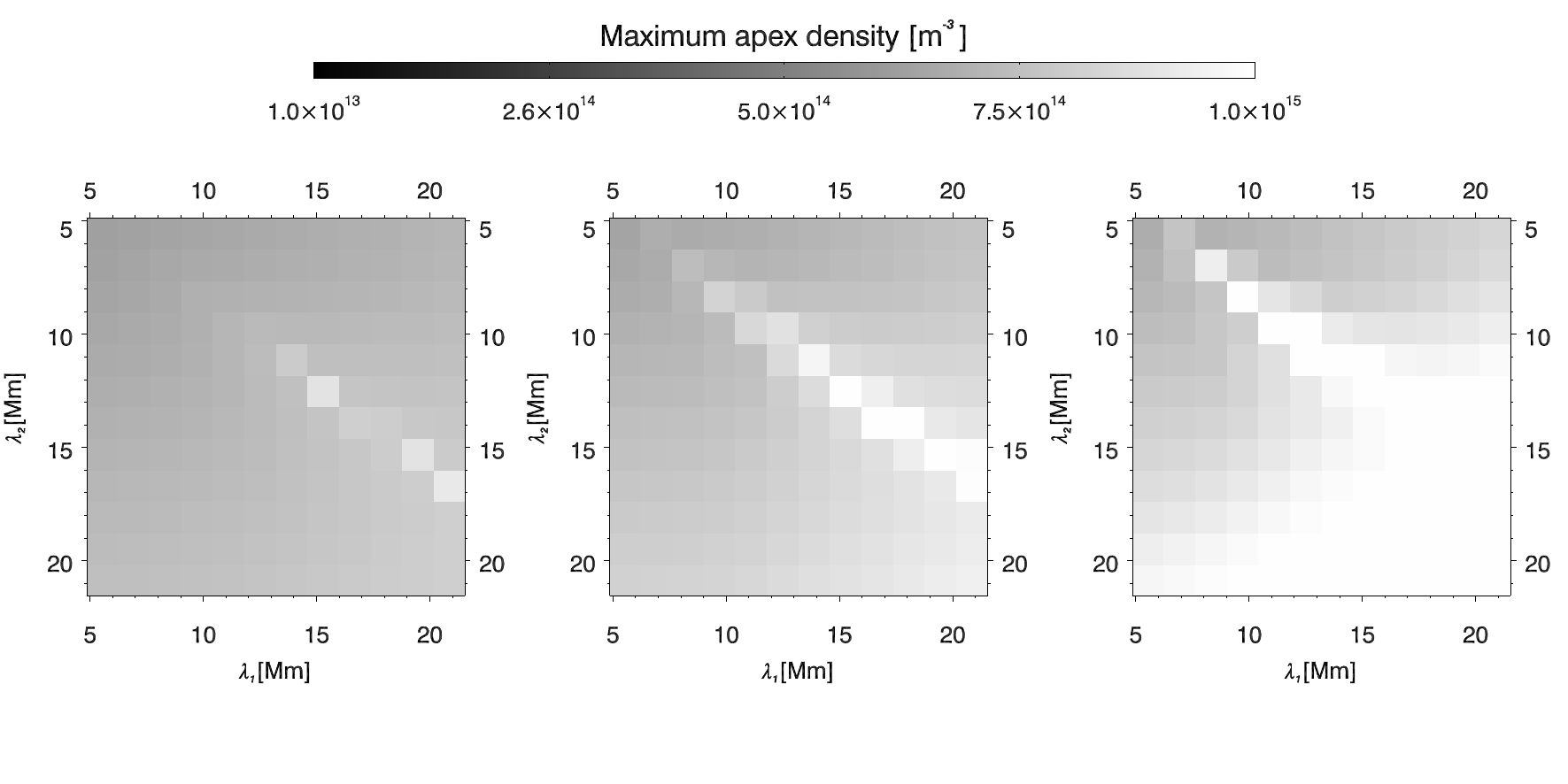}{0.5\textwidth}{$H_1=4. \times 10^{-5} \, \mathrm{W} \, \mathrm{m}^{-3}$; $H_1=8. \times 10^{-5} \, \mathrm{W} \, \mathrm{m}^{-3}$; $H_1=16. \times 10^{-5} \, \mathrm{W} \, \mathrm{m}^{-3}$}
	}
	\gridline{\fig{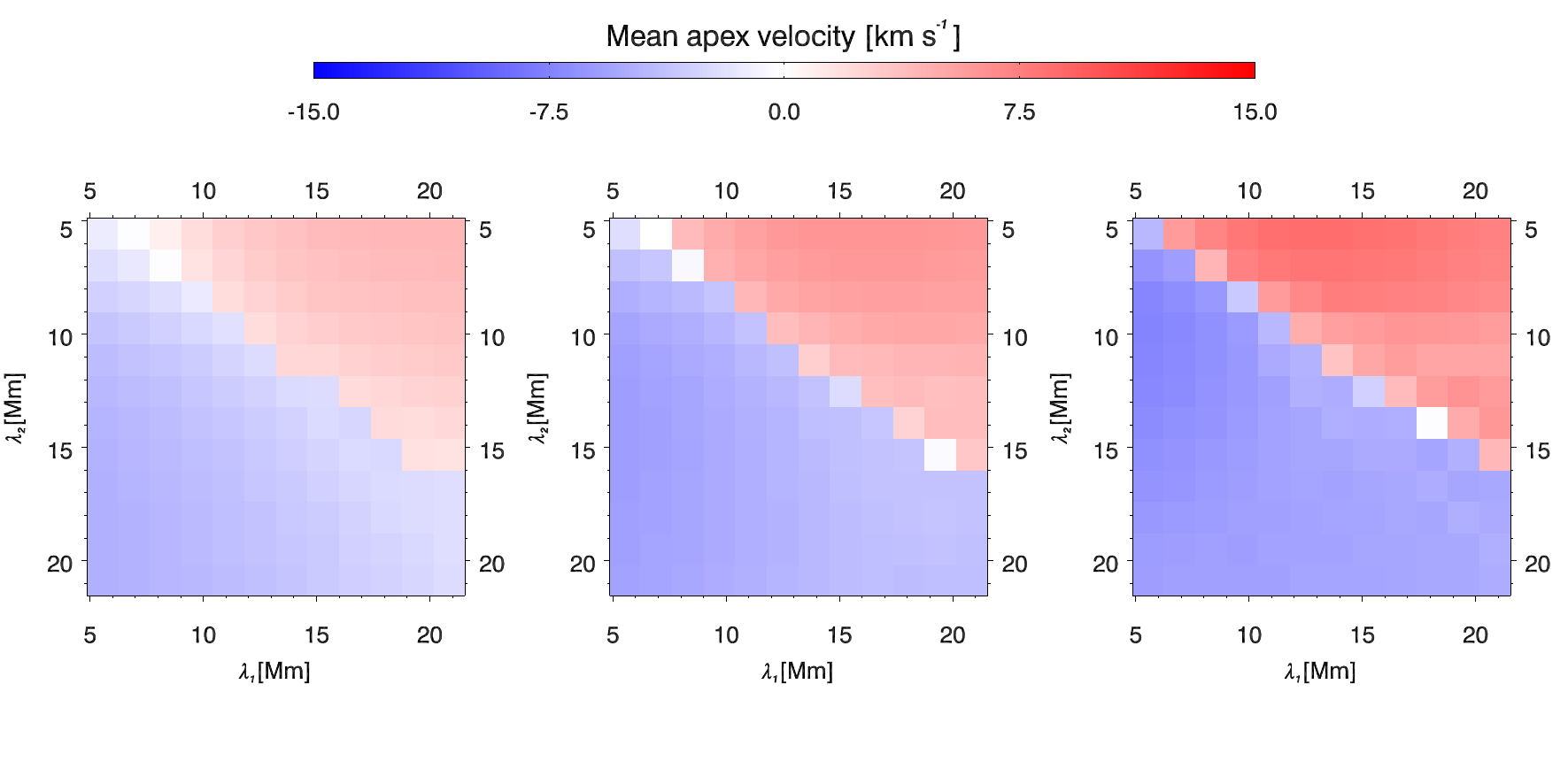}{0.5\textwidth}{$H_1=4. \times 10^{-5} \, \mathrm{W} \, \mathrm{m}^{-3}$; $H_1=8. \times 10^{-5} \, \mathrm{W} \, \mathrm{m}^{-3}$; $H_1=16. \times 10^{-5} \, \mathrm{W} \, \mathrm{m}^{-3}$}
	}
	\caption{Same as Figure~\ref{fig:te_ne_simu_loop_a} for the simulations using the loop~C geometry.The velocity is displayed between $\pm15 \, \mathrm{km} \, \mathrm{s}^{-1}$.} 
	\label{fig:te_ne_simu_loop_c}

\end{figure}    

We use the same $H_0$, i.e. $1 \times 10^{-7} \, \mathrm{W} \, \mathrm{m}^{-3}$, for the simulations with this loop geometry. We scan two values\footnote{In our analysis, we scanned a third value of $H_1$: $320 \, H_0$, however we detected no TNE cases for this scan.} of $H_1$: $640 \, H_0$, as presented in Figure~\ref{fig:scan_loop_b_0}, and $1280\, H_0$, as presented in Figure~\ref{fig:scan_loop_b_1}. Note that the temperature scale between these two Figures is different. It will be the same for the plots concerning the loop~C.
For each value of $H_1$, $\lambda_1$ is scanned between $7\%$ and $18\%$ of $L$, i.e. we test 12 values between 25.7~Mm and 66.1~Mm and $\lambda_2$ is scanned between $2\%$ and $13\%$ of $L$, i.e. 12 values between 7.3~Mm and 47.7~Mm, for a total of $2 \times 12 \times 12 = 288$ simulations. 

The TNE cycles are also located in a restricted domain but are now shifted to the region where $\lambda_1 > \lambda_2$, i.e. asymmetric heating cases when the eastern footpoint (leg) is heated more than the western one. Compared to loop~A, the sub-field of the parameter space presented here is thus not centered around the symmetric heating cases (indicated by the black dashed line).
The scale heights for the energy deposition have to be larger than for loop~A to reach TNE conditions. 

Only one simulation shows TNE cycles with a symmetric heating function. For this simulation, $\lambda_1 = \lambda_2 = 33.0$~Mm (for $H_1 = 1280 H_0$, see Figure~\ref{fig:scan_loop_b_1}). However, this simulation is at the edge of the TNE domain. The envelope of the TNE domain is roughly restricted to the following values: $1.0 < \lambda_{1} / \lambda_{2} < 3.4$, though the exact shape of the TNE domain is more complex. We notice that the range of $\lambda_{1}$ taken by TNE cases (between $7\%$ and $17\%$ of $L$) is wider than the one of $\lambda_{2}$ (between $2\%$ and $9\%$ of $L$). This is probably due to the asymmetry of the loop geometry, the field line from the LFFF extrapolations being skewed toward one footpoint.

As for Loop~A, the more the loop is heated, the wider the TNE domain is.
Looking now at the condensations in these TNE cases for the two $H_1$ values scanned, we notice that respectively $56 \%$ and $45 \%$ of them have cycles with incomplete condensations.
Moreover, these IC cases tend to be at the edges of the TNE domain.

In the same way as for loop~A, the maximum of the averaged apex temperature and density and the mean apex velocity are displayed in Figure~\ref{fig:te_ne_simu_loop_b}. The values reached for both temperature and density are similar to the ones for loop~A. We also see a larger apex temperature in the TNE domain, compared to the surrounding SF cases, as was the case for loop~A. The velocities at the apex are much smaller than for loop~A (maximum 15 $\mathrm{km} \, \mathrm{s}^{-1}$). But we observe the same pattern of velocity evolution within the parameter space, i.e. higher velocities when the heating is highly stratified, $\lambda_1 + \lambda_2 < 60 \, \mathrm{Mm}$ in that case. Moreover, as for loop~A, velocities at the apex are lower in the TNE domain.

At each side of the TNE domain, the simulations are dominated by siphon flows, the direction depending on the asymmetry of the heating. We also find a few SE cases (see location in Figures~\ref{fig:scan_loop_b_0} and~\ref{fig:scan_loop_b_1}). Note that the white pixels in Figure~\ref{fig:te_ne_simu_loop_b} does not necessarily indicate SE cases as TNE cases can show zero velocities at the apex. SE cases point out no flows along the loop for most of the three days of the simulation.

Finally, it is worth noting the presence of high-frequency fluctuations (wavy pattern) in these simulations, especially around the eastern footpoint. We surmise that it is probably due to a combination of the thick chromosphere at this footpoint and the numerical treatment of the transition region. Loop~B has a portion that is almost tangent to the photosphere at its eastern leg (i.e., with small projected gravity, see Figure~\ref{fig:geometries_b_c} and Paper~I). Indeed, this sawtooth pattern is not observed for the other loop geometries or for the high-resolution simulations in Figure~\ref{fig:detailed_simu}.

	\subsubsection{Loop C}

For this last loop geometry, we parameterize the heating function with $H_0=2 \times 10^{-6} \, \mathrm{W} \, \mathrm{m}^{-3}$. We scan three values of $H_1$: $20 H_0$, as presented in Figure~\ref{fig:scan_loop_c_0}, $40 H_0$, as presented in Figure~\ref{fig:scan_loop_c_1}, and $80 H_0$, as presented in Figure~\ref{fig:scan_loop_c_2}. 
For each value of $H_1$, $\lambda_1$ and $\lambda_2$ are scanned between $4\%$ and $15\%$ of $L$, i.e. we test 12 values between 5.6~Mm and 20.9~Mm. Every combination is tested, so we have in total $3 \times 12 \times 12 = 432$ simulations. 

With this loop geometry, we notice that TNE cycles appear first for symmetric or slightly asymmetric heating conditions ($\lambda_{1}>\lambda_{2}$) when $H_1=20 H_0$. Then when we increase $H_1$, more TNE appear for asymmetric heating conditions, specially for $\lambda_{1}>\lambda_{2}$. Finally, when $H_1=80 H_0$, we can notice that the TNE domain becomes much larger than for loop~A and B. In particular, there does not appear to be a limit to TNE for large scale heights.
However, the TNE domain remains limited as barely any TNE cases appear for a very high stratification of the heating ($\lambda_{1}$ or $\lambda_{2}$ smaller than 8.3~Mm).

We notice also that most of the TNE cases have CC cycles (respectively $0\%$ of the TNE events for the two first value of $H_1$ and $1 \%$ for the last one). The siphon flow cases surround the TNE domain, with IC cycles starting to appear on the boundary of this domain. 

Figure~\ref{fig:te_ne_simu_loop_c} shows the maximum of the averaged apex temperature and density and the mean apex velocity. The temperature and density reached are similar to those of the other loops but we clearly see the large range covered by the CCs events (high temperature and high density). The velocities are close to the ones observed for loop~B but the pattern we observed for loops~A and~B, i.e. higher velocities for small heating scale heights is not as clear here.

\section{The occurrence of TNE}\label{discussion}

\subsection{Conditions that favor TNE and constraints on the heating}

\begin{figure*}
	\centering
	
	\gridline{\fig{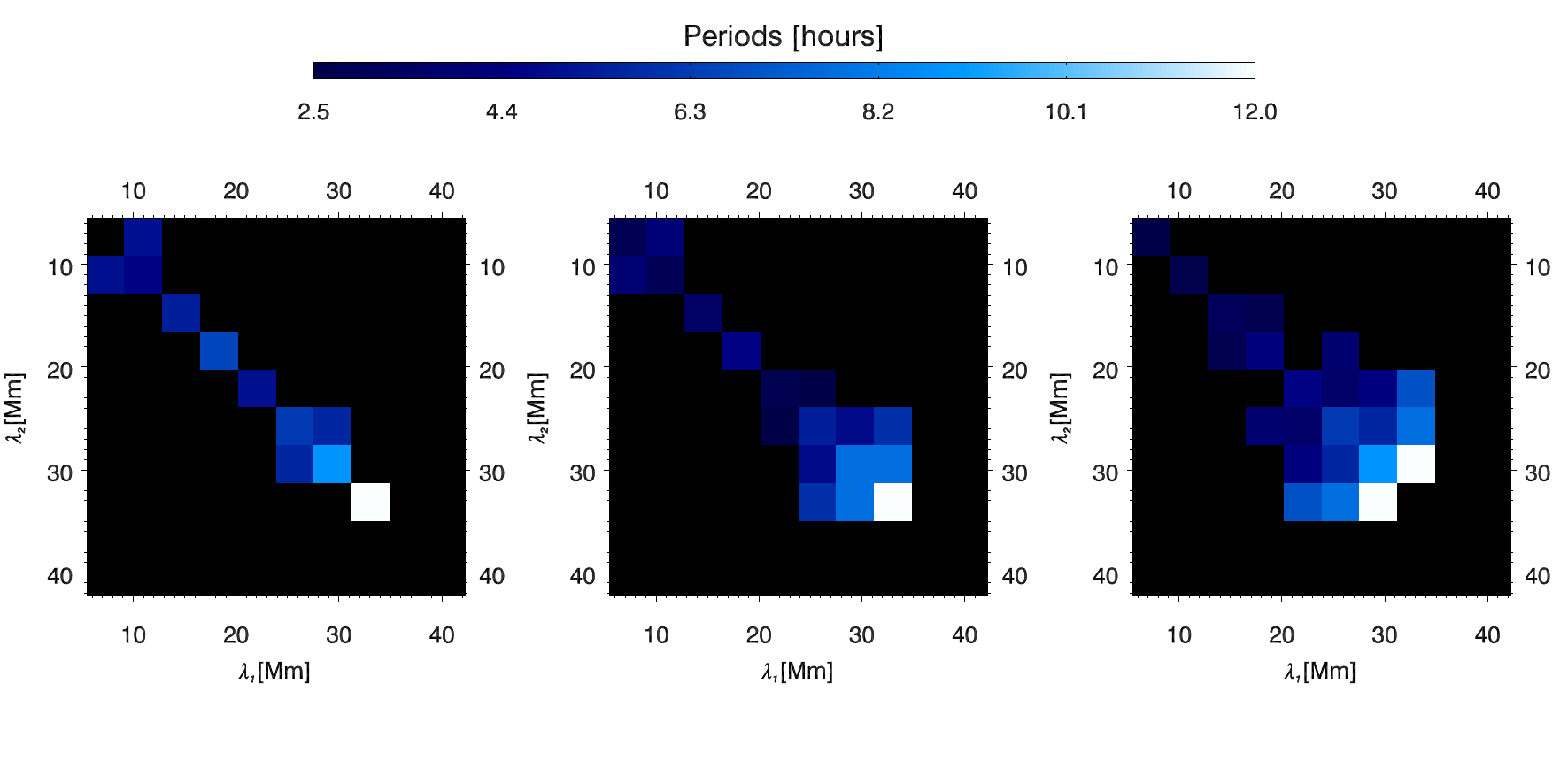}{0.65\textwidth}{$H_1=6.4 \times 10^{-5} \, \mathrm{W} \, \mathrm{m}^{-3}$; $H_1=12.8 \times 10^{-5} \, \mathrm{W} \, \mathrm{m}^{-3}$; $H_1=25.6 \times 10^{-5} \, \mathrm{W} \, \mathrm{m}^{-3}$}
	}
	\gridline{\fig{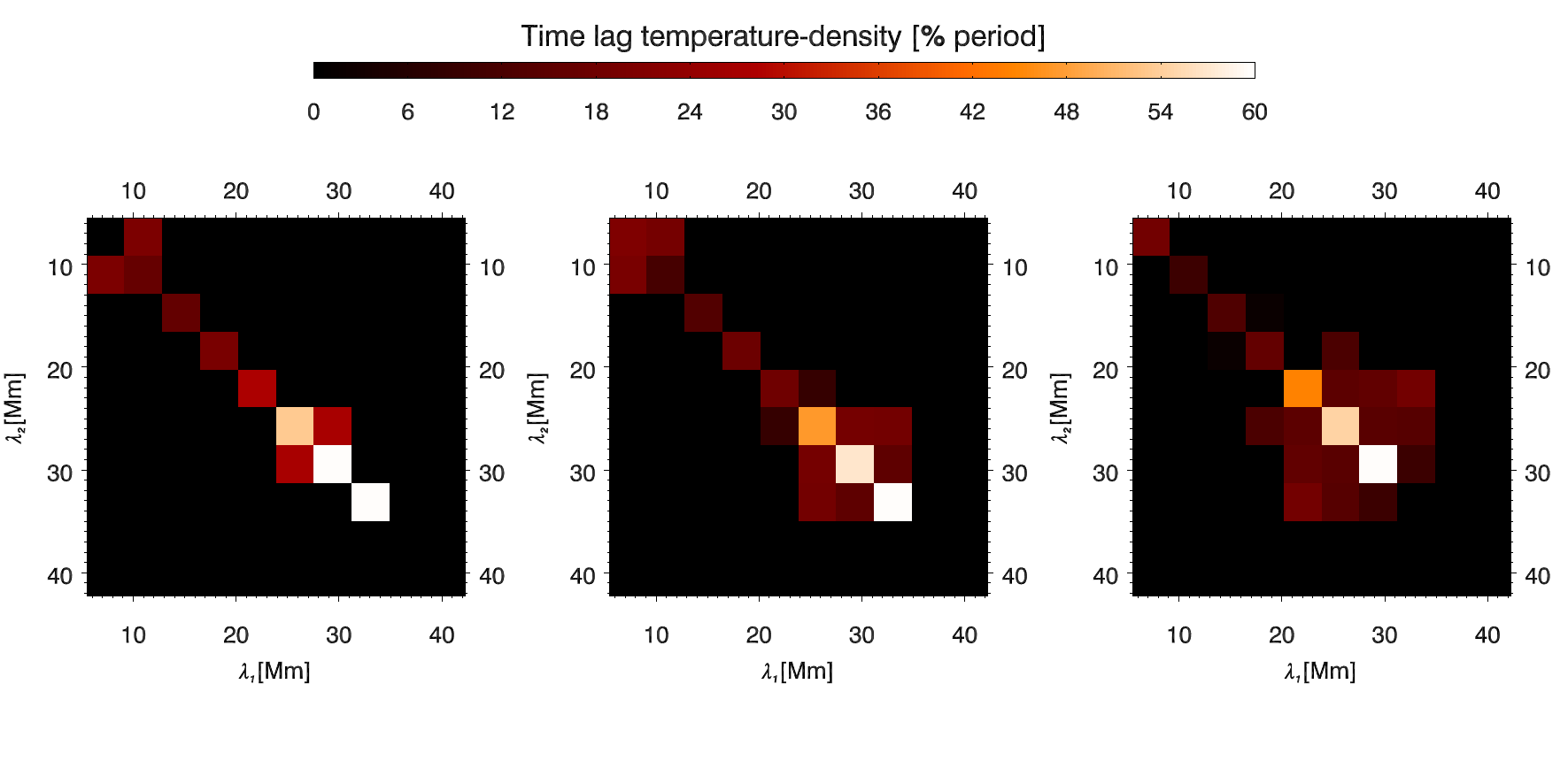}{0.65\textwidth}{$H_1=6.4 \times 10^{-5} \, \mathrm{W} \, \mathrm{m}^{-3}$; $H_1=12.8 \times 10^{-5} \, \mathrm{W} \, \mathrm{m}^{-3}$; $H_1=25.6 \times 10^{-5} \, \mathrm{W} \, \mathrm{m}^{-3}$}
	}
	\caption{Evolution of different properties of the TNE cases within the heating parameter-space ($\lambda_{1}$, $\lambda_{2}$, and $H_1$) for the simulations with the loop~A geometry. The black areas designate simulations where we do not detect TNE cycles. Top: Period of the TNE cycles. Bottom: Time lag between $T_e$ and $n_e$ around the loop apex, displayed as a percentage of the cycle period.} 
	\label{fig:parameters_simu_loop_a}
\end{figure*}    


\begin{figure*}
	\centering
	
	\gridline{\fig{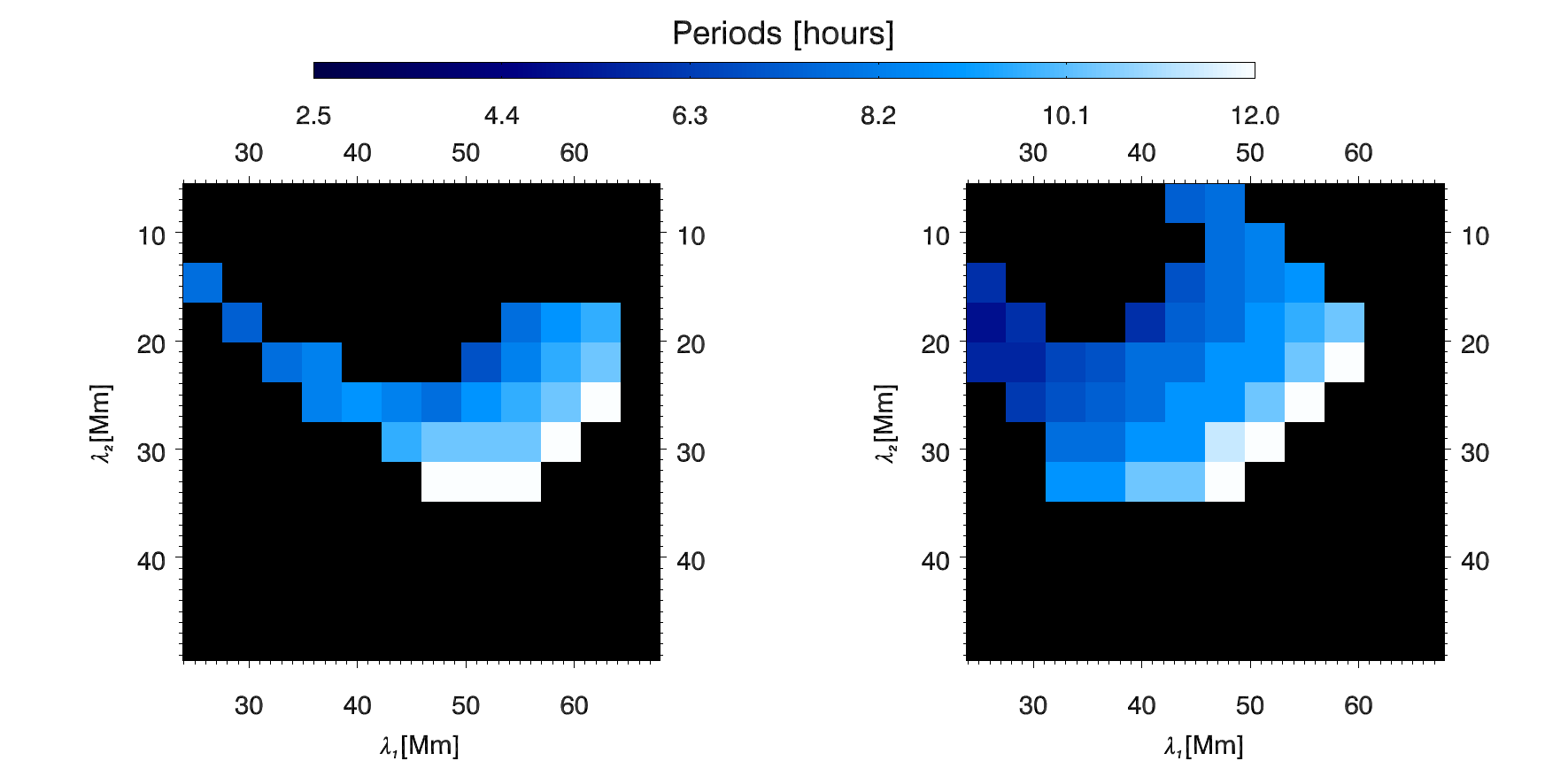}{0.65\textwidth}{$H_1=12.8 \times 10^{-5} \, \mathrm{W} \, \mathrm{m}^{-3}$; $H_1=25.6 \times 10^{-5} \, \mathrm{W} \, \mathrm{m}^{-3}$}
	}
	\gridline{\fig{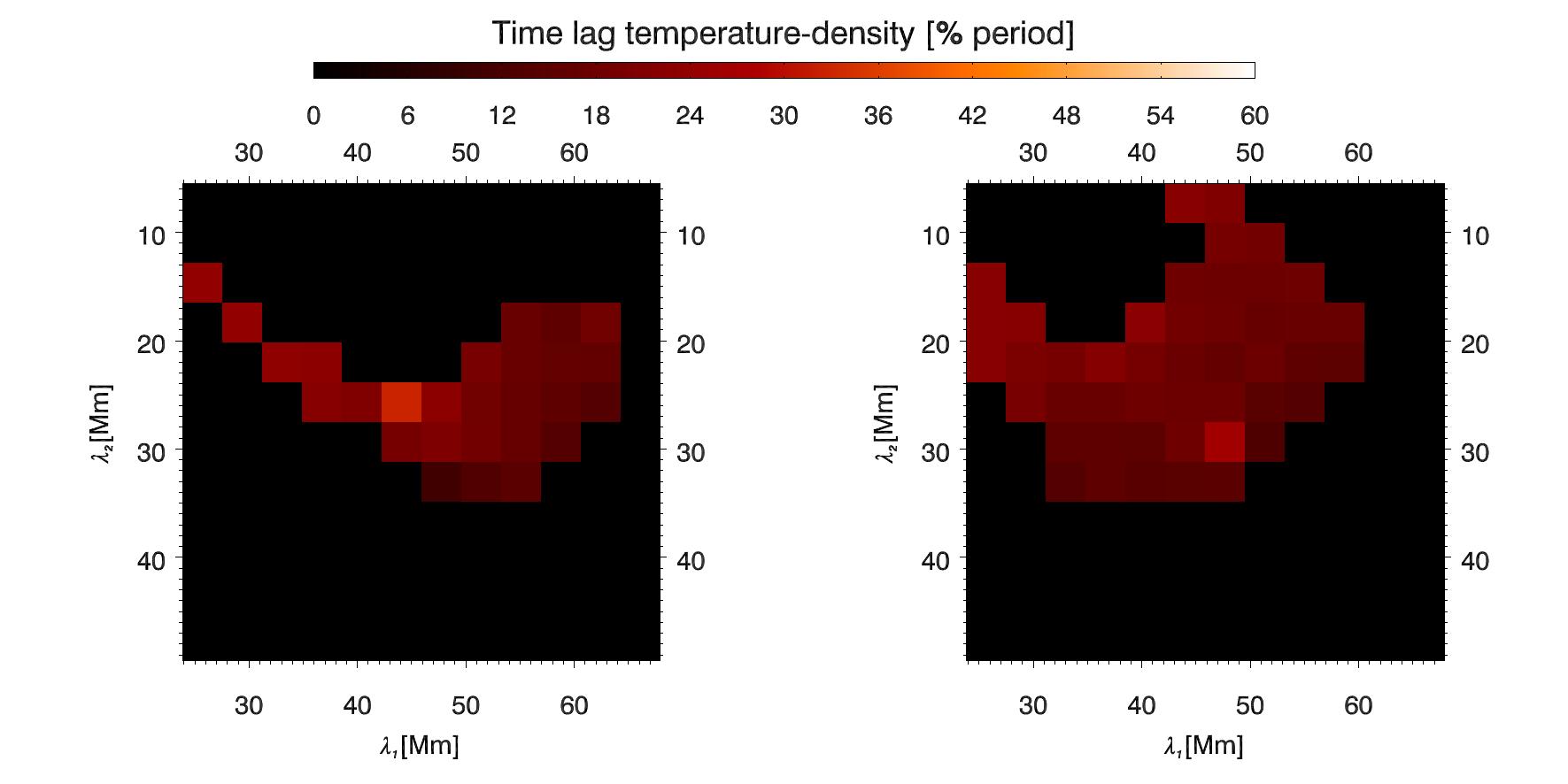}{0.65\textwidth}{$H_1=12.8 \times 10^{-5} \, \mathrm{W} \, \mathrm{m}^{-3}$; $H_1=25.6 \times 10^{-5} \, \mathrm{W} \, \mathrm{m}^{-3}$}
	}    
	
	\caption{Same as Figure~\ref{fig:parameters_simu_loop_a} for the simulations using the loop~B geometry.} 
	\label{fig:parameters_simu_loop_b}
\end{figure*}    


\begin{figure*}
	\centering
	
	\gridline{\fig{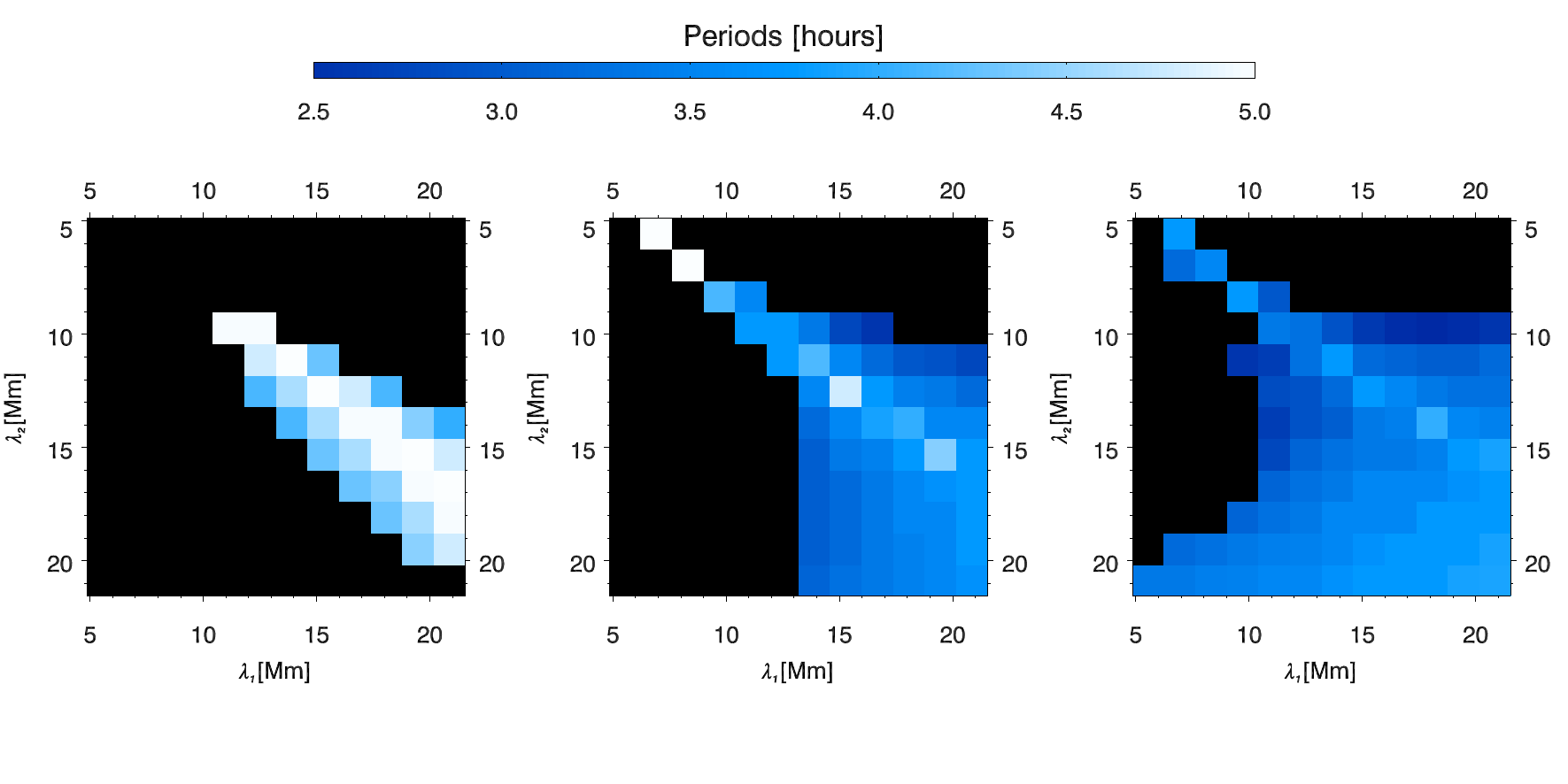}{0.65\textwidth}{$H_1=4. \times 10^{-5} \, \mathrm{W} \, \mathrm{m}^{-3}$; $H_1=8. \times 10^{-5} \, \mathrm{W} \, \mathrm{m}^{-3}$; $H_1=16. \times 10^{-5} \, \mathrm{W} \, \mathrm{m}^{-3}$}
	}
	\gridline{\fig{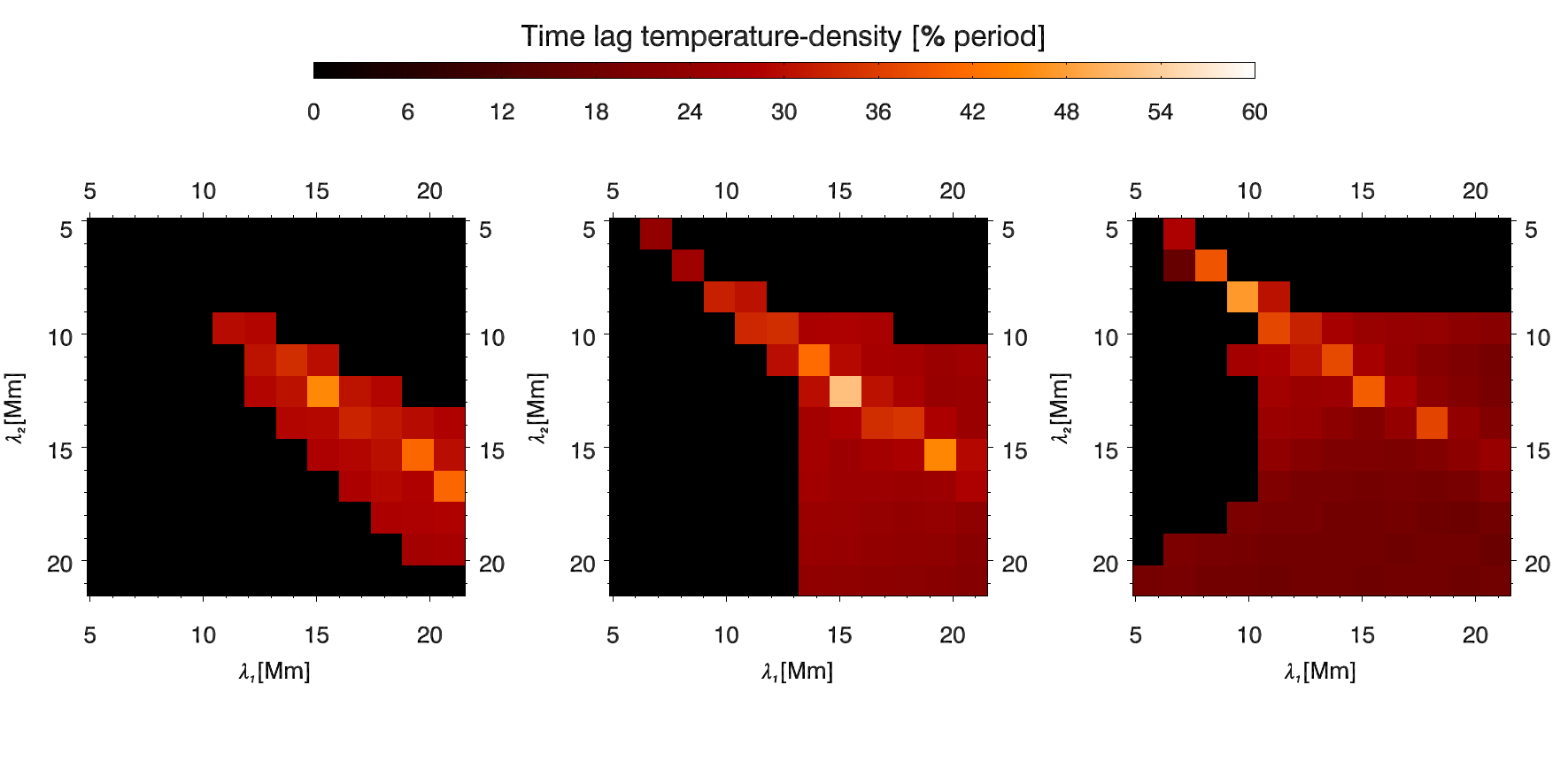}{0.65\textwidth}{$H_1=4. \times 10^{-5} \, \mathrm{W} \, \mathrm{m}^{-3}$; $H_1=8. \times 10^{-5} \, \mathrm{W} \, \mathrm{m}^{-3}$; $H_1=16. \times 10^{-5} \, \mathrm{W} \, \mathrm{m}^{-3}$}
	}	
	
	\caption{Same as Figure~\ref{fig:parameters_simu_loop_a} for the simulations using the loop~C geometry.} 
	\label{fig:parameters_simu_loop_c}
\end{figure*}    

Scanning the parameter space of heating configurations for different loop geometries, we have noticed that the distribution of the occurrence of TNE depends on the loop geometry. 
However, from this study, it seems that we are able to produce TNE-favorable conditions for any loop geometry. TNE will occur if the heating strength is sufficient to produce a loop dense enough to create a thermal runaway at high altitudes, and if this heating is deposited on specific scale heights. 

From the heating parameter-space scan that we conducted with three different loop geometries, we can conclude that a stratified heating is a necessary condition to produce TNE, but it is not sufficient, as already found by many authors \citep[e.g.,][]{muller__2004,susino2010, mikic2013}. 
For each loop geometry, the system undergoes TNE cycles for specific heating stratifications:
\begin{itemize}
	\item $\lambda_{1} \simeq \lambda_{2}$ for loop~A;
	\item $\lambda_{1} >\lambda_{2}$ for loop~B;
	\item For loop~C, we observe two behaviors:  $\lambda_{1} \sim \lambda_{2}$ when $H_1$ is small, $\lambda_{1}$ or $\lambda_{2} > $ 8.3 Mm when $H_1$ is large.

\end{itemize}

As stated before, in this paper we present only a sub-field of the parameter-space that we scanned. For each loop geometry, we also tested smaller and larger values of $H_1$ than the values presented here. However, when $H_1$ is too small, we do not reach typical coronal loops temperatures and densities and a large majority of the loops do not show any TNE cycles. When the $H_1$ are too large, the temperature of the loops is too high ($>4$~MK) compared to the warm pulsating loops observed with AIA.

For all the loop geometries, the more heating ($H_1$) is applied at the footpoints the less the system requires heating symmetry to achieve TNE cycles. In other words, higher $H_1$ leads to more TNE within the parameter space. Higher $H_1$ induces more chromospheric evaporation, which results in denser plasma at coronal altitudes, favoring the thermal runaway.
Indeed, the input heating $H_1$ has to be sufficient to inject the density required for triggering TNE events.
The loops that we are modeling in this paper are quite large (367~Mm and 139~Mm) and thus a large chromospheric evaporation is needed to injected enough density, which can explain the large $Q_0$ values required to have TNE cycles (see values in Table~\ref{table:scan_loop}).

It is worth remembering that the loop~C geometry is extracted from a field-line corresponding to a loop bundle that is not undergoing cycles in the observations. Interestingly, it is for this loop geometry that the TNE cycles have the highest probability\footnote{With respect to the explored volume of the parameter space.} to occur, according to our simulations results with the highest $H_1$ tested. If our model (quasi-steady stratified heating) is indeed correct, it would mean that we can constraint the heating for these particular loops. That would mean that loop~C, which is not showing any pulsations in the AIA data, is not heated enough at the footpoints to inject the excess of density needed in the loop bundle to trigger the thermal runaway and/or that the heating is very stratified. In the same way, loop~B is showing pulsations in the AIA data so we can guess that for this loop bundle the heating is asymmetric, stratified and relatively important.

\subsection{Exploration de-correlated from the magnetic field strength}
For each of the geometries tested, not all the stratified heating configurations lead to TNE. The area where TNE occurs is limited to some range in the heating parameter space. This leads to the question as to whether the area explored within the parameter space is realistic.
In particular, the heating is somewhat correlated to the magnetic field strength \citep[see e.g. turbulent models, in e.g.][]{rappazzo_coronal_2007} and therefore we may have explored heating parameters that are unrealistic.

On the other hand, the strength of the magnetic field along the loops does not take into account the magnetic topology, which necessarily influences the heating as well \citep[formation of separatrices, preferential reconnection sites, e.g.][]{aly_current_1997,pariat_current_2009, parnell_structure_2010, wyper_spine-fan_2012}.
The heating parameter-space scan for loop~B shows that we can produce TNE cycles for this loop geometry only with asymmetric heating profiles. This heating configuration was validated \textit{a posteriori} in Paper~I by the magnetic topology found around the eastern footpoints of this loop bundle, that can favor continuous reconnection and thus enhanced heating.

\subsection{Common characteristics of TNE events}

From the analysis of the flows, using in particular the averaged apex velocity (see Section~\ref{sec:main_simu}), we noticed that the siphon flows are more intense for short heating scale heights. Moreover, they tend to be weaker close to the TNE conditions (see Figures~\ref{fig:parameters_simu_loop_a} to~\ref{fig:parameters_simu_loop_c} ). 

We examine also some characteristics of the cycles of the TNE cases. Figures~\ref{fig:parameters_simu_loop_a}, \ref{fig:parameters_simu_loop_b}, and \ref{fig:parameters_simu_loop_c} show, for each simulation, the periods of the cycles and time lags between the temperature and density averaged around the apex. 

\paragraph{Period of the TNE cycles} We can notice an evident dependence on the loop length. The periods are from 2.5~hr to 15.5~hr for loop~A, and from 5.5~hr to 15.5~hr loop~B, which are both 367~Mm long, and from 2.4~hr to 5.9~hr for loop~C, which is 139~Mm long. This dependence has already been seen in the EIT event statistics of \citet{auchere2014}, the AIA event statistics of \citet{froment_pulsations_2016}, and in the three events of \citet{froment2015}.
For loop~A, the period of the cycles increases along the diagonal of each $H_1$-constant grid and between two grids, i.e. when $Q_0$, and thus the maximum $T_e$ at the apex, is increasing. 
For the other loops, we find the same general dependence, but the detailed behavior is more complex.
The period increases with $Q_0$ and~$L$. 

\paragraph{Time lag between $T_e$ and $n_e$}
We compute the time lag between the temperature and density evolution. This delay is also a characteristic of TNE cycles; it is a signature of TNE when combined with the periodicity. It also explains the systematic cooling pattern observed between EUV channels, the intensity peaking first in the hotter channels and then in the cooler ones \citep[e.g.,][]{viall&klimchuk2012}. In case of TNE events this observed cooling can be explained by a faster rise of the temperature than the temperature fall combined (or only if) the density is low during the heating phase compared to the cooling phase (see Section 3.2.3 in Paper~I for more details). 

This time lag is given here by the peak of the cross-correlation between the average temperature and density curves around the loop apex. We choose to display them as a fraction of the period in order to compare the cases more easily. We explore systematically positive time delays between 0-60\% of the TNE cycles period. Indeed, in our knowledge no TNE simulations have been reported to show an increasing of the density before the temperature and the signal being periodic we avoid in that way to detected spurious negative time lags between $T_e$ and $n_e$.
For loop~A, we notice that there are very long delays when there are very strong CCs. In some cases this delay is close to the period (see the notes in Table~\ref{table:scan_loop}). Moreover, for some CC cases the shape of the temperature and density curves are very different which leads to poor cross-correlations values and underestimated time lags (not catching the strongest density peak).
For loops~B~and~C the time lags tend to be maximum close to symmetric heating cases, otherwise becoming quite uniform within the TNE domain, i.e. about 20-30\% of the period which is what was observed in \citet{froment2015}.


\begin{table*}
	
	\caption{Summary of the main characteristics of the TNE cases from the heating parameter scan for loops~A, B, and C. The different ranges are corresponding to the ranges covered by each grid plot presented in Figures~\ref{fig:scan_loop_a} to~\ref{fig:scan_loop_c_2}. $T_e$, $n_e$ and $v$ values are the mean values around the loop apex. The percentage of IC is given compared to the number of TNE cases for each group of simulations.}
	\label{table:scan_loop}
	\centering
	\begin{tabular}{c c c c c c c c c c}
		\hline\hline
		& L & $H_1$  & $Q_0$ &$<T_{e}>$ &$<n_{e}>$ &$<v>$ & Periods & Time lag $T_{e}$-$n_{e}$ & $\%$ IC   \\
		& [Mm] & $10^{-5} [\mathrm{W} \, \mathrm{m}^{-3}]$ & $10^{4} [\mathrm{W} \, \mathrm{m}^{-2}$]  & [MK] &  $10^{14} [\mathrm{m}^{-3}]$ &  [$\mathrm{km} \, \mathrm{s}^{-1}$] & [hr] & $\%$ period &  \\
		\hline
		
		\multirow{3}{*}{Loop A} & \multirow{3}{*}{367} & $6. 4$  & $0.2-0.5$  & $0.7-1.4$  &  $0.3- 3.5 $ & $-22$/$+24$ & $4.6-15.5$ &  $16- 60$ \footnote{\label{note1}We saturated the delay exploration at 60\% of the period. Further analysis shown that some time lags are close to close 100\% of the period. For some other cases the time lag is underestimated because the cross-correlation technique captures only the first peak of density. It is happening in case of very strong condensations formed at the loop apex.} & 18 \\
		& & $12.8$   &$0.3 - 0.9$ & $0.5 - 2.1$   & $0.2 - 6.9 $  & $-32$/$+39$ & $2.5-15.5$  & $9-60$\footref{note1} & 13 \\   
		& & $25.6$   & $0.6 - 1.9$ &  $0.6 - 2.7$  &  $0.4 - 12.0 $ & $-68$/$+68$ &  $2.5-12.4$ & $2-60$\footref{note1} & 25 \\
		
		\hline
		
		\multirow{2}{*}{Loop B} & \multirow{2}{*}{367} & $6. 4$  & $2.2 - 7.8$  & $0.8 - 2.4 $  &  $0.9 - 4.3 $ & $-8$/$+6$ & $6.9-12.4$ &$11-33$ & 56 \\
		& & $12.8$   & $4.4-14.0$ & $1.1 - 2.8$  &  $1.4 - 6.7 $ & $-12$/$+11$ & $5.0-12.4$ & $13-26$ & 45
		\\
		
		\hline
		
		\multirow{3}{*}{Loop C} & \multirow{3}{*}{139} & $4. 0$  & $2.6 - 5.6$  & $0.8 - 1.2 $ &  $2.5 - 4.0 $ & $-3$/$+3$  & $4.0-5.6$ & $27-45$ & 0 \\
		& & $8.0$   & $2.5 - 10.7$ & $0.7 - 1.7 $   &  $2.0-6.1 $ & $-4$/$+5$  & $2.5-5.9$ & $22-52$ & 0 \\
		& & $16.0$   & $3.8 - 20.3$ & $0.8 - 2.5 $   &  $2.4-10.3 $ & $-6$/$+7$  & $2.4-4.0$ & $17-47$ & 1 \\
		
		\hline
	\end{tabular}
\end{table*}

\section{Loop behaviors in these simulations and comparison with the observations}\label{sec:comparaison_obs}


\begin{figure*}
	\centering
	
	\gridline{\fig{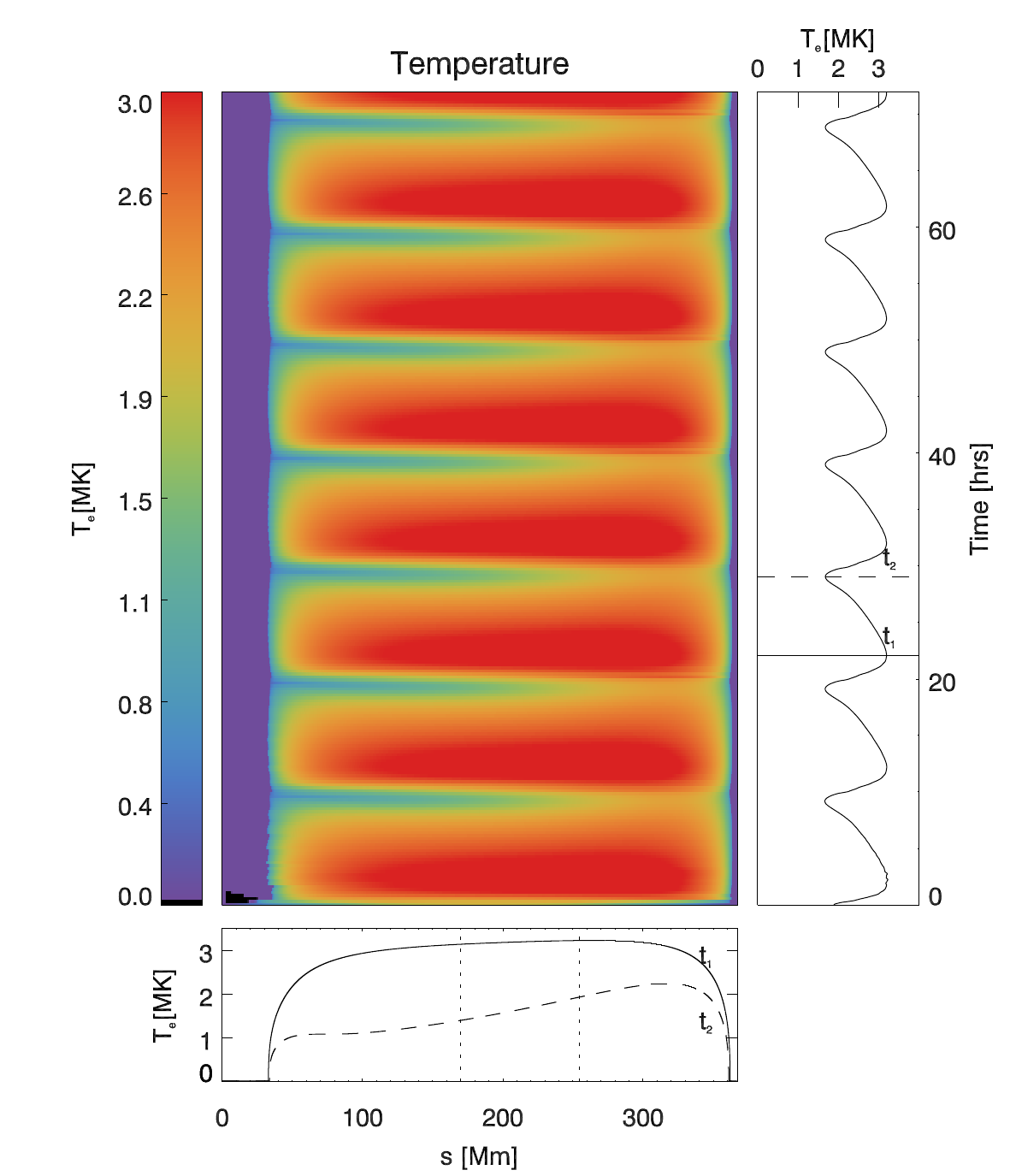}{0.28\textwidth}{}
	\fig{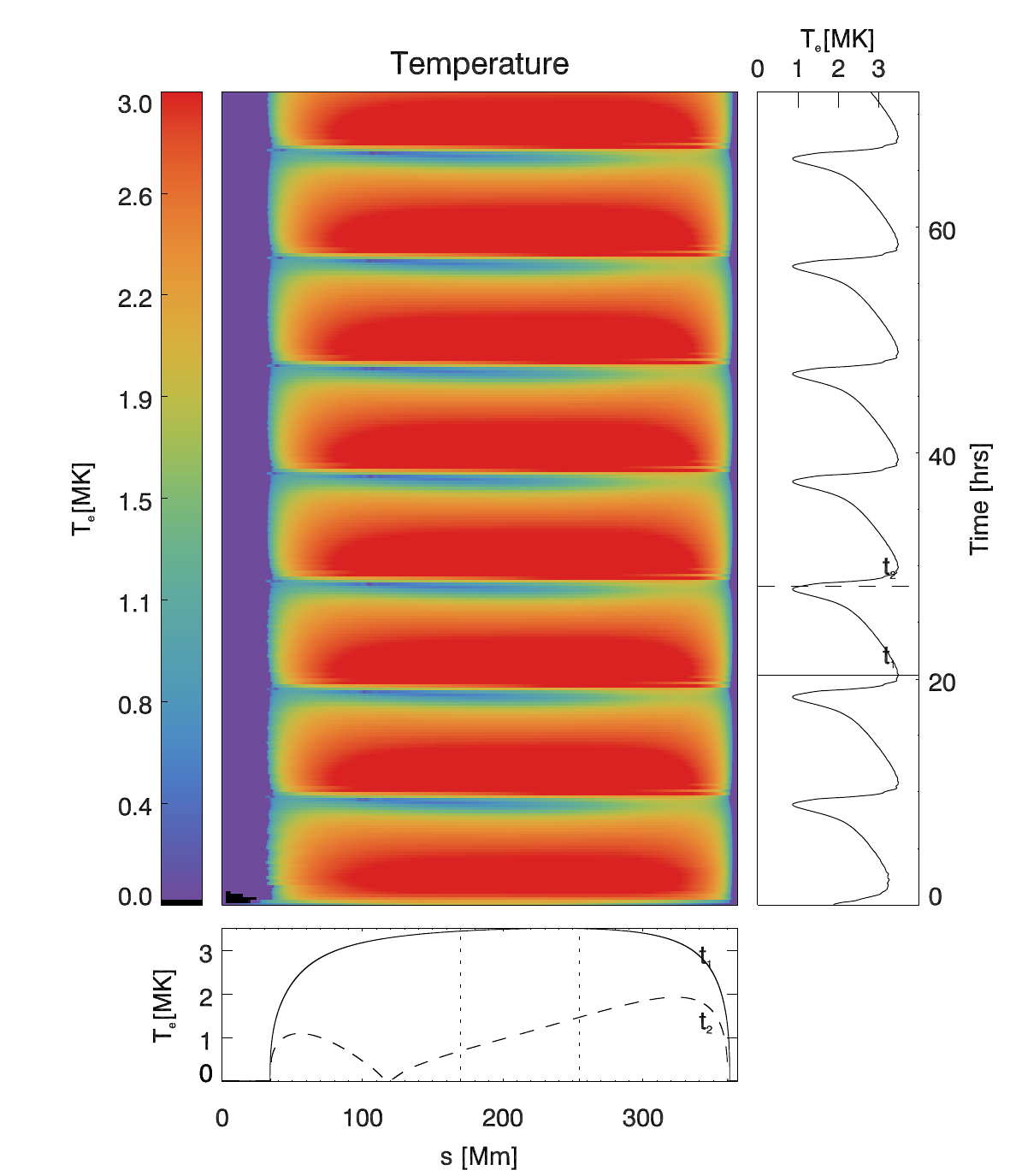}{0.28\textwidth}{}
	\fig{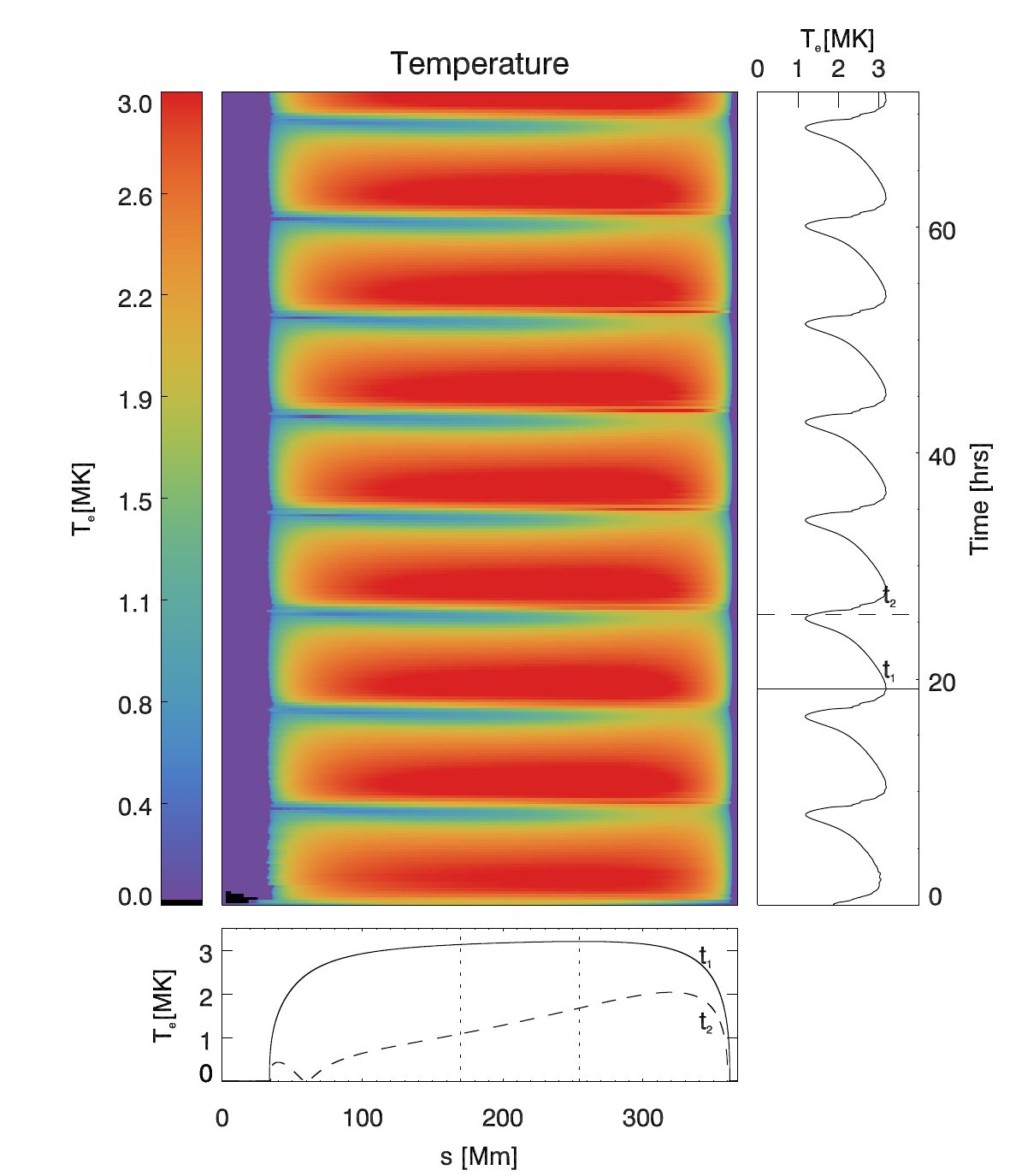}{0.28\textwidth}{}
	}
	
	\gridline{\fig{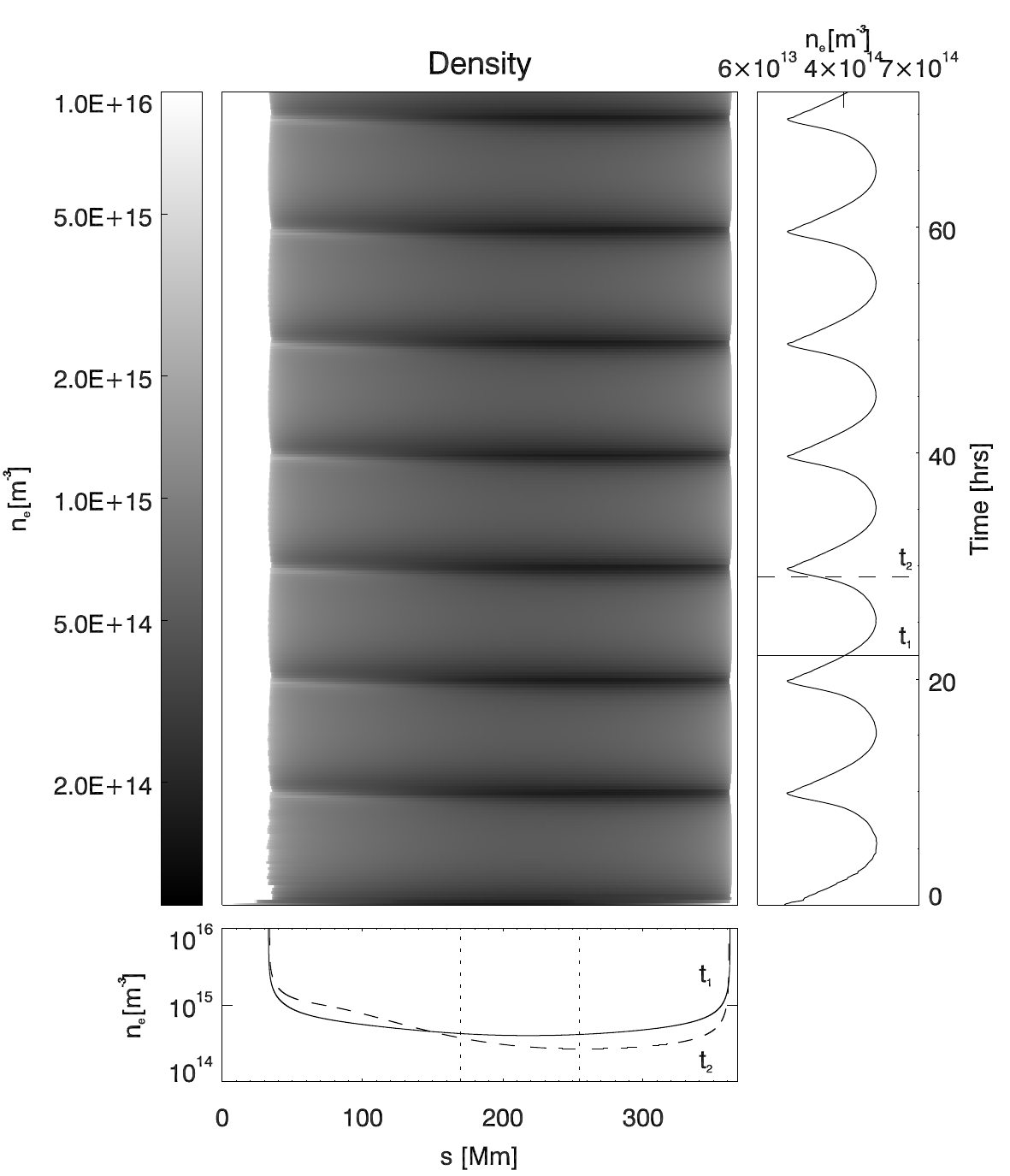}{0.28\textwidth}{}
		\fig{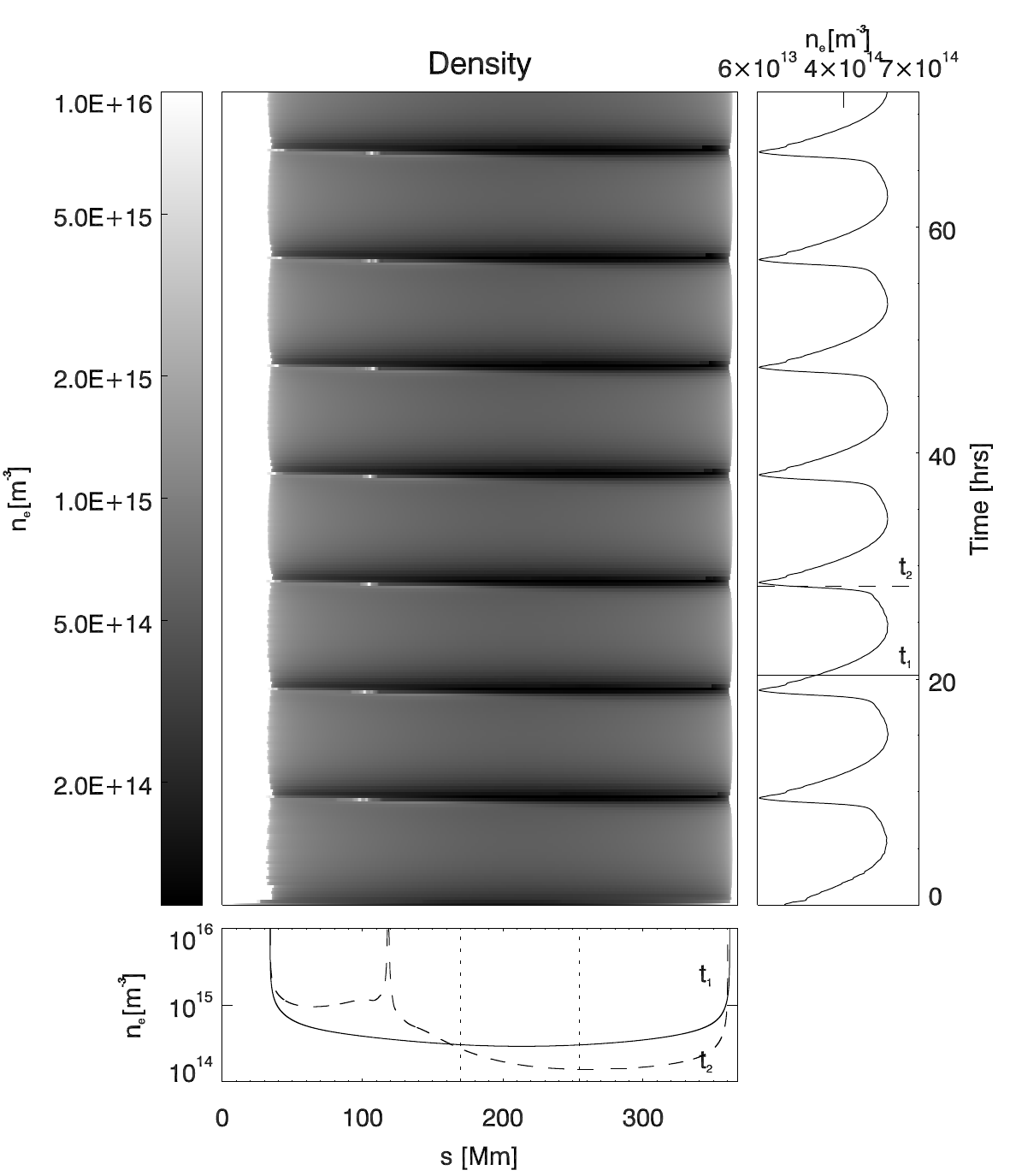}{0.28\textwidth}{}
		\fig{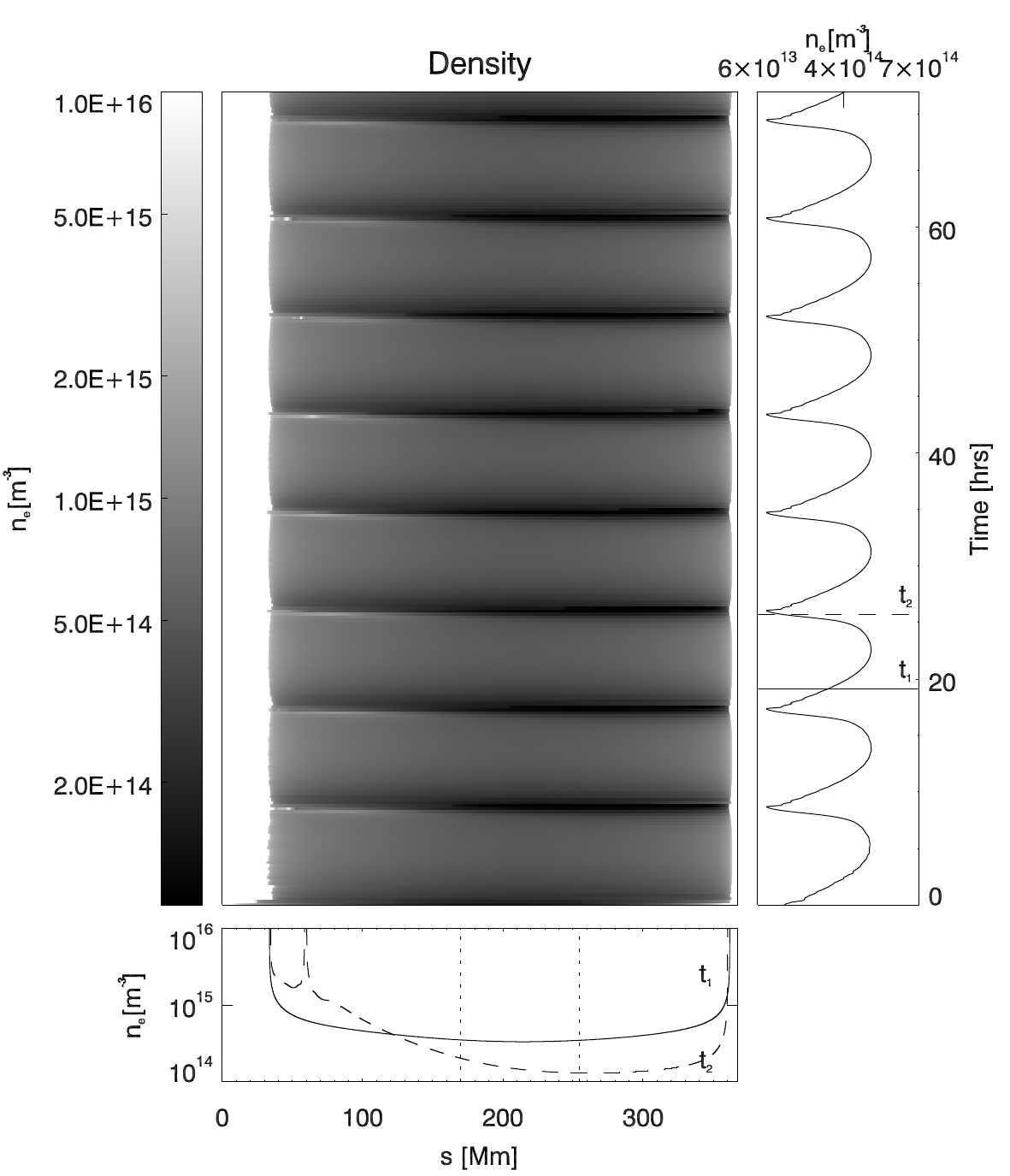}{0.28\textwidth}{}
	}
	
	\gridline{\fig{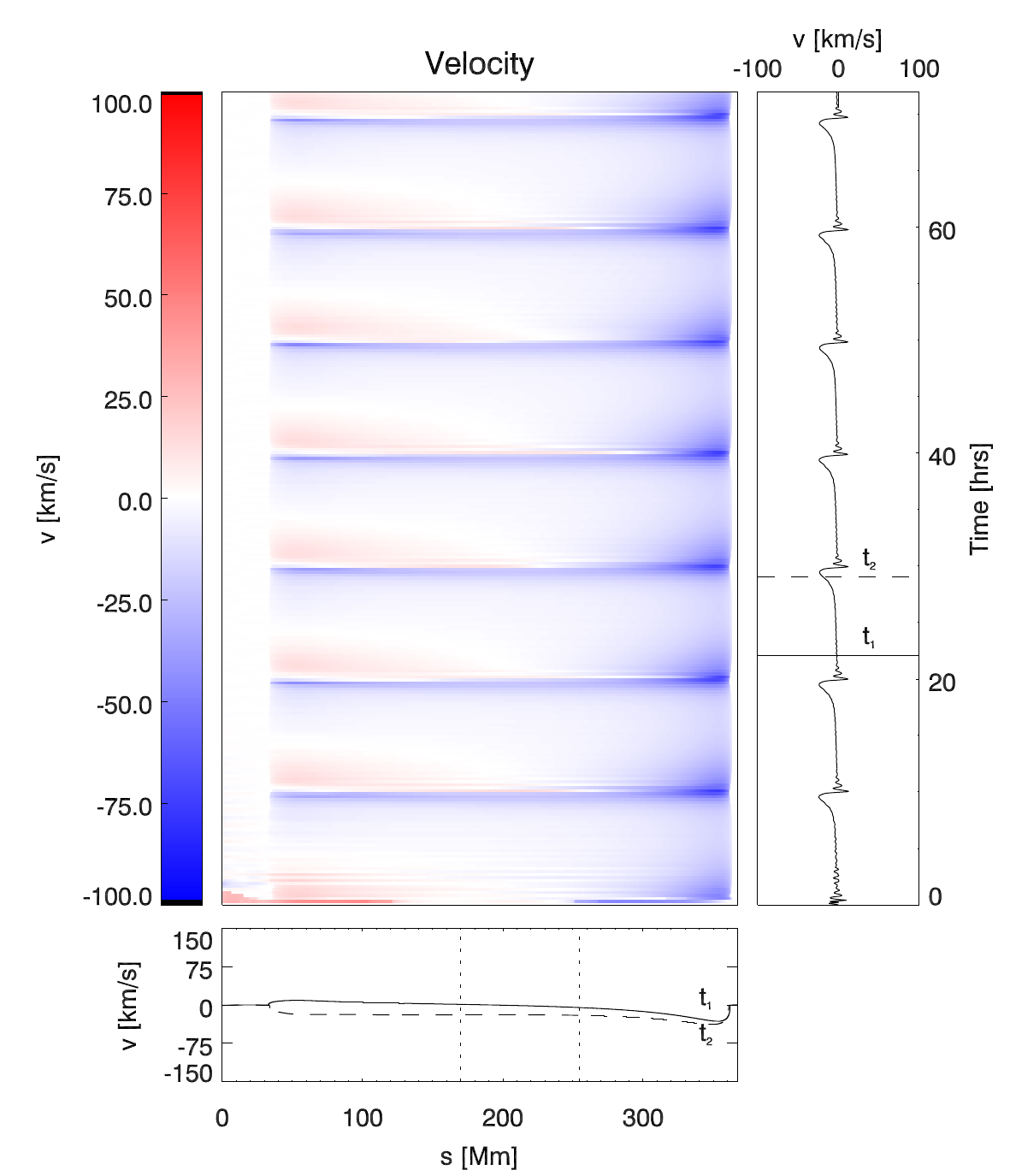}{0.28\textwidth}{IC}
		\fig{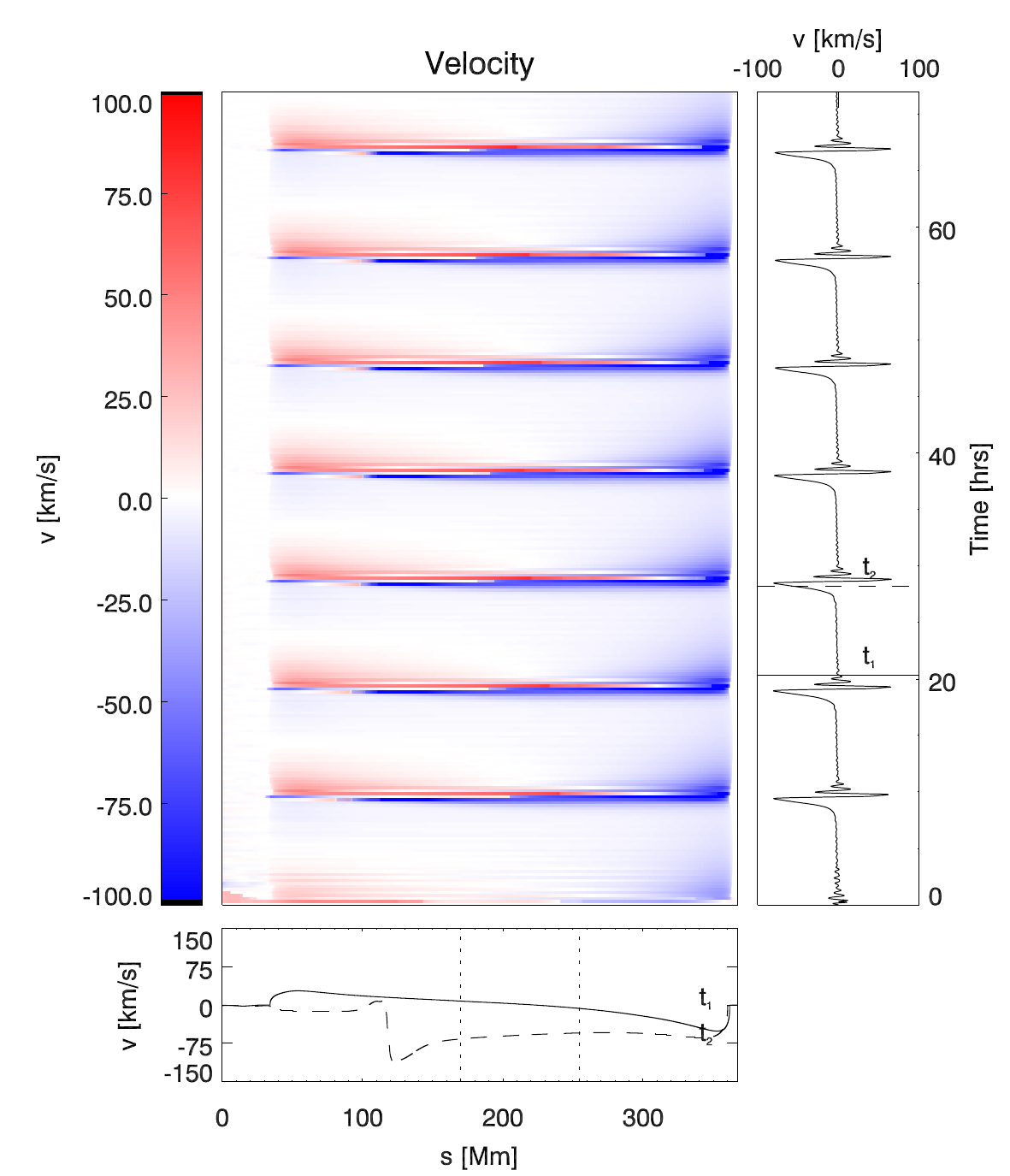}{0.28\textwidth}{CC~1}
		\fig{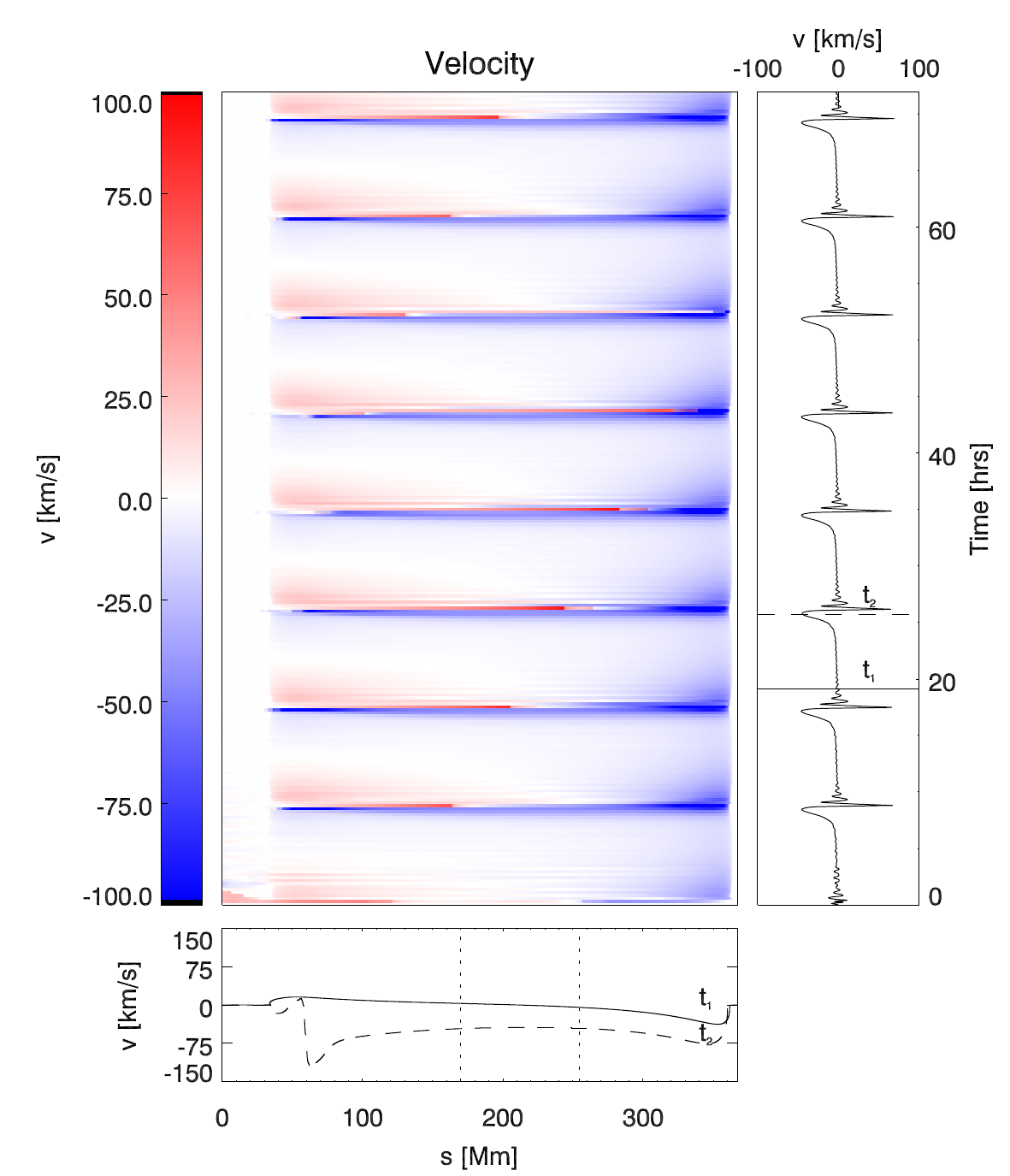}{0.28\textwidth}{CC~2}
	}

	\caption{Evolution of the temperature $T_e$, density $n_e$ and the longitudinal velocity $v$ for three simulations using the loop~B geometry. These simulations use the same heating parameters as the simulations indicated by red dots in Figure~\ref{fig:scan_loop_b_1}  ($H_1=12.8 \times 10^{-5} \, \mathrm{W} \, \mathrm{m}^{-3}$). They are repeated here using the classic Spitzer conductivity and 100, 000 mesh points. First column: Incomplete condensations simulation (IC), with $\lambda_{1}=40.4$~Mm and $\lambda_{2}=33.0$~Mm. Second column: Complete condensations simulation with $\lambda_{1}=44.0$~Mm and $\lambda_{2}=29.4$~Mm (CC 1). Third column: Complete condensations simulation with $\lambda_{1}=40.4$~Mm and $\lambda_{2}=29.4$~Mm (CC 2). Each line represents respectively the evolution of  $T_e$,  $n_e$ and $v$ along the loop during the 72 hr of the simulation (in the style of Figure~4 of Paper~I). On the right of the 2D plots, we display the evolution of respectively  $T_e$,  $n_e$ and $v$ around the loop apex (mean value between the two dotted bars in the bottom panel). On the bottom of the 2D plots, we show two  profiles (solid and dashed lines, corresponding respectively to the hot phase at $t_1$ and the cool phase at $t_2$, indicated by the solid and dashed lines on the right plots. Note that $t_1$ and $t_2$ are different for each simulations. For the velocity, the red (positive) is for flows from the eastern footpoint to the western one and the opposite for blue. On the 2D plots and the loop profiles, $n_e$ is shown in logarithmic scale. However, the apex time series $n_e$ are in linear scale.}
	\label{fig:detailed_simu}
	
\end{figure*}    


\begin{figure*}
	\centering
	\gridline{\fig{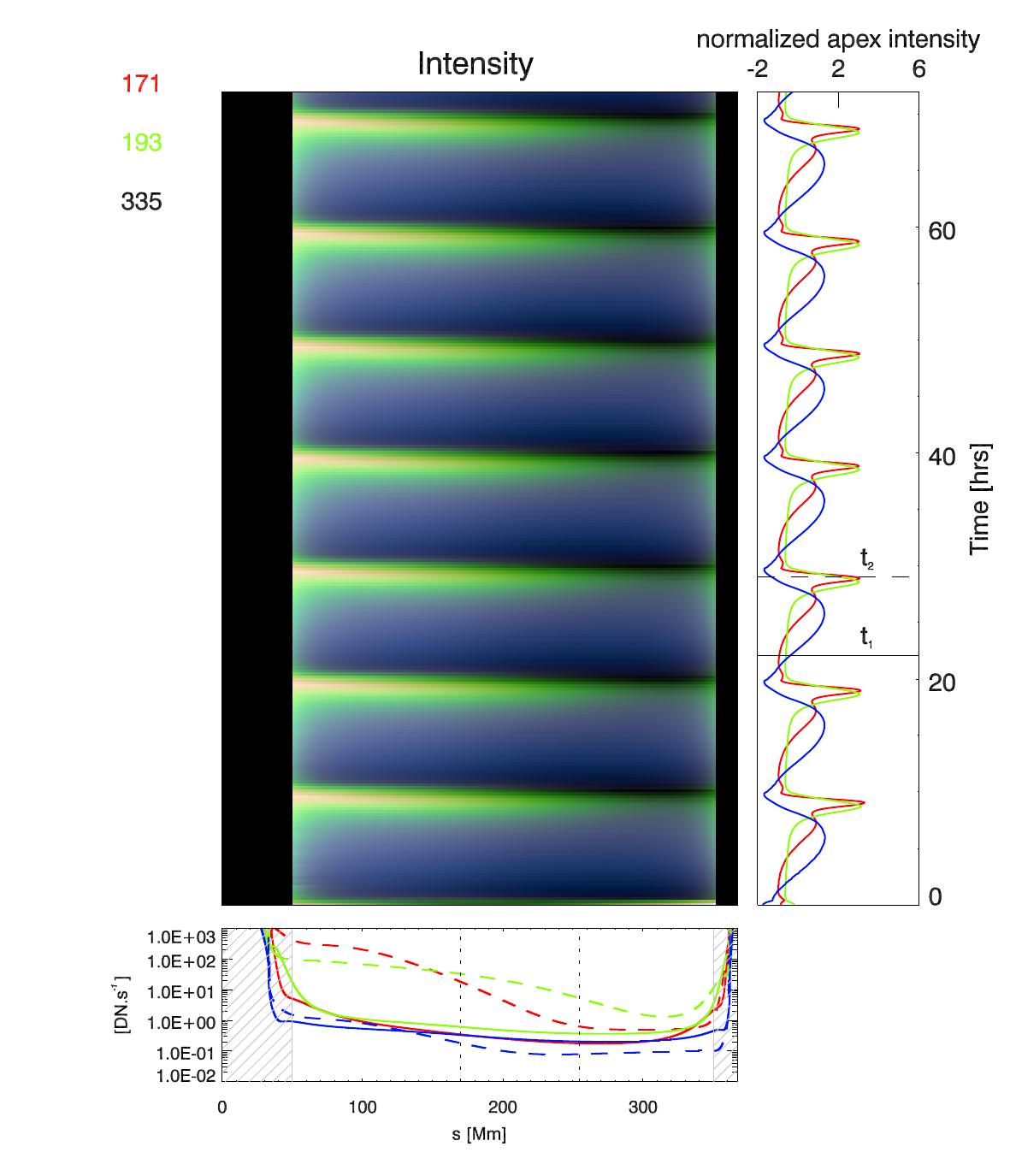}{0.45\textwidth}{IC}
		\fig{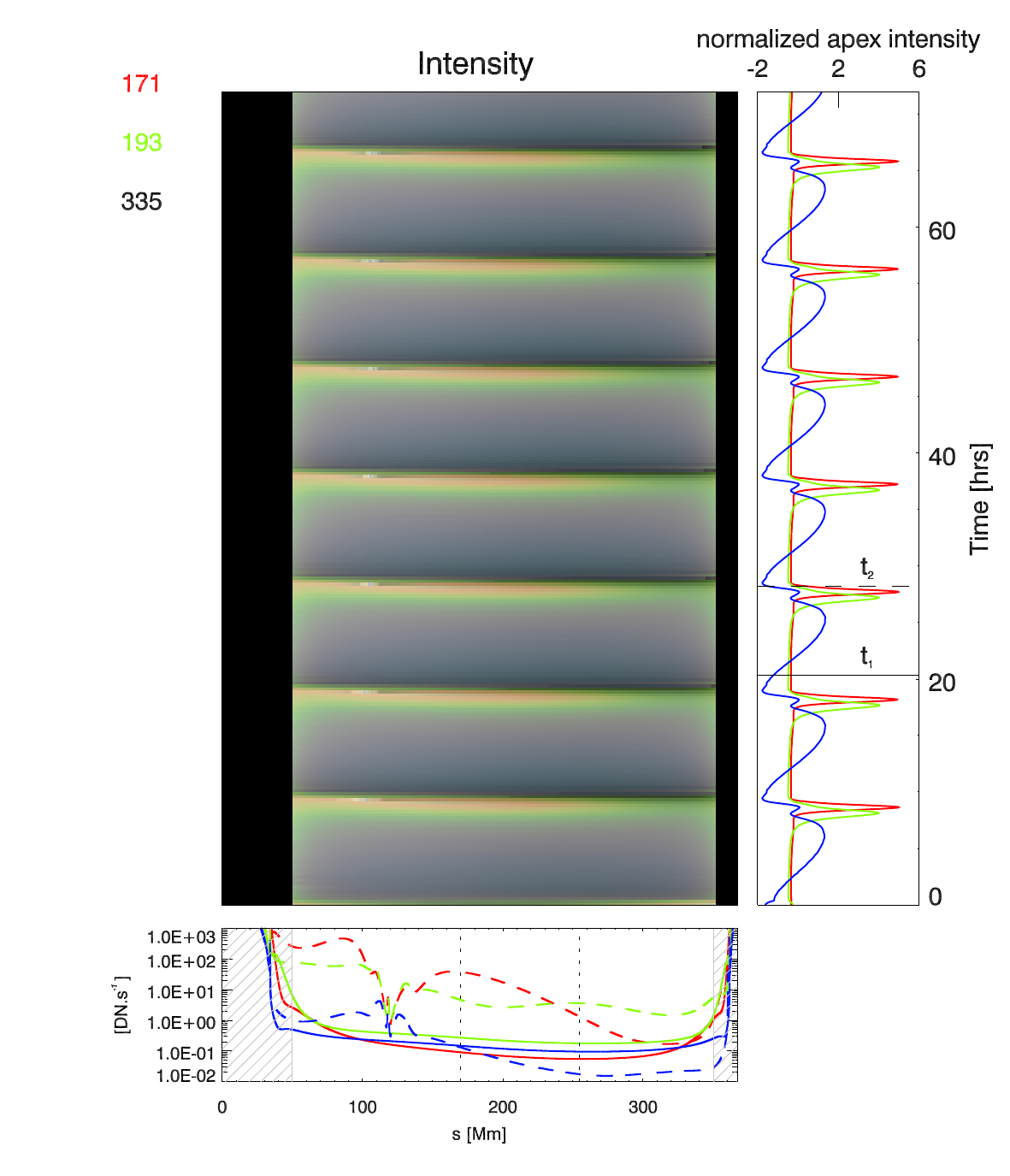}{0.45\textwidth}{CC~1}
	}
	\gridline{\fig{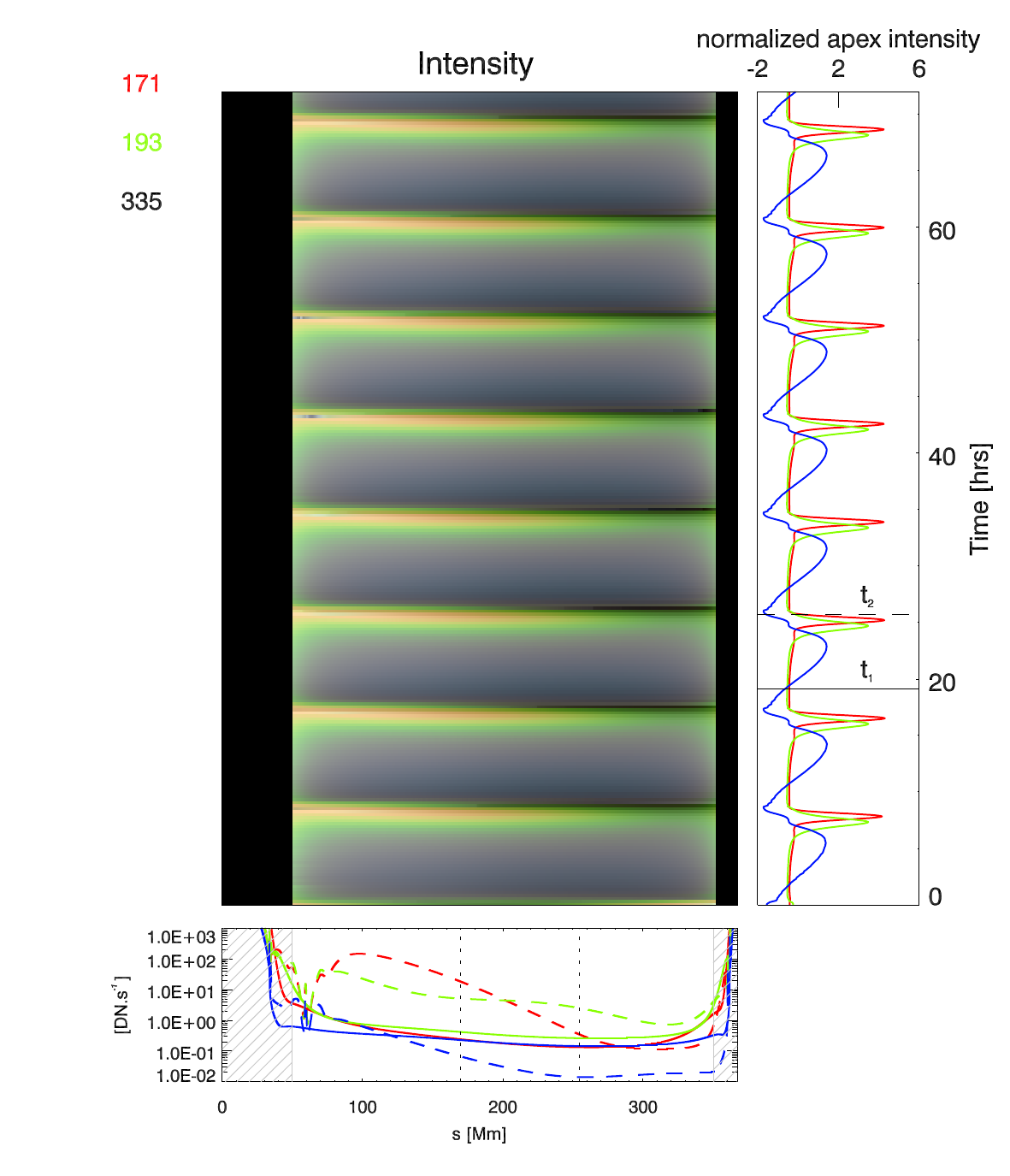}{0.45\textwidth}{CC~2}
		\fig{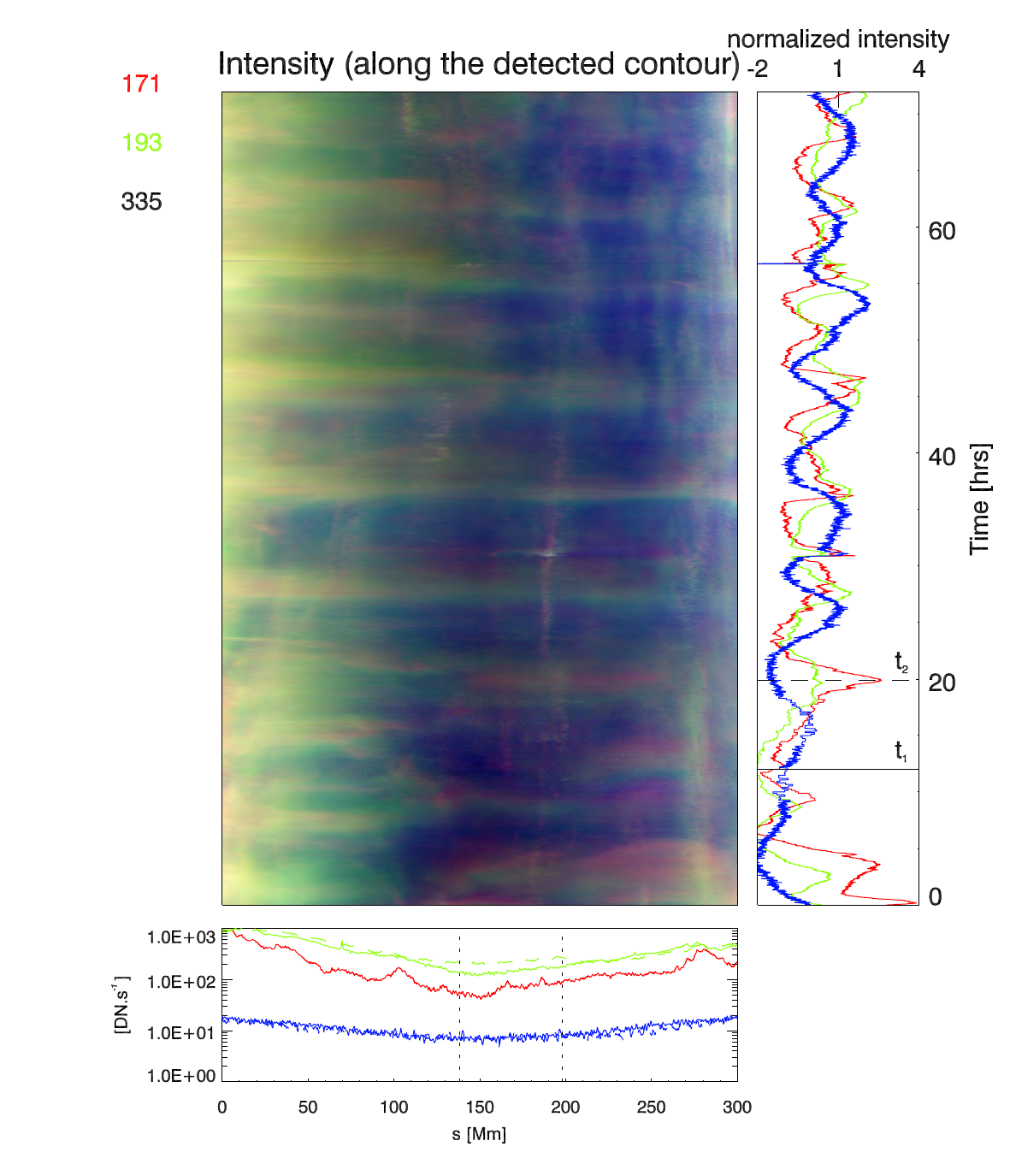}{0.45\textwidth}{AIA observations}
	}
	\caption{Comparison between the synthetic AIA intensities for the IC and CC simulations presented in Figure~\ref{fig:detailed_simu} and the observed intensity evolution along the pulsating loop bundle. The 171~\AA~channel is plotted in red, 193~\AA~in green and 335~\AA~in blue. The average intensities, normalized to variance, are plotted on the right of each 2D plot. The $t_1$ and $t_2$ profiles are plotted under each 2D plot, in the same way as in Figure~\ref{fig:detailed_simu}. The black area on the 2D plots and the gray hashed regions on the loop profiles, i.e. the parts of the loop under $s = 70$~Mm and above $s = 350$~Mm, are not considered as the simulations and the intensity calculation is only correct in the coronal part of the loop (see Section~3.2.2. in Paper~I.) The actual AIA intensities are extracted from a smoothed version of the orange contour of Figure~\ref{fig:carte_extrapol}.  We trace as well the evolution of the intensity in a portion around the loop apex (looking at the profiles along the contour), and the intensity profiles in the same way as for the simulations. Note that the range of intensities displayed is not the same as for the profiles from the simulations.}
	\label{fig:intensity_tricolor_plots}
\end{figure*}    

\subsection{EUV pulsations and coronal rain}\label{sec:condensations_details}

Figure~\ref{fig:detailed_simu} shows three TNE cases for loop~B. We display the temperature, density and velocity evolution along the loop, for three days of simulation, i.e. about 8 evaporation/condensation cycles \citep[giving thus a period close to the one detected for event~1  in][]{froment2015}.  
These simulations are extracted from Figure~\ref{fig:scan_loop_b_1} and thus correspond to $H_1 = 12.8 \times 10^{-5} \, \mathrm{W} \, \mathrm{m}^{-3}$. They are indicated by a red dot in Figure~\ref{fig:scan_loop_b_1}. They all have similar heating conditions. However, and as discussed earlier, these simulations are repeated using the unmodified Spitzer conductivity and 100,000 mesh points, as in Paper~I.

The first simulation is an incomplete condensation case for which $\lambda_{1}=40.4$~Mm and $\lambda_{2}=33.0$~Mm. Note that this simulation is not the same as the one presented in Paper~I (with $\lambda_{1}=50$~Mm and $\lambda_{2}=20$~Mm) that has the same loop geometry.
The two other simulations present complete condensations, with different locations of the condensations. For the middle simulation: $\lambda_{1}=44.0$~Mm and $\lambda_{2}=29.4$~Mm, and the condensations form close to the apex.
For the last simulation: $\lambda_{1}=40.4$~Mm and $\lambda_{2}=29.4$~Mm, and the condensations form closer to the eastern footpoint. 
At $t_1$, we display the loop profiles when the temperature reaches a maximum at the apex for one of the cycles, and at $t_2$ the profiles when the temperature reaches a local minimum. 
These three loops have a maximum apex temperature around 3~MK. During the cooling phase, when the condensations are established, the temperature drops to $\sim 1$~MK in the eastern leg of the IC simulation, while the density increases by a factor of $\sim 2$. For the CC simulations, the temperature drops locally to 0.01~MK and the density increases by a factor of 10. 

We also notice a larger velocity for CC (up to $140\, \mathrm{km} \,  \mathrm{s}^{-1}$)  compared to the IC case (about $10 \, \mathrm{km} \,  \mathrm{s}^{-1}$ at the footpoints), probably due to the increase of the density of the condensation that falls compared the density of the loop itself. The velocity is also higher when the CC start closer to the apex, due to the longer acceleration time. We notice periodic siphon flows for both the IC and the CCs cases. The ones for CCs being stronger. 

As already indicated before, around the apex, the amplitudes of the temperature (and even the density) evolution are not dramatically different for the complete and incomplete cases. The complete condensation, which is triggered in the eastern leg, outside of the apex area, has a minimal effect on the temperature evolution around the apex. It is thus not possible to use the evolution of the parameters at the apex to distinguish between complete and incomplete cases. 

In Figure~\ref{fig:intensity_tricolor_plots}, we trace the evolution of the EUV intensity as it would have been seen in the 171~\AA, 193~\AA, and 335~\AA~channels of AIA (see Section 3.2.1 of Paper~I for computation details), for these three loop systems (noted as IC, CC~1 and CC~2). As in Paper~I, we used the AIA response functions to isothermal plasma for each channels, calculated with CHIANTI version 8.0 \citep{delzanna2015}. For comparison, we add the same plot from the AIA observations. The intensity is given along the loop defined by a smoothed version of the orange contour\footnote{Note that the length of the observed loop is then a bit shorter that the length of the simulated loops (derived form LFFF extrapolations). It does not affect our analysis as we discard the synthetic intensities from the simulated footpoints.} displayed in Figure~\ref{fig:carte_extrapol}. 
However, as we already pointed out in detail in Paper~I, the synthetic intensity can only be examined in the coronal part of the loop (due to the limitations of the model, we exclude the chromosphere and  the low-transition region of the intensity analysis). We will discuss in more detail the intensity variation and values along the observed contour in the next section.

The overall pulsating behavior is well reproduced in the three simulations. In both complete and incomplete cases we can find the same global cooling pattern, with the intensity peaking first in 335, then 193 and finally 171, following the order of the peak responses of the channels. Note that we choose these three simulations because they are showing condensations to the eastern footpoint and thus matching the intensity pattern (higher intensities close to the eastern footpoint) along the observed loop bundle. The simulation presented in Paper~I was showing condensations to the western leg of the loop. This heating case was showing the most convincing intensity light curves at the apex (and directly comparable to the observed light curves) among the simulations of the parameter space explored. However the IC case chosen in the present paper is showing a more convincing pattern along the loop for the reason detailed above (asymmetry of the intensity between the two loop legs). 

Between the IC and CCs simulations, the biggest differences\footnote{We notice also smaller intensity values for the cooling phases of the CCs cases than for the ones of the IC case but with only one loop it can delicate be to focus on absolute values.} occur at the location of the triggering of the condensations, where we can see another 335 peak, corresponding to the 0.2~MK peak of that band\footnote{The peak at lower temperature (O~III to O~V lines) has been accounted in the AIA response function since February 2013 (version 4), following the measurement of \citet{soufli_band_2012}}.
However, it seems quite challenging to look for this smaller peak in AIA data, as it is probably hidden in the line-of-sight integration to distinguish between CC and IC cases.

The pulsations that we observed in \citet{froment2015} may also include co-spatial and simultaneous coronal rain events. However, it is  quite difficult to distinguish between complete and incomplete cases using only the coronal channel of AIA, as discussed in \citet{froment_long-period_2017}. It remains also difficult to conclude firmly for on-disk observations. However, the time lag between the 171 and 131 channels can help to identify the nature of the condensations even if we only have access to the mean behavior of the loop bundles. In \citet{auchere2018}, the authors detect long-period EUV pulsations coincident with coronal rain, in a region observed off-limb. In this study the 131 intensity peaks after 171, which was not the case for the events studied on-disk in \citet{froment2015}. We found no time lag between these channels, which indicates that the temperature of the plasma did not decrease on average below the peak response of 171 (around 0.8~MK).

\subsection{Are all the results of these simulations realistic?}

We have previously seen that some of the TNE cases can reproduce very well the average behavior observed with AIA, in case of long-period intensity pulsations (see also the results of Paper~I). 
Looking beyond the TNE cases, we can ask ourselves if the non TNE cases produced in the parameter space are realistic. Only a few simulations are hydrostatic and most of the non-TNE simulations are dominated by continuous siphon flows lasting for most of the three days of the simulations. For these case the simulated intensity would not show any temporal variations, which is inconsistent with observed EUV loops.
In this regard, we have to bear in mind that in this simplified analysis we have modeled the average behavior of loop systems, and have not included the variability of the heating that is likely to occur in the corona. Further studies, including an exploration of the temporal variations of the heating and/or loop geometry are needed to quantify the dynamics of these systems and their stability.

One way to check if the densities produced by our simulations are consistent with observations is to compare the observed AIA intensities with the simulated ones. In Figure~\ref{fig:intensity_tricolor_plots}, we display the intensity along loop~B for three different simulations, as it would be seen with AIA, considering a radius of the cross-section of the loop bundle of 100~km at $s=0$. The intensity values in $\mathrm{DN} \, \mathrm{s}^{-1}$ are between 0.1 and 10, except during the cooling phases. Considering that the loop intensity is about $10\%$ above the background emission \citep{del_zanna_solar_2003,viall&klimchuk2011}, this is consistent with the intensity counts derived froms AIA observations (between about 1 and 100 $\mathrm{DN} \, \mathrm{s}^{-1}$) displayed in the same Figure. The fact that we model only one loop explains as well why we can easily identify the condensations in the profiles. In contrast, in the AIA observations the difference of intensity between the heating and cooling phase profiles chosen at $t_1$ and $t_2$ is quite small. The signature of the condensations is probably hidden by the background and foreground emission.

\section{Summary}\label{summary}

In this paper, we explored a large range of dynamics, scanning different regimes of thermal non-equilibrium (TNE) and other behaviors in coronal loops.
Several parameter-space studies regarding TNE cycles have already been conducted \citep[e.g.,][]{muller__2004,susino2010, mikic2013}. Our study takes into account the recent discovery that long-period intensity pulsations are commonly observed in coronal loops.

The 1D hydrodynamic description of loops we used allows us to rapidly scan the parameter space. The model presented is rather simple, but captures the highly nonlinear dynamics of coronal loops. Indeed, we are able to nicely summarize the thermodynamic evolution of loops, even though the transition region and chromospheric behavior cannot be examined in detail.

For this extensive study we chose to explore a broad range of heating configurations, without explicitly limiting the heating profiles to a function of the magnetic field strength. We present in this paper a subset of this study, showing the results of 1,020 simulations.

We found TNE events in specific regions of the parameter space explored. 
With the different loop geometries (one semi-circular and two from LFFF extrapolations) used for the heating parameter scan, we conclude that:

\begin{itemize} 

	\item Any loop geometry seems suitable for a loop system to undergo TNE cycles.
	\item However for each loop geometry the heating requirements to obtain TNE cycles are not the same.
	\item A stratified heating is a necessary condition but it is not sufficient to produce TNE. For each loop geometry, the heating parameter domain where we obtain TNE is different.
	\item The domain where we find TNE in the heating parameter space is limited.
	\item The more the heating is important at the footpoints the more the loop is likely to undergo TNE cycles and in particular complete condensations (CCs), rather than incomplete condensations (ICs), due to the high density of the plasma injected in the loop from chromospheric evaporations.
	
\end{itemize} 

These conclusions might at first sight imply that any loop system could undergo condensation and evaporation cycles. However, this is not the case. In reality, the geometry and heating conditions vary from point to point. For a given loop, only one set of heating parameters exists. TNE is only possible when there is a specific match between the loop geometry and the heating conditions.
 
Indeed, the long-period intensity pulsations reported by \citet{auchere2014} and identified as TNE cycle by \citet{froment2015,froment_long-period_2017} are widely observed in the corona but not in every loop bundle.
There are probably many more cases of such cycles in loops, with heating conditions that change too much over time, producing more limited and irregular cycles.
The \citet{auchere2014} technique was designed to detect regular intensity pulsations, and is thus most sensitive to TNE events with stable pulsations. The detection of the probably more frequent cases in which coronal conditions evolve with time would require a different method.

Our work present several limitations, in particular: simple input heating, poor treatment in the chromosphere and the transition region, simulation of a single loop, no time dependence of the heating. However, it aims to be a first step towards the exploration of the complex parameter space we only merely touched upon. More parameters could play an important role to trigger and maintain evaporation and condensation cycles. Eventually, elaborate simulations, possibly multi-dimensional, with a proper forward modeling could help to constrain the heating of the loop observed by comparing their behavior (cycles or not, period, time lag between the temperature and the density evolution, ...) with the results of such simulations.

This extensive parameter space study allowed us as well to explore some characteristics of the TNE events.
These characteristics are summarized in Table~\ref{table:scan_loop} and Figures~\ref{fig:parameters_simu_loop_a} to \ref{fig:parameters_simu_loop_c}. We found that the period (from 2.4 to 15.5~hrs) is increasing with the length of the loop and with the maximum temperature reached. These periods tend also to be longer for CC compared to IC for the same loop geometry. The time delay between the temperature and density evolution, characteristic of TNE events when combined with the periodicity, is constant to within 20-30\% of the period for most of the simulations (strong CC cases are an exception). We found as well that some loop geometries are more favorable to CC cases (see loop~C). Moreover, looking at IC and CC cases in more details we show that both are showing siphon flows during the cooling phases. For CCs these flows are stronger. This is consistent with 2.5D magneto-hydrodynamic simulations of \citet{fang2013, fang_coronal_2015}.

To conclude,  we presented a unified picture of numerical simulations of cooling/heating in loops.
We reaffirm in particular that coronal rain and long-period intensity pulsations are two manifestations of the same phenomenon, as demonstrated observationally by \citet{auchere2018}.

 \acknowledgements
This work is an outgrowth of the work presented during the VII Coronal Loop Workshop and at Hinode 9. The authors acknowledge useful comments from attendees of these conferences. The authors would like to thank Jim Klimchuk and Patrick Antolin for fruitful discussions on thermal nonequilibrium and long-period pulsations in loops. The SDO/AIA and SDO/HMI data are available by courtesy of NASA/SDO and the AIA and HMI science teams. This work used data provided by the MEDOC data and operations centre (CNES / CNRS / Univ. Paris-Sud), \url{http://medoc.ias.u-psud.fr/}. Zoran Miki\'c was supported by NASA Heliophysics Supporting Research Grant NNX16AH03G. This research was supported by the Research Council of Norway, project number 250810, and through its Centres of Excellence scheme, project number 262622.

\bibliographystyle{aasjournal}                        
\bibliography{paper_simulation_parameter_scan_arxiv}

\end{document}